\documentclass[10pt,onecolumn]{IEEEtran}

\usepackage{amsmath}
\usepackage{amsthm}
\usepackage{amssymb}
\usepackage{graphicx}
\usepackage{subfig}
\usepackage{rotating}
\usepackage{flushend}
\usepackage{verbatim}
\allowdisplaybreaks

\newsavebox{\tempbox}

\usepackage{xcolor}

\newcommand{\logt}{\log_{2}}

\newcommand{\abs}[1]{\left| #1 \right|}

\newcommand{\defn}{\triangleq}

\newcommand{\g}[3]{g^{n}_{ #1  }\left( #2,#3 \right)}
\newcommand{\bg}[3]{\bar{g}^{n}_{ #1  }\left( #2,#3 \right)}

\newcommand{\mbf}[1]{\mathbf{#1}}
\newcommand{\mcf}[1]{\mathcal{#1}}

\newcommand{\aexp}[1]{\frac{1}{n} \logt \left| {#1} \right| }

\newcommand{\cexp}[1]{\frac{1}{n} \logt {#1} }
\newcommand{\card}[1]{\logt \left| \mcf{#1} \right| }
\newcommand{\set}[1]{\left\{ #1 \right\}}

\DeclareMathOperator{\markov}{\setlength{\unitlength}{.5cm} \begin{picture}(1,1)  \put(0,.22){\line(1,0){1}}  \put(.5,.22){\circle{.3}}   \end{picture}}

\newtheorem{theorem}{Theorem}
\newtheorem{define}[theorem]{Definition}
\newtheorem{lemma}[theorem]{Lemma}
\newtheorem{cor}[theorem]{Corollary}

\newtheorem{remark}[theorem]{Remark}
\let\oldremark\remark
\renewcommand{\remark}{\oldremark\normalfont}

\IEEEoverridecommandlockouts

\begin{document}

\title{Equal-image-size source partitioning: \\
  Creating strong Fano's inequalities for multi-terminal discrete memoryless channels}

\author{Eric Graves and Tan F. Wong
\thanks{Eric Graves is with Army Research Lab,  Adelphi, MD
  20783, U.S.A. \texttt{ericsgra@ufl.edu} }%
\thanks{Tan F. Wong is with Department of Electrical and Computer Engineering,
 University of Florida,
 Gainesville, FL 32611, U.S.A. 
\texttt{twong@ufl.edu}}%
\thanks{This research was presented in part at the \emph{2014 IEEE International Symposium on Information Theory} in Honolulu, HI, USA. T. F. Wong was supported by the National Science Foundation
under Grant CCF-1320086. Eric Graves was supported by a National
Research Council Research Associateship Award at Army Research Lab.}
}

\maketitle

\begin{abstract}
  This paper introduces equal-image-size source partitioning, a new tool for analyzing channel and joint source-channel coding in a multi-terminal discrete memoryless channel environment. Equal-image-size source partitioning divides the source (combination of messages and codewords) into a sub-exponential number of subsets. Over each of these subsets, the exponential orders of the minimum image sizes of most messages are roughly equal to the same entropy term.  This property gives us the strength of minimum image sizes and the flexibility of entropy terms.  Using the method of equal-image-size source partitioning, we prove separate necessary conditions for the existence of average-error and maximum-error codes. These necessary conditions are much stronger than the standard Fano's inequality, and can be weakened to render versions of Fano's inequality that apply to codes with non-vanishing error probabilities. To demonstrate the power of this new tool, we employ the stronger average-error version of Fano's inequality to prove the strong converse for the discrete memoryless wiretap channel with decaying leakage, which heretofore has been an open problem.  \end{abstract}


\section{Introduction}

A minimum $\mu$-image of a set $A \subseteq \mcf{X}^n$ over a discrete memoryless channel (DMC), specified by the conditional distribution $P_{Y|X}$, is the smallest set $B \subset \mcf{Y}^n$ such that $P_{Y|X}^n(B|x^n) \geq \mu$ for all $x^n \in A$. 
For any $\epsilon$-maximum error, $n$-length code over $P_{Y|X}$, the decoding subset of $\mcf{Y}^n$ for a particular message value must constitute a $\mu$-image of the subset of $\mcf{X}^n$ corresponding to the message value, for some small $\mu$.
Noting this, the intuitive sphere packing argument for channel capacity naturally extends by interpreting the minimum $\mu$-image as the ``sphere'' of the smallest size mapped to from  the codewords of an $\epsilon$-error code (see section~\ref{sec:background_image} for more details). 
Expressing capacity results in terms of minimum image sizes has many advantages, such as allowing for expressions of channel capacity as a function of~$\epsilon$. 
Furthermore because images sizes are not functions of the distribution of $X^n$, they are apt for use in joint source-channel problems for which messages may not be uniformly distributed.
Unfortunately there are also significant drawbacks to analysis by minimum image size. 
For instance, there is no currently known method by which to calculate the minimum image size of any arbitrary set other than a singleton. 
This is perhaps why it is common to instead employ ``spheres'' whose sizes can be expressed in terms of entropies in the sphere packing argument. Entropies allow for simple algebraic manipulations and hence lead to simple representations of the capacity of many basic channels.
These two different types of characterizations are often referred to as image size characterization and entropy size characterization, and the sets of possible image size characterizations and entropy size characterizations are referred to as the achievable exponent and achievable entropy regions, respectively. 
In order to take advantage of image size characterizations, we need to express minimum image sizes in terms of entropies. As Csisz{\'a}r and K{\"o}rner note in \cite[p.~339]{CK} though 
\begin{quote}
\emph{We shall see in Chapter 16 that the corresponding image size characterizations can be used to prove strong converse
results for source networks and also to solve channel network
problems. In this respect, it is important that the sets of achievable
entropy resp. exponent triples have the same two dimensional
projections see Theorems 15.11 and 15.20. The two sets, however, need not be equal; their relationship is described by Corollary 15.13. }
\end{quote} 
The primary motivation of this work is to rectify this incongruity, and in doing so provide new stronger necessary conditions for reliable communications that have both the robustness of image size techniques while maintaining the algebraic flexibility of entropies.

In a three-terminal setting with a single message, it has been well established that the two-dimensional projections of image size characterization and the entropy characterization are equal~\cite[Theorem~15.11]{CK}. Results beyond three terminals are rare and partial. In addition, in multi-terminal settings there typically exist multiple receivers which are only required to decode a subset of the messages. In an earlier paper~\cite{graves2014equating}, we have shown that every source set may be partitioned into $\mcf{O}(n)$ subsets, within each the entropy and image size characterizations are equal.  The first significant contribution of the current paper is to extend this partitioning method to simultaneously account for multiple messages and multiple receivers. Over every partitioning subset, the image size characterization and the entropy characterization are equal in that the exponential orders of the minimum image sizes for nearly all messages are equal to the same entropy quantity. Furthermore, the partition results in the distribution of the messages being nearly uniform over every partitioning subset, while the number of partitioning subsets remains polynomial in $n$ $(\mcf{O}(n^5))$.

Our second significant contribution, new necessary conditions for reliable communications over multi-terminal DMCs, then follows. These necessary conditions 
(see Theorems~\ref{thm:maxerr_fano} and~\ref{thm:isfd}) 
are direct consequences of the equal-image-size partitions described above. More specifically, by the blowing up lemma~\cite[Ch.~5]{CK}, the exponential order of the minimum image size is effectively invariant to the value of $\epsilon$.  Due to the equality between image size and entropy characterizations by our partitioning approach, the entropy terms in the sphere packing argument for codes with small error probabilities are nearly equivalent to those for codes with larger error probabilities.  This suggests that the necessary conditions of reliable communications expressed in terms of these entropies may be made effectively invariant to the decoding error probabilities.  Another way to look at these necessary conditions is that they imply all codes may only increase their rates by allowing transmissions which have nearly zero probability of decoding. Errors of this type have previously been considered by Effros et al. in regards to composite channels, where the probability of an error of this type occurring was deemed the outage probability~\cite{effros2010gen}.

From our new necessary conditions we may obtain more traditional, stronger versions of Fano's inequality. The strong inequalities with regards to average probability of error work for nearly uniform messages (see Corollary~\ref{cor:uniform}) and information-stable messages (see Corollary~\ref{cor:stable}), while the maximum-error version (see Corollary~\ref{cor:max_fano}) applies universally. We deem these particular results as strong Fano's inequalities because we may write them in the form of the standard Fano's inequality except for the error term being replaced by a term which almost universally vanishes. Much of the complexity in regards to this paper revolves around crafting necessary conditions which are easy to apply, and apply directly to many active research problems. To demonstrate the power of the results, we present as an application example a simple solution to the strong converse problem for the discrete-memoryless wiretap channel (DM-WTC) with vanishing leakage, which heretofore has been an open problem. 

We organize the rest of the paper as follows. Background on the methods used and similar approaches will be discussed first in section~\ref{sec:background}. A preview of our main results will be provided in section~\ref{sec:preview} with an example showing application of the strong average-error Fano's inequality to prove the strong converse for the DM-WTC. 
The mathematical machinery that we employ to establish equal-image-size source partitioning will be developed in sections~\ref{sec:partition} and~\ref{sec:lemmas}. The proposed equal-image-size source partition will be developed in section~\ref{sec:mt}. The new necessary conditions for reliable communications and strong Fano's inequalities will come in section~\ref{sec:fanos}. Finally we will conclude this paper in section~\ref{sec:conclusion} with a brief list of some basic multi-terminal DMCs to which our results immediately apply.

\subsection{A note on notation}\label{sec:notation}
The notation used in this paper mostly follows that employed in~\cite{CK}, except for example the mutual information between a pair of random variables $X$ and $Y$ is written in the more common notation of $I(X;Y)$. Moreover, the notation for conditional entropy will be slightly abused
throughout the paper. Within, when a quantity such as $H(Y^n | X^n \in
A)$ is expressed it will mean $H(Y^n | E=1)$, where $E$ is an indicator random
variable taking the value $1$ if $X^n \in A$ and $0$ if not.

To simplify writing, let $[i:j]$ denote the set of integers starting at $i$ and ending at $j$, inclusively. When we refer to $\mcf{M}$ as an index set, we restrict $\mcf{M}$ to be discrete. A random index is a random variable distributed over an index set. Let $M_1, M_2, \ldots, M_J$ be $J$ random indices joint distributed over $\mcf{M}_1 \times \mcf{M}_2, \times \cdots \times \mcf{M}_J$. For any $S \subseteq [1:J]$, we write $\mcf{M}_{S}$ and $M_S$ as shorthand forms of $\text{\large $\times$}_{j \in S} \mcf{M}_j$ and $(M_j)_{j \in S}$, respectively. 

Consider a pair of discrete random variables $X$ and $Y$ over alphabets $\mcf{X}$ and $\mcf{Y}$, respectively.  For any $A \subseteq \mcf{X}^n$ such that $P_{X^n}(A)>0$, whenever there is no ambiguity we use $P_{Y^n|X^n \in A}(y^n)$ to denote $\Pr\{Y^n = y^n | X^n \in A\}$ for brevity.
For any $\eta \in [0,1]$, a set $B \subseteq \mcf{Y}^n$ is called an \emph{$\eta$-image} of $A \subseteq \mcf{X}^n$ over the DMC $P_{Y|X}$ \cite[Ch. 15]{CK} if $P^n_{Y|X}(B| x^n) \geq \eta$ for every $x^n \in A$. On the other hand, $B$ is called an \emph{$\eta$-quasi-image} of $A$ over $P_{Y|X}$ \cite[Problem~15.13]{CK} if $P_{Y^n | X^n \in A}(B) \geq \eta$.  The minimum size of $\eta$-images of $A$ over $P_{Y|X}$ will be denoted by $g^n_{Y|X}(A,\eta)$, while the minimum size of $\eta$-quasi-images of $A$ over $P_{Y|X}$ will be denoted by $\bar g^n_{Y|X}(A,\eta)$.


\section{Background}\label{sec:background}

\subsection{Fano's inequality}

Fano's inequality is one of the most widely used inequalities in the field of information theory. First appearing in Fano's class notes~\cite{fano1952class}, the inequality can be used to relate the entropy of a message $M$, distributed over an index set $\mcf{M}$, conditioned on a reconstruction $\hat M$ with the probability of error of that reconstruction $\epsilon$. The exact inequality
\[
\frac{1}{n}H(M|\hat M) \leq \frac{\epsilon}{n} \card{M} + \frac{1}{n},
\]
can be tight for specific $M$, $\hat M$, and $\epsilon$. It is most commonly used in proving converses of coding theorems, where when combined with the data processing inequality~\cite[Lemma~3.11]{CK}, results in
\[
\frac{1}{n} H(M) \leq \frac{1}{n} I(M;Y^n) + \frac{\epsilon}{n} \card{M} + \frac{1}{n}.
\]
We then can say if $\epsilon \rightarrow 0$ and $\aexp{\mcf{M}}=R$ is a finite constant, $\frac{1}{n} I(M;Y^n)$ asymptotically upper bounds $\frac{1}{n} H(M)$. In channel coding problems, the message $M$ is uniform and so $\frac{1}{n} I(M;Y^n)$ asymptotically upper bounds the code rate $R$.
Fano's inequality also works in joint source-channel coding problems, as is used in proving the source-channel separation theorem for the two-terminal DMC~\cite[Pg.~221]{CT}. The most general form of Fano's inequality to date is due to Han and Verd{\'u}~\cite{han1994generalizing}, who removed the constraint that at least one of the random variables involved in the inequality be discrete.

As Wolfowitz first showed, even with a non-vanishing decoding error probability, the upper bound on the rate of messages that can be transmitted through a two-terminal DMC is asymptotically equal to that with a vanishing error probability~\cite{wolfowitz1957}. Wolfowitz introduced the concept of capacity dependent upon error, usually denoted by $C(\epsilon)$. Following the terminology of Csisz{\'a}r and K{\"o}rner~\cite[Pg.~93]{CK}, a converse result showing $C(\epsilon) = \lim_{\epsilon' \rightarrow 0} C(\epsilon')$ for all $\epsilon \in (0,1)$ is called a \emph{strong converse}. Verd{\'u} and Han~\cite{verdu94} showed the stronger assertions that this is true for all finite $n$, and that all rates larger must have error probability approaching unity hold for all two-terminal DMCs. 

Clearly though the bound in Fano's inequality is influenced by the probability of error $\epsilon$.  This dependence makes Fano's inequality ill-suited for application to channel codes with non-vanishing error probabilities. This in turn has lead to other different methods of proving strong converses, such as the meta-converse proposed by Polyanskiy et al.~\cite{polyanskiy2010channel}. The meta-converse leverages the idea that any decoder can be considered as a binary hypothesis test between the correct codeword set and the incorrect codeword set. Bounding the decoding error by the best binary hypothesis test, new bounds, which are relatively tight even for small values of $n$, can be established.  In contrast to the original version of Fano's inequality, the stronger versions presented in Corollaries~\ref{cor:max_fano},~\ref{cor:uniform}, and~\ref{cor:stable} directly apply to codes with non-vanishing decoding error probabilities over multi-terminal DMCs.

Fano's inequality is also problematic when used in regards to characterizing joint source-channel coding (JSCC) problems. Using Fano's inequality for JSCC problem necessitates either the restriction of vanishing error probabilities, or that messages (sources) whose probability exponents converge to the sources' entropy rates. Both of these restrictions are limiting as results by Kostina et al.~\cite{Kostina15} suggest that allowing non-vanishing error probabilities in conjunction with compression may lead to increased rates. In contrast to the original version of Fano's inequality, the necessary conditions supplied by Theorems~\ref{thm:maxerr_fano} and~\ref{thm:isfd} can be used to upper bound such rate gains in JSCC problems over multi-terminal DMCs.

\subsection{Image size characterizations}\label{sec:background_image}
Image size characterizations, originally introduced in G{\'a}cs and K{\"o}rner~\cite{gacs73} and Ahlswede et al.~\cite{Ahlswede76image}, are of particular importance for DMCs due to the blowing up lemma~\cite[Ch. 5]{CK}. 
Margulis~\cite{Margulis74} first introduced the blowing up lemma to study hop distance in hyper-connected graphs. 
In the context of DMCs, it can be used to show that any $\alpha_n$-image with $\alpha_n$ not decaying too fast is close in size to a $\beta_n$-image with $\beta_n$ not approaching unity too fast (see \cite[Lemma~6.6]{CK} or Lemma~\ref{lem:6.6}). 
Ahlswede~\cite{Ahlswede76} used the blowing up lemma to prove  a local strong converse for maximal error codes over a two-terminal DMC, showing that all bad codes have a good subcode of almost the same rate. 
Using the same lemma, K{\"o}rner and Martin~\cite{KM77} developed a general framework for determining the achievable rates of a number of source and channel networks. On the other hand, many of the strong converses for some of the most fundamental multi-terminal DMCs studied in literature were proven using image size characterization techniques. K{\"o}rner and Martin~\cite{Korner77} employed such a technique to prove the strong converse of a discrete memoryless broadcast channel with degraded message sets. Dueck~\cite{dueck1981strong} used these methods to prove the strong converse of the discrete memoryless multiple access channel with independent messages. 

For a detailed overview of image size characterization techniques, see~\cite[Chs.~5, 6, 15, 16]{CK}.
Here we briefly summarize the sphere packing argument in~\cite[Ch. 6]{CK} to motivate the development of the results in this paper. Consider sending a uniform message $M$ from the message set $\mcf{M}$ over a two-terminal DMC specified by $P_{Y|X}$ using
a $(n,\epsilon)$-maximal error channel code $(f^n,\varphi^n)$ with $\epsilon \in (0,1)$.
For the purposes of simple discussion here, assume that the encoder $f^n: \mcf{M} \rightarrow \mcf{X}^n$ and the decoder $\varphi^n : \mcf{Y}^n \rightarrow \mcf{M}$ are both deterministic. Let $A \defn \{f^n(m) : m \in \mcf{M}\}$ denote the set of codewords used by $f^n$. 
Pick $\mu>0$ such that $\mu+\epsilon<1$ and let $B \subseteq \mcf{Y}^n$ be a minimum $(\mu+\epsilon)$-image of $A$ over $P_{Y|X}$. That is, $g^n_{Y|X}(A, \mu+\epsilon) = \abs{B}$. Let $\varphi^{-n}(m)$ denote the decoding region for the message $m \in \mcf{M}$. The 
maximum error requirement implies that $P^n_{Y|X} \left(\varphi^{-n}(m) \middle| f^n(m) \right) \geq 1 - \epsilon$ for all $m \in \mcf{M}$. Hence we have $P^n_{Y|X} \left(\varphi^{-n}(m) \cap B \middle| f^n(m) \right) \geq \mu$. In other words, this means that $\varphi^{-n}(m) \cap B$ is a $\mu$-image of the singleton $\{f^n(m)\}$, and hence $\abs{\varphi^{-n}(m) \cap B} \geq  g^n_{Y|X}(f^n(m), \mu)$ for every $m \in \mcf{M}$. It is clear now that the subsets $\varphi^{-n}(m) \cap B$ for $m \in \mcf{M}$ serve as the ``spheres'' in the sphere packing argument. More specifically,
\[
g^n_{Y|X}(A, \mu+\epsilon) = \abs{B} = \sum_{m \in \mcf{M}} \abs{\varphi^{-n}(m) \cap B}
\geq \abs{\mcf{M}} \cdot \min_{m \in \mcf{M}}  g^n_{Y|X}(f^n(m), \mu)
\]
which implies 
\begin{equation} \label{eq:dmc_rate_bound}
\aexp{\mcf{M}} \leq \cexp{g^n_{Y|X}(A, \mu+\epsilon)} - \min_{m \in \mcf{M}} \cexp{g^n_{Y|X}(f^n(m), \mu)}.
\end{equation}
As a result, we have just obtained an upper bound on the rate of the $(n,\epsilon)$-maximal error channel code in terms of minimum image sizes. Moreover as a consequence of the blowing up lemma (see \cite[Lemma~6.6]{CK} or Lemma~\ref{lem:6.6}), the terms on the right hand side of~\eqref{eq:dmc_rate_bound} remain roughly the same regardless of the value of $\epsilon$ within the range of $(0,1)$. Thus, unlike the standard Fano's inequality, this bound may be used to establish the strong converse of the DMC. 

Nevertheless usefulness of code rate bounds expressed in terms of minimum image sizes, like~\eqref{eq:dmc_rate_bound}, depends upon the availability of simple image size characterizations. As mentioned before, while such characterizations exist for the two-terminal DMC (see \cite[Ch. 6]{CK}) and the three-terminal DMC with a single message (see \cite[Ch. 15]{CK}), simple image size characterizations for more general channels have been largely missing. This motivates us to develop the proposed tool of equal-image-size source partitioning (see Theorem~\ref{thm:cond}) to solve the more general image size characterization problem and to apply this tool to obtain more general necessary conditions of reliable communications over multi-terminal DMCs (see section~\ref{sec:fanos}).

\section{Preview of main results} \label{sec:preview}
The main result of this paper is the proposed (nearly) equal-image-size  partitioning of a source simultaneously over a number of DMCs. Consider a set $A \subseteq \mcf{X}^n$ of nearly uniform sequences are mapped to $\mcf{Y}^n_k$ by the DMC $P_{Y_k|X}$ for $k \in [1:K]$. The set $A$ can be partitioned by $J$ indices $M_1, M_2, \ldots, M_J$. Then we may partition $A$ in another way into at most $n^2$ subsets. Index this new partition by $V^*$. Consider the intersection of this new partition and any old partition indexed by $M_{S}$ where $S \subseteq [1:J]$, and denote each partitioning subset in the intersection by $A_{M_S=m_S,V^*=v}$. Fixing any~$v$, the minimum image sizes of ``most'' of the partitioning subsets $A_{M_S=m_S,V^*=v}$ are approximately of the same exponential order. More specifically, $\cexp{g^n_{Y_k|X}(A_{M_S=m_S,V^*=v},\eta)} \approx \frac{1}{n} H(Y_k^n|M_S, V^*=v)$ for ``most'' $m_S$. The qualifier ``most'' above may again be quantified in terms of exponential order. The more precise statement of this source partitioning method will be developed in the following sections, culminating in the results described in Theorem~\ref{thm:cond}.

As mentioned in the previous section, one main application of image size characterizations is to find outer bounds on the capacity regions of multi-terminal DMCs. With the aid of equal-image-size source partitioning, we are able to develop strong versions of Fano's inequality for multi-terminal DMCs that do not require the decoding error probabilities to vanish. These stronger versions of Fano's inequality provide us an easy-to-use tool to find outer bounds of capacity regions for codes with non-vanishing error probabilities.  Consider the multi-terminal communication scenario in which a set of $J$ messages $M_1, M_2, \ldots, M_J$ ranging over $\mcf{M}_1, \mcf{M}_2, \ldots, \mcf{M}_J$, respectively, are to be sent to $K$ receivers through a set of $K$ DMCs $P_{Y_1|X}, P_{Y_2|X}, \ldots, P_{Y_K|X}$\footnote{Because only marginal decoding errors made at individual receivers are of concern,  the marginal conditional distributions $P_{Y_1|X}, P_{Y_2|X}, \ldots, P_{Y_K|X}$ are sufficient in specifying all such error events. As a result, we speak of ``a set of $K$ DMCs'' rather than ``the multi-terminal DMC specified by $P_{Y_1, Y_2, \ldots, Y_K|X}$.''}.  The set of possible codewords is denoted by $A$, which can be any subset of $\mcf{X}^n$. Let $S_1, S_2, \ldots, S_K$ be any $K$ non-empty subsets of $[1:J]$ with the interpretation that the $k$th receiver is to decode the message $M_{S_k}$.  Let $F^n : \mcf{M}_1 \times \dots \times \mcf{M}_J \rightarrow A$ denote the (possibly stochastic) encoding function and $\Phi^n_{k}: Y_k^n \rightarrow \mcf{M}_{S_k}$ denote the (possibly stochastic) decoding function employed by the $k$th receiver, for $k \in [1:K]$. Note that distributed encoding is allowed in this model. For example, if there are $L$ distributed encoders, each generates a codeword in $\mcf{X}_l^n$ for $l \in [1:L]$, we may set $\mcf{X}^n = \mcf{X}_1^n \times \cdots \times \mcf{X}_L^n$, $A = A_1 \times \cdots \times A_L$, where $A_l \subseteq \mcf{X}_l^n$, and disjointly distribute the $J$ messages to the $L$ encoders. Then the following two stronger versions of Fano's inequality are some of the main results that we will present in section~\ref{sec:fanos}:
\smallskip
\subsubsection*{\textbf{Strong maximum-error Fano's inequality}} 
If the encoder-decoder pairs $(F^n,\Phi_k^n)$ have
maximum errors
\begin{equation*}
\max_{(m_{S_k},x^n) \in \mcf{M}_{S_k} \times A : P_{M_{S_k},X^n}(m_{S_k},x^n) > 0}
\Pr\left\{ \Phi^n_k(Y^n_k) \neq M_{S_k} \,\middle|\, M_{S_k} = m_{S_k},X^n = x^n \right\}
\leq \epsilon < 1 
\end{equation*}
for all $k \in [1:K]$, then there exists $\mu_n \rightarrow 0$ such that
\[
\frac{1}{n}H(M_{S_k}) \leq \frac{1}{n}I(M_{S_k};Y_k^n) + \mu_n,
\]
for all $k \in [1:K]$.
\smallskip
\subsubsection*{\textbf{Strong average-error Fano's inequality}} 
If the encoder-decoder pairs $(F^n,\Phi_k^n)$ have average errors
$\Pr\{ \Phi^n_k(Y^n_k) \neq M_{S_k}\} \leq \epsilon < 1$ for all $k \in [1:K]$,
then there exist $\mu_n \rightarrow 0$, a random index $Q$ over an index set $\mcf{Q}$ with at most $\Gamma n^5$ elements for some $\Gamma >0$, and $\mcf{Q}_k^* \subseteq \mcf{Q}$ satisfying $P_Q(\mcf{Q}_k^*) \geq \frac{1-\epsilon}{4}$ 
such that $Q \markov X^n \markov Y_k^n$ and 
\begin{equation*}
\aexp{\mcf{M}_{S_k}} \leq \frac{1}{n}I(M_{S_k};Y_k^n | Q=q) +\mu_n,
\end{equation*}
for all $q \in \mcf{Q}_k^*$, as long as $M_{S_k}$ is uniformly distributed.

\medskip
\noindent Stronger and more thorough results (Theorem~\ref{thm:maxerr_fano}--Corollary~\ref{cor:stable}) than the two strong Fano's inequalities stated above will be  developed and presented in section~\ref{sec:fanos}.

\smallskip
\subsubsection*{\textbf{Application example}}
The strong converse for the general discrete memoryless wiretap channel (DM-WTC) is a heretofore open problem. The best known results were derived by Tan and Bloch~\cite{tan15wiretap} and independently by Hayashi et al.~\cite{Hayashi14}, and only pertain to the case where the wiretap channel is degraded. Such a scenario reduces the complexity by not requiring an auxiliary random variable to characterize the secrecy capacity. In particular, Tan and Bloch accomplish their result using an information spectrum approach, while Hayashi et al consider the question in regards to active hypothesis testing\footnote{It should be noted that their result simultaneously applied to both secret message transmission and secret key agreement, and allows for arbitrary leakage.}. As a simple application example for our results, we employ the strong average-error Fano's inequality to provide a strong converse for the general DM-WTC.

The DM-WTC $(\mcf{X}, P_{Y,Z|X},\mcf{Y}\times \mcf{Z})$ consists of a sender $(X)$, a legitimate receiver $(Y)$, and an eavesdropper $(Z)$. For any $R>0$, a uniformly distributed message $M$ over the message set $\mcf{M}=[1:2^{nR}]$ is to be sent reliably from $X$ to $Y$ and discreetly against eavesdropping by $Z$. For any $A \subseteq \mcf{X}^n$, consider the encoding function $F^n: \mcf{M} \rightarrow A$ and the decoding function $\Phi^n: \mcf{Y}^n \rightarrow \mcf{M}$. For any $\epsilon \in (0,1)$ and $l_n>0$, a $(n,R,\epsilon,l_n)$-code for the DM-WTC is any code $(F^n,\Phi^n)$ which meets the following two requirements:
\begin{itemize}
\item{\emph{Reliability:}}  $\Pr \left\{ \Phi^n(Y^n) \neq M \right\} \leq \epsilon$, 
\item{\emph{Leakage:}}  $I(M; Z^n) \leq l_n$.
\end{itemize}
Like \cite{tan15wiretap}, we impose the decaying leakage requirement of $\frac{l_n}{n} \rightarrow 0$.

Apply the strong average-error Fano's inequality above to the DM-WTC with the reliability requirement, we obtain an index set $\mcf{Q}$ with $\abs{\mcf{Q}} \leq \Gamma n^5$ for some $\Gamma>0$, a random index $Q$ over $\mcf{Q}$, $\mcf{Q}^* \subseteq \mcf{Q}$, and $\mu_n \rightarrow 0$ such that $P_Q(\mcf{Q}^*) \geq \frac{1-\epsilon}{4}$ and
\begin{equation} \label{eq:ex1}
R \leq \frac{1}{n} I(M; Y^n | Q = q) + \mu_n,
\end{equation}
for all $q \in \mcf{Q}^*$. Let $Q^*$ be a random index over $\mcf{Q}^*$ defined by the conditional distribution $P_{Q|Q \in \mcf{Q}^*}$. Since $Q \markov X^n \markov (Y^n,Z^n)$, we also have $Q^* \markov X^n \markov (Y^n,Z^n)$. From~\eqref{eq:ex1}, we obtain
\begin{equation} \label{eq:ex2}
R \leq \frac{1}{n} I(M; Y^n | Q^*) + \mu_n.
\end{equation}
On the other hand, from the leakage requirement 
\begin{equation*}
\frac{1-\epsilon}{4} \cdot I(M;Z^n|Q\in \mcf{Q}^*) 
\leq   I(M;Z^n|Q\in \mcf{Q}^*) P_Q(\mcf{Q}^*) 
\leq I(M;Z^n) + 1 \leq l_n + 1.
\end{equation*}
Hence
\begin{align}\label{eq:wtc_leak}
\frac{1}{n} I(M;Z^n|Q^*)
\leq 
\frac{1}{n}I(M;Z^n| Q \in \mcf{Q}^*) + \aexp{\mcf{Q}^*} 
\leq
\frac{4(l_n+1)}{(1-\epsilon)n} + \cexp{\Gamma n^5}
\end{align}
Thus, combining~\eqref{eq:ex2} and~\eqref{eq:wtc_leak} results in
\begin{equation*}
R \leq \frac{1}{n} I(M;Y^n| Q^*) - \frac{1}{n} I(M;Z^n|Q^*)+ \mu_n + \frac{4(l_n+1)}{(1-\epsilon)n} 
+\cexp{\Gamma n^5} .
\end{equation*}
Noting that $\varepsilon_n \defn \mu_n + \frac{4(l_n+1)}{(1-\epsilon)n}
+\cexp{\Gamma n^5} \rightarrow 0$,
and following the steps of~\cite[Section~22.1.2]{GNIT}, we may obtain
\begin{equation*}
R \leq I(U;Y) - I(U;Z) +  \varepsilon_n,
\end{equation*}
for some $U$ over $\mcf{U}$ with $\abs{\mcf{U}} \leq \abs{\mcf{X}}$ such that $U\markov X \markov (Y,Z)$. 
This proves the strong converse for the general DM-WTC with decaying leakage.

\section{Partitioning Index and Entropy Spectrum Partition} \label{sec:partition}

In this section, we describe the notions of partitioning index, entropy spectrum partition (slicing)~\cite{han2003}, and nearly uniform distribution. They provide the basic machinery that we will employ in later sections to develop source partitioning results. The entropy spectrum partition method that we use here is a slight variant within the class of information/entropy spectrum slicing methods developed in~\cite{han2003}. This class of methods find many different applications in information theory (see~\cite{han2003} for more detailed discussions).

While the definitions and results are stated for the sequence space $\mcf{X}^n$, they clearly extend to other sequence spaces. When we say $X^n$ is distributed over $A \subseteq \mcf{X}^n$, it is assumed with no loss of generality that $P_{X^n}(x^n) > 0$ for all $x^n \in A$. Otherwise we may just remove the zero-probability sequences from $A$.

\begin{define}\label{def:blocks}
  Let $A \subseteq \mcf{X}^n$ and $\mcf{M}$ be an index set. Let $X^n$
  and $M$ be jointly distributed random variables over $A$ and
  $\mcf{M}$, respectively.  For each $m \in \mcf{M}$, define
  $A_{M=m} \defn \{ x^n \in A: P_{M,X^n}(m,x^n) > 0\}$. 
  Then $M$ is called a partitioning index of $A$  with respect to
  (w.r.t.) $P_{X^n}$ if
  $A = \bigcup_{m \in \mcf{M}} A_{M=m}$ and
  $A_{M=m}\cap A_{M=\hat m} = \emptyset$ for all $m\neq \hat m$. 
  We may simply say $M$ partitions $A$ when the underlying distribution 
  $P_{X^n}$ of $X^n$ over $A$ is clear from the context.
\end{define}
 
\begin{lemma}\label{lem:joint_dist}
Consider any $A \subseteq \mcf{X}^n$ and partitioning indices
w.r.t. $P_{X^n}$ over $A$.
\begin{enumerate}
\item Suppose that $M$ partitions $A$. Then $P_M(m)=0$ if and only if
  $A_{M=m} = \emptyset$.  In addition, for all $m \in \mcf{M}$ such that
  $P_M(m)>0$, $\{M=m\} = \{X^n \in A_{M=m}\}$. Equivalently, $M=h(X^n)$
  where $h(x^n) = m$ if $x^n \in A_{M=m}$.

\item Suppose that $M$ partitions $A$.  Then for every non-empty $A' \subseteq A$, $M$ partitions $A'$ w.r.t $P_{X^n|X^n \in A'}$ with $A'_{M=m} = A' \cap A_{M=m}$.  As is clear from the context, we may simply say $M$ also partitions $A'$.

\item $(M_1,M_2)$ is a partitioning index of $A$ if and only if $M_1$ and $M_2$ are both partitioning indices of $A$.

\item Let $M_1$ and $M_2$ be partitioning indices of $A$. 
  Then
  $A_{(M_1,M_2)=(m_1,m_2)} = A_{M_1=m_1} \cap A_{M_2=m_2}$. Thus
  we may write $A_{M_1=m_1,M_2=m_2}$ in place of
  $A_{(M_1,M_2)=(m_1,m_2)}$ or $A_{M_1=m_1} \cap A_{M_2=m_2}$.
  Furthermore, for each $m_1\in\mcf{M}_1$ such that
  $A_{M_1=m_1} \neq \emptyset$, $M_2$ is a partitoning index of
  $A_{M_1=m_1}$ w.r.t. $P_{X^n|X^n \in A_{M_1=m_1}}$ (or equivalently
  $P_{X^n|M_1=m_1}$) with
  $(A_{M_1=m_1})_{M_2=m_2} = A_{M_1=m_1,M_2=m_2}$. Hence
  $\{M_2=m_2|M_1=m_1\} = \{X^n \in A_{M_1=m_1, M_2=m_2} | X^n \in
  A_{M_1=m_1}\}$ if the latter event is non-empty.

\item Let $M_1$ be a partitioning index of $A$ w.r.t. $P_{X^n}$. Let
  $\{ M_2(m_1)\}_{m_1\in \mcf{M}_1}$ be a collection of random indices
  such that $M_2(m_1)$ is a partitioning index of $A_{M_1=m_1}$
  w.r.t. $P_{X^n|M_1=m_1}$, distributed over $\mcf{M}_2$. Then
  $(M_1,M_2(M_1))$ is a partitioning index w.r.t. $P_{X^n}$.
\end{enumerate}
\end{lemma}
\begin{IEEEproof}
\begin{enumerate}
\item First, it is obvious from the definition of $A_{M=m}$ that
  $P_M(m) =0$ if and only if $A_{M=m}=\emptyset$. 
  Consider now for each $m \in \mcf{M}$ such that $P_M(m)>0$. 
  Then $\Pr\{X^n \in A \setminus A_{M=m}| M=m\}=
  \frac{P_{M,X^n}(m,  A \setminus A_{M=m})}{P_M(m)} = 0$, again
  due to the very definition of $A_{M=m}$. Hence $M=m$ implies
  $X^n \in A_{M=m}$. On the other hand, 
\[
\Pr\{ M \neq m | X^n \in A_{M=m}\} 
= \frac{\sum_{\hat m \neq m} P_{M,X^n}(\hat m, A_{M=m})}{P_{X^n}(
  A_{M=m})} = 0
\]
where the last equality is due to the fact that $M$ is a partitioning index of $A$. Hence $X^n \in A_{M=m}$ implies $M=m$. Hence by setting $h(x^n) = m$ if $x^n \in A_{M=m}$, $M=h(X^n)$ with probability one.

\item Note that $P_{M,X^n | X^n \in A'}(m,x^n) = \frac{P_{M,X^n}(m,x^n)}{P_{X^n}(A')}$,
  and hence $A'_{M=m} = A' \cap A_{M=m}$. Thus $M$ partitions $A'$ w.r.t.
  $P_{M,X^n | X^n \in A'}$.

\item First, suppose that $(M_1,M_2)$ is a partitioning index of
  $A$. Clearly
  $A_{M_1=m_1} = \bigcup_{m_2 \in \mcf{M}_2}
  A_{(M_1,M_2)=(m_1,m_2)}$.
  Hence
  $\bigcup_{m_1 \in \mcf{M}_1} A_{M_1=m_1} = \bigcup_{(m_1,m_2) \in
    \mcf{M}_1\times\mcf{M}_2} A_{(M_1,M_2)=(m_1,m_2)} = A$.
  In addition, for any $\hat m_1 \neq m_1$,
  $A_{M_1=m_1} \cap A_{M_1=\hat m_1} = \bigcup_{m_2,\hat m_2 \in
    \mcf{M}_2} A_{(M_1,M_2)=(m_1,m_2)} \cap A_{(M_1,M_2)=(\hat
    m_1,\hat m_2)} = \emptyset$.
  Therefore $M_1$ is a paritioning index of $A$. The same argument
  also applies to show that $M_2$ is a paritioning index of $A$.

  On the other hand, suppose that both $M_1$ and $M_2$ are
  partitioning indices of $A$.  Clearly we have
  $A_{(M_1,M_2)=(m_1,m_2)} \subseteq A_{M_1=m_1} \cap
  A_{M_2=m_2}$. For any $(m_1,m_2) \neq (\hat m_1, \hat m_2)$,
  $A_{(M_1,M_2)=(m_1,m_2)} \cap A_{(M_1,M_2)=(\hat m_1, \hat m_2)}
  \subseteq A_{M_1=m_1} \cap A_{M_2=m_2} \cap A_{M_1=\hat m_1} \cap
  A_{M_2= \hat m_2} = \emptyset$. But for
  every $x^n \in A$, since $P_{X^n}(x^n) >0$,
  $P_{M_1,M_2,X^n}(m_1,m_2,x^n) > 0$ and hence $x^n \in A_{(M_1,M_2)=(m_1,m_2)}$ for some
  $(m_1,m_2) \in \mcf{M}_1\times\mcf{M}_2$. This means that $A =
  \bigcup_{(m_1,m_2) \in \mcf{M}_1\times\mcf{M}_2}
  A_{(M_1,M_2)=(m_1,m_2)}$. Therefore $(M_1,M_2)$ is a  partitioning
  index of $A$.

\item Suppose that $M_1$ and $M_2$ are partitioning indices of $A$
  w.r.t. $P_{X^n}$. Then by part 3) of the lemma, $(M_1,M_2)$ is also
  a partitioning index of $A$ w.r.t. $P_{X^n}$. Moreover, by part 1) of the lemma, $M_1$ and $M_2$ are conditionally independent given $X^n$. That is,
  we have $P_{M_1,M_2,X^n}(m_1,m_2,x^n) =
  P_{M_1|X^n}(m_1|x^n)P_{M_2|X^n}(m_2|x^n) P_{X^n}(x^n)$, which
  implies
  $A_{M_1=m_1} \cap A_{M_2=m_2} \subseteq A_{(M_1,M_2)=(m_1,m_2)}$. As
  above, we also clearly have
  $A_{(M_1,M_2)=(m_1,m_2)} \subseteq A_{M_1=m_1} \cap A_{M_2=m_2}$.
  Therefore
  $A_{M_1=m_1,M_2=m_2} \defn A_{M_1=m_1} \cap A_{M_2=m_2}=
  A_{(M_1,M_2)=(m_1,m_2)}$.

  Consider any fixed $m_1 \in \mcf{M}_1$ such that
  $A_{M_1=m_1} \neq \emptyset$ (i.e., $P_{M_1}(m_1)>0$). 
  By
  part 1) of the lemma, we have
  $P_{X^n|X^n\in A_{M_1=m_1}}=P_{X^n|M_1=m_1}$. Further, by part 2) of the lemma, 
  we have
$M_2$ partitions $ A_{M_1=m_1}$
w.r.t. $P_{X^n|M_1=m_1}$ with $(A_{M_1=m_1})_{M_2=m_2} =
A_{M_1=m_1} \cap A_{M_2=m_2} = A_{M_1=m_1,M_2=m_2}$. The final assertion then
results directly from part 1).

\item For convenience in notation, write $Q=(M_1,M_2(M_1))$, which
  distributes over $\mcf{M}_1\times \mcf{M}_2$. Note that
  $P_{Q,X^n}((m_1,m_2),x^n) = P_{M_2(M_1),X^n|M_1} (m_2,x^n|m_1)
  P_{M_1}(m_1)$.
  Hence $A_{Q=(m_1,m_2)} = (A_{M_1=m_1})_{M_2(m_1)=m_2}$. Clearly
  \[
  \bigcup_{(m_1,m_2) \in
    \mcf{M}_1\times \mcf{M}_2} A_{Q=(m_1,m_2)} =A
  \]
  and $A_{Q=(m_1,m_2)} \cap A_{Q=(\hat m_1,\hat m_2)}$ for all
  $(m_1,m_2) \neq (\hat m_1,\hat m_2)$. Therefore $Q$ is a
  partitioning index of $A$ w.r.t. $P_{X^n}$.
\end{enumerate}
\end{IEEEproof}
All parts of Lemma~\ref{lem:joint_dist} will be used repetitively many times in the rest of the paper. To avoid prolixity, we will not explicitly refer to each use of the lemma.

\begin{define}\label{def:infspec}
  Let $A \subseteq \mcf{X}^n$. Let $P_{X^n}$ be a distribution on $A$, and
  $i_{X^n} = -\cexp{P_{X^n}}$ be the corresponding entropy spectrum. For
  any $0 < \delta, \delta_n < 1$, define
  $K_{\delta_n, \delta}(A) \defn \left\lceil \frac{\delta+
      \aexp{A}}{\delta_n} \right\rceil$
  and the $(\delta_n, \delta)$-entropy spectrum partition of $A$
  w.r.t.  $i_{X^n}$ as $\{A_k\}_{k=0}^{K_{\delta_n, \delta}(A)}$, where
\[
A_k \defn 
\begin{cases}
\{ x^n \in A : k\delta_n\leq i_{X^n}(x^n) < (k+1)\delta_n\} 
& \text{ for } k\in [0:K_{\delta_n, \delta}(A) -1] \\
\{ x^n \in A :  K_{\delta_n, \delta}(A)\delta_n\leq i_{X^n}(x^n) < \infty \}
&\text{ for } k = K_{\delta_n, \delta}(A).
\end{cases}
\]
Clearly $A = \bigcup_{k=0}^{K_{\delta_n, \delta}(A)} A_k$ because of our convention that $P_{X^n}(x^n) > 0$ for all $x^n \in A$. 
\end{define}
Suppose $\{A_k\}$ is the $(\delta_n, \delta)$-entropy spectrum
partition of $A$ w.r.t. $-\cexp{P_{X^n}}$, and $\upsilon(x^n) \defn k$
if $x^n \in A_k$.  The random variable $\upsilon(X^n)$ is clearly a partitioning
index of $A$ w.r.t. $P_{X^n}$, and is conditionally independent of any other 
partitioning index of $A$ given $X^n$.

\begin{lemma}\label{lem:bk_size}
  Let $\{ A_k \}$ be the $(\delta_n,\delta)$-entropy spectrum
  partition of $A \subseteq \mcf{X}^n$ w.r.t. $-\cexp{P_{X^n}}$. Then for
  every $A_{k}$, $k \in [0:K_{\delta_n,\delta}(A)]$,
\[
  \aexp{A_k} < (k+1) \delta_n. 
\]
In addition if $P_{X^n}(A_k) > 2^{-n \delta_n}$, then
\[
 \left| \aexp{A_k} - k \delta_n \right| < \delta_n.
\]
\end{lemma}
\begin{IEEEproof}
Trivially we have 
\[
\aexp{A_{K_{\delta_n, \delta}(A)}} \leq \aexp{A} < (K_{\delta_n,
  \delta}(A)+ 1) \delta_n.
\]
For $k \in [0:K_{\delta_n,\delta}(A) - 1]$,
\[
1 \geq \sum_{x^n \in A_{k}} P_{X^n}(x^n) > |A_{k}| 2^{-n (k+1) \delta_n},
\]
and therefore $\aexp{A_k} < (k+1) \delta_n$.  Similarly suppose that
$P_{X^n}(A_k) = \sum_{x^n \in A_k} P_{X^n}(x^n) > 2^{-n \delta_n}$. Then
\[
2^{-n \delta_n} < \sum_{x^n \in A_{k} } P_{X^n}(x^n) \leq 2^{-n k \delta_n} |
A_k|,
\]
and therefore $\aexp{A_k} > (k-1)\delta_n$. Combining both results gives
us $ \left| \aexp{A_k} - k \delta_n \right| < \delta_n$.
\end{IEEEproof}
\begin{lemma}\label{lem:bk_prob}
Let $\{A_k\}$ be the $(\delta_n,\delta)$-entropy spectrum
  partition of $A \subseteq \mcf{X}^n$ w.r.t. $-\cexp{P_{X^n}}$. Then for
  each $A_{k}$, $k \in [0:K_{\delta_n,\delta}(A)-1]$, satisfying $P_{X^n}(A_k) > 0$, 
\[
\frac{2^{-n\delta_n}}{\abs{A_k}} < \Pr\{ X^n = x^n | X^n \in A_k\}
< \frac{2^{n\delta_n}}{\abs{A_k}}
\]
for all $x^n \in A_k$.
\end{lemma}
\begin{IEEEproof}
Suppose $P_{X^n}(A_k)>0$. Obviously
$\Pr\{ X^n=x^n | X^n \in A_k \} 
= \begin{cases} 
  \frac{P_{X^n}(x^n)}{P_{X^n}(A_k)} & \text{if } x^n \in A_k \\
  0 & \text{if } x^n \notin A_k
\end{cases}$. Since 
\[
\abs{A_k} \cdot \frac{2^{-n (k+1) \delta_n}}{P_{X^n}(A_k)} <
\sum_{x^n \in A_k} \Pr \{ X^n = x^n| X^n \in  A_k \}  = 1
\leq \abs{A_k} \cdot \frac{2^{-n k \delta_n}}{P_{X^n}(A_k)},
\]
we  arrive at the bound 
\[
2^{n k \delta_n} \leq \frac{\abs{A_k}}{P_{X^n}(A_k)} < 2^{n (k+1)
  \delta_n}. 
\]
Thus for all $x^n \in A_k$, 
\[
\frac{2^{-n\delta_n}}{\abs{A_k}} = \frac{1}{\abs{A_k}}  2^{-n (k+1) \delta_n} 2^{n k \delta_n} 
< \Pr \{ X^n = x^n| X^n \in  A_k \} 
=\frac{P_{X^n}(x^n)}{P_{X^n}(A_k)} \frac{\abs{A_k}}{\abs{A_k}}
< \frac{1}{\abs{A_k}}  2^{-n k \delta_n} 2^{n (k+1) \delta_n} 
=\frac{2^{n\delta_n}}{\abs{A_k}}.
\]
\end{IEEEproof}

\begin{define} \label{def:uniform} Let $\gamma_n \geq 1$ and $P_{n}$ be
  a distribution on any arbitrary set $A_n$.  If
  $\frac{\max_{a \in A_n} P_{n}(a)}{\min_{a \in A_n} P_{n}(a)} \leq
  \gamma_n$,
  then $P_{n}$ is referred to as a $\gamma_n$-uniform distribution on
  $A_n$.
\end{define}

\begin{lemma}\label{lem:htoa}
Let $X^n$ be 
  $\gamma_n$-uniform distributed over any $A \subseteq \mcf{X}^n$. Then 
\[
\logt\abs{A} - \logt\gamma_n 
\leq H(X^n) \leq \logt\abs{A}. 
\]
\end{lemma}
\begin{IEEEproof}
Let $P_{X^n}$ be the distribution of $X^n$ over $A$. Clearly $P_{X^n}(x^n)
= \frac{P_{X^n}(x^n)}{P_{X^n}(A)} = \frac{P_{X^n}(x^n) }{\sum_{\hat x^n \in A}
  P_{X^n}(\hat x^n)}$ for all $x^n \in A$. Since $P_{X^n}$ is
$\gamma_n$-uniform, we must also have $\frac{\max_{x^n\in A}
  P_{X^n}(x^n)}{\min_{x^n\in A} P_{X^n}(x^n)} \leq \gamma_n$.
But then
\[
\frac{1}{\gamma_n\abs{A}} 
\leq \frac{\min_{x^n\in A} P_{X^n}(x^n)}{\abs{A} \max_{x^n\in A} P_{X^n}(x^n)} 
\leq P_{X^n}(x^n) 
\leq \frac{\max_{x^n\in A} P_{X^n}(x^n)}{\abs{A} \min_{x^n\in A} P_{X^n}(x^n)} 
\leq \frac{\gamma_n}{\abs{A}}.
\]
As a result,
\[
H(X^n) = - \sum_{x^n \in A} P_{X^n}(x^n) \logt P_{X^n|A}(x^n) 
\geq - \sum_{x^n \in A} P_{X^n}(x^n) \logt\frac{\gamma_n}{\abs{A}}
= \logt\abs{A} - \logt\gamma_n.
\]
The upper bound on $H(X^n)$ is the standard upper bound on
entropy.
\end{IEEEproof}

An obvious consequence of Lemma~\ref{lem:bk_prob} is that for any $P_{X^n}$ over $A \subseteq \mcf{X}^n$, $X^n$ is conditionally $2^{\frac{2}{n}}$-uniformly distributed over each of the $(\frac{1}{n^2}, \delta)$-entropy spectrum partition set, except $A_{K_{\frac{1}{n^2},\delta}(A)}$. We can dismiss this exception set for many of our results stated later because it is clear from the definition of entropy spectrum partition that $P_{X^n}(A_{K_{\frac{1}{n^2},\delta}(A)}) \leq 2^{-n\delta}$.  Furthermore it is important to note that there are at most $(\delta+\card{X}) n^2$ partition sets in this case.  Thus the number of partition sets is sub-exponentially small compared to the number of sequences in $\mcf{X}^n$. In fact, a more general result can be obtained as shown in the following lemma: \begin{lemma}\label{lem:x_slice}
  Let $X^n$ be randomly distributed over $A \subseteq \mcf{X}^n$ and
  $M$ be a partitioning index of $A$ w.r.t. $P_{X^n}$. For any
  $\delta>0$ and $\rho\geq 0$, there exist
  a random variable $W$
  over an index set $\mcf{W}$ satisfying the following
  properties:
\begin{enumerate}
\item $\abs{\mcf{W}} \leq \Gamma n^{\rho+2}$,
\item there is a $w_0 \in \mcf{W}$ with $P_{W}(w_0) \leq 2^{-n\delta}$,
\item $X^n$ is conditionally $2^{\frac{2}{n^{\rho+1}}}$-uniform given
  $W = w$ for each $w \in \mcf{W}$ except $w_0$,
\item $M$ is conditionally $2^{\frac{6}{n^{\rho+1}}}$-uniform given
  $W = w$ for each $w \in \mcf{W}$ except $w_0$, 
\item $W$ partitions $A$ w.r.t. $P_{X^n}$, and
\item $W$ partitions $\mcf{M}$ w.r.t. $P_{M}$.
\end{enumerate}
Note that the same $\Gamma$ can be
employed uniformly for all $A \subseteq \mcf{X}^n$.
\end{lemma}
\begin{IEEEproof}
  Write
  $K_n \defn K_{\frac{1}{n^{\rho+2}},2\delta}(A) = \left\lceil 2\delta n^{
\rho+2} + n^{\rho+1}
    \logt\abs{A} \right\rceil \leq n^{\rho+2}(2\delta+\card{X}) + 1$
  for convenience.  Let $\{ A_k\}_{k=0}^{K_n}$ be the
  $(\frac{1}{n^{\rho+2}},2\delta)$-entropy spectrum partition of $A$
  w.r.t. $-\cexp{P_{X^n}}$.  For each $k \in [0:K_n]$ such that $A_k
  \neq \emptyset$, define
\[
\mcf{M}_{k,l} \defn \begin{cases}
\left\{m\in \mcf{M}:
 2^{-\frac{l+1}{n^{\rho+1}}} < \frac{\abs{A_k \cap A_{M=m}}}{\abs{A_k}}
 \leq 2^{-\frac{l}{n^{\rho+1}}} \right\} 
& \text{for } l \in [0:K_n-1] \\
\left\{m\in \mcf{M}: 
 \frac{\abs{A_k \cap A_{M=m}}}{\abs{A_k}} \leq 2^{-\frac{K_n}{n^{\rho+1}}}
\right\} 
& \text{for } l = K_n,
\end{cases}
\]
and
$\hat A_{k,l} \defn \bigcup_{m \in \mcf{M}_{k,l}} A_k \cap A_{M=m}$.
It is easy to see that $\hat A_{k,l} \cap \hat A_{k',l'} = \emptyset$
for all $(k,l) \neq (k',l')$ and
$\bigcup_{(k,l) \in [0:K_n]^2} \hat A_{k,l} = A$.  By
Lemma~\ref{lem:bk_prob}, for each $k \in [0:K_n-1]$ with
$A_k \neq \emptyset$,
\begin{equation} \label{eq:PX=}
\frac{2^{-\frac{1}{n^{\rho+1}}}}{\abs{A_k}} 
\leq\Pr\{X^n =x^n | X^n \in A_k\} 
\leq\frac{2^{\frac{1}{n^{\rho+1}}}}{\abs{A_k}}
\end{equation}
for all $x^n \in A_k$. This implies that for each
$(k,l) \in [0:K_n-1]^2$ with $\hat A_{k,l} \neq \emptyset$,
\begin{align} 
\frac{ 2^{-\frac{3}{n^{\rho+1}}}}{ \abs{\mcf{M}_{k,l}}}
&\leq 
\frac{ 2^{-\frac{l+1}{n^{\rho+1}}} \cdot 2^{-\frac{1}{n^{\rho+1}}}}{ \abs{\mcf{M}_{k,l}}
   \cdot 2^{-\frac{l}{n^{\rho+1}}} \cdot 2^{\frac{1}{n^{\rho+1}}}}
\leq
\frac{ \abs{A_k \cap A_{M=m}} \cdot \frac{2^{-\frac{1}{n^{\rho+1}}}}{\abs{A_k}}}{
   \sum_{m' \in \mcf{M}_{k,l}} \abs{A_k \cap A_{M=m'}} \cdot
   \frac{2^{\frac{1}{n^{\rho+1}}}}{\abs{A_k}}} 
\notag \\
&\leq \Pr\{M=m | X^n \in \hat A_{k,l}\} 
= 
\frac{\sum_{x^n \in A_k \cap A_{M=m}} 
  \Pr\{ X^n=x^n|X^n \in A_k\}}{\sum_{m' \in \mcf{M}_{k,l}}\sum_{x^n
    \in A_k \cap A_{M=m'}} 
  \Pr\{ X^n=x^n|X^n \in A_k\}}
\notag \\
 & \leq 
\frac{ \abs{A_k \cap A_{M=m}} \cdot \frac{2^{\frac{1}{n^{\rho+1}}}}{\abs{A_k}}}{
   \sum_{m' \in \mcf{M}_{k,l}} \abs{A_k \cap A_{M=m'}} \cdot
   \frac{2^{-\frac{1}{n^{\rho+1}}}}{\abs{A_k}}} 
\leq
\frac{ 2^{-\frac{l}{n^{\rho+1}}} \cdot 2^{\frac{1}{n^{\rho+1}}}}{ \abs{\mcf{M}_{k,l}}
   \cdot 2^{-\frac{l+1}{n^{\rho+1}}} \cdot 2^{-\frac{1}{n^{\rho+1}}}}
= \frac{ 2^{\frac{3}{n^{\rho+1}}}}{ \abs{\mcf{M}_{k,l}}}
   \label{eq:PM=}
\end{align}
for all $m \in \mcf{M}_{k,l}$.  Further, let
$\tilde A \defn \bigcup_{(k,l) \in [0:K_n-1]^2} \hat A_{k,l}$.  Then
by \eqref{eq:PX=},
\begin{align*}
P_{X^n}(A\setminus \tilde A) 
&\leq 
 \sum_{x^n\in A_{K_n}} P_{X^n}(x^n) + \sum_{k=0}^{K_n-1} \sum_{m \in
  \mcf{M}_{k,K_n}} \sum_{x^n \in A_k \cap A_{M=m}} P_{X^n}(x^n) 
\notag \\
&\leq 
\abs{A_{K_n}} \cdot  2^{-\frac{K_n}{n^{\rho+1}}} + \sum_{k=0}^{K_n-1} \sum_{m \in
  \mcf{M}_{k,K_n}} 2^{\frac{1}{n^{\rho+1}}}\frac{\abs{A_k \cap A_{M=m}}}{\abs{A_k}} P_{X^n}(A_k) 
\notag \\
&\leq
  \abs{A} \cdot 2^{-\frac{K_n}{n^{\rho+1}}}  + 
  \sum_{k=0}^{K_n-1} \abs{\mcf{M}_{k,K_n}} \cdot 
  2^{-\frac{K_n-1}{n^{\rho+1}}} P_{X^n}(A_k)
\notag \\
&\leq 2^{-2n \delta}  + 2^{-2n \delta+1}
\leq 2^{-n\delta}
\end{align*}
where the second last inequality is valid when $n$ is sufficiently large, and
it is due to the fact that $\abs{\mcf{M}} \leq \abs{A}$ as $M$
partitions $A$. 

Now define the random index $W$ over $\mcf{W} \defn [0:K_n-1]^2 \cup \{w_0\}$ by setting $W = (k,l)$ if $X^n \in \hat A_{k,l}$ and $W=w_0$ if $X^n \in A\setminus \tilde A$.  Then $W$ is a partitioning index of $A$ w.r.t. $P_{X^n}$. Also $P_W(w_0) = P_{X^n}(\tilde A) \leq 2^{-n\delta}$, and~\eqref{eq:PX=} and \eqref{eq:PM=} imply that $X^n$ and $M$ are conditionally $2^{\frac{2}{n^{\rho+1}}}$- and $2^{\frac{6}{n^{\rho+1}}}$-uniform given $W = (k,l)$ for each $(k,l) \in [0:K_n-1]^2$, respectively.  To show $W$ partitions $\mcf{M}$ w.r.t. $P_M$, observe that $P_{M,W}(m,w) = P_{X^n,W}(A_{M=m},w)$. This implies that $A_{M=m} \subseteq A_{W=w}$ if $m \in \mcf{M}_{W=w}$. But then as $W$ partitions $A$, for any $m \in \mcf{M}_{W=w} \cap \mcf{M}_{W=w'}$ where $w' \neq w$, $A_{M=m} = \emptyset$ and hence $P_M(m) = 0$, contradicting the assumption that $m \in \mcf{M}_{W=w}$. As a result, we must have $\mcf{M}_{W=w} \cap \mcf{M}_{W=w'} = \emptyset$. Further by convention (removing zero-probability elements from $\mcf{M}$ if necessary), $P_{M}(m)>0$ for all $m \in \mcf{M}$, and thus $\bigcup_{w \in \mcf{W}} \mcf{M}_{W=w} = \mcf{M}$.  
\end{IEEEproof}

\section{Image size bounds}\label{sec:lemmas}

In this section, we develop upper and lower bounds on the minimum image size. The main results are Lemmas~\ref{lem:daco} and~\ref{lem:subset}, which play a major role in enabling the source partitioning results in the next section. We also note that weaker versions of the lemmas in this section have been presented in our earlier work~\cite{graves2014equating}.

Fix any non-empty $A \subseteq \mcf{X}^n$, $\delta \geq 0$, $\delta_n \in (0,1)$ satisfying $n\delta_n \rightarrow \infty$. Let $X^n$ be distributed over $A$ and $Y^n$ be conditionally distributed given $X^n$ according to $P^n_{Y|X}$. Throughout this section, let $\{B_k\}$ be the $(\delta_n, \delta)$-entropy spectrum partition of $\mcf{Y}^n$ w.r.t. $-\cexp{P_{Y^n}}$. Note that we adopt the convention from section~\ref{sec:partition} that $P_{Y^n}(y^n) > 0$ for all $y^n \in B_k \subseteq \mcf{Y}^n$ and all $k$. Also write $K_{n} \defn K_{\delta_n,\delta}(\mcf{Y}^n)$ for notational simplicity.

\begin{lemma}\label{lem:barg_bound}
For any $\eta \in (0,1]$ and sufficiently large $n$, there exists a
  $k'_n \in [0:K_{n}]$ such that
\[
\cexp{\bar g^n_{Y|X}(A,\eta) } \leq 
\aexp{ \bigcup_{k=0}^{k'_n} B_{k} }
\leq (k'_n+2) \delta_n.
\]
Furthermore if $P_{Y^n}(B_{k'_n-1}) > 2^{-n \delta_n}$, then
\[
\cexp{\bar g^n_{Y|X}(A,\eta) } \geq \aexp{B_{k'_n-1}} \geq (k'_n-2)
\delta_n.
\]
\end{lemma}
\begin{IEEEproof}
  Let $\eta_{-1} \defn 0$ and $\eta_{k'} \defn P_{Y^n}
  \left(\bigcup_{k=0}^{k'} B_k \right)$ for $k' \in [0: K_{n}]$. Note
  then that $0=\eta_{-1} \leq \eta_0 \leq \eta_1 \leq \cdots \leq
  \eta_{K_{n}} = 1$. In addition $\eta_{k-1} = \eta_k$ implies
  $B_k = \emptyset$. Write $B' = \bigcup_{k=0}^{k'} B_k$ to simplify
  notation below. Clearly $B'$ is an $\eta_{k'}$-quasi-image of
  $A$. We claim that $B'$ is in fact the unique
  $\eta_{k'}$-quasi-image of $A$ that achieves the minimum size $\bar
  g^n_{Y|X}(A,\eta_{k'})$.  To show the claim, consider a set $\hat B
  \subseteq \mcf{Y}^n$ such that $\hat B \neq B'$ and $|\hat B| \leq
  |B'|$. For $k'=K_{n}$, $\hat B$ clearly cannot be the
  $\eta_{k'}$-quasi-image of $A$. On the other hand, for $k' \in
  [0:K_{n}-1]$,
\begin{align*}
P_{Y^n}(\hat B)
 &= P_{Y^n}(B') -  P_{Y^n}(B' \setminus \hat B) + P_{Y^n}(\hat B \setminus B') \\
& =  \eta_{k'} - \sum_{k=0}^{k'} \sum_{y^n \in B_k \setminus \hat B}
P_{Y^n}(y^n) 
+ \sum_{k \geq k'+ 1} \sum_{y^n \in B_k \cap \hat B} P_{Y^n}(y^n) \\
&<  \eta_{k'} 
- \left( |B'| - |B' \cap \hat B| \right) 2^{-n(k'+1) \delta_n} 
+ \left( |\hat B| - |B' \cap \hat B| \right) 2^{-n(k'+1) \delta_n}  
\leq \eta_{k'}.
\end{align*}
Thus $\hat B$ cannot be an $\eta_{k'}$-quasi-image of $A$. 

Next it is clear that for any $\eta \in (0,1]$ there exists a $k'_n$ such that
$ \eta_{k'_n-1} < \eta \leq \eta_{k'_n}$ which gives us that
\begin{equation}
\aexp{ \bigcup_{k=0}^{k'_n-1} B_{k} } \leq \cexp{\bar g^n_{Y|X}(A,\eta) }
 \leq \aexp{ \bigcup_{k=0}^{k'_n} B_{k} }.
\label{eq:lem3_1}
\end{equation}
Lemma~\ref{lem:bk_size} and the upper bound in \eqref{eq:lem3_1} give
us that
\[
\cexp{\bar g^n_{Y|X}(A,\eta) } 
\leq \cexp{ \sum_{k=0}^{k'_n} 2^{n(k+1) \delta_n} }
\leq (k'_n+2) \delta_n
\]
when $n$ is sufficiently large.  Furthermore if $P_{Y^n|X^n \in A}(B_{k'_n-1}) > 2^{-n
  \delta_n}$, then combining the lower bound of \eqref{eq:lem3_1} and
Lemma~\ref{lem:bk_size} again we have
\[
\cexp{\bar g^n_{Y|X}(A,\eta) } \geq \aexp{B_{k'_n-1}} \geq (k'_n-2) \delta_n.
\]
\end{IEEEproof}

\begin{lemma}\label{lem:6.6}
For any $\alpha_n$ and $\beta_n$ such that $0 < \alpha_n < \beta_n < 1$ and $\frac{-\logt \min(\alpha_n, 1- \beta_n) }{n} \rightarrow 0$
there exists $\tau_n \rightarrow 0$ such that
\[
0 \leq \cexp{ \g{Y|X}{A'}{\beta_n}} - \cexp{ \g{Y|X}{A'}{\alpha_n}}
\leq \tau_n
\]
for every $A' \subseteq \mcf{X}^n$. Furthermore the same $\tau_n$ can be used
uniformly for all $\min(\alpha_n, 1-\beta_n) \geq \frac{1}{n^2}$. 
\end{lemma}
\begin{IEEEproof}
  This is a slightly strengthened version of \cite[Lemma~6.6]{CK},
  whose proof (along with the proofs of \cite[Ch.~5]{CK}) directly applies to the
  current lemma. 
\end{IEEEproof}

\begin{lemma} \label{lem:ApsubA} 
 
  Let $X^n$ be $\gamma_n$-uniform distributed over $A$.
  Then for any $\gamma_n \leq n$ and $\alpha_n \in (0,1]$ with
  $\frac{-\logt \alpha_n}{n} \rightarrow 0$,
  there exist $A' \subseteq A$, $\tau_n \rightarrow 0$, and $\beta_n
  \rightarrow 1$ such that 
  $\frac{\abs{A'}}{\abs{A}} > \left(\frac{1}{\gamma_n}-\frac{1}{n}
  \right) \alpha_n$ and
  \[
  \cexp{\g{Y|X}{A'}{ \beta_n}} \leq \cexp{\bg{Y|X}{A}{\alpha_n}} +
  \tau_n,
  \]
  whenever $n$ is sufficiently large. Neither $\tau_n$ nor $\beta_n$
  depends on $A$. Furthermore neither depends on $\alpha_n$ if
  $\alpha_n \geq \frac{1}{n}$.
\end{lemma}
\begin{IEEEproof}
  Let $B\subseteq \mcf{Y}^n$ be an $\alpha_n$-quasi-image of $A$ that
  achieves $\bg{Y|X}{A}{\alpha_n}$.  Define
\[
A' \defn \left\{x^n \in A: P^n_{Y|X}(B | x^n) \geq
\frac{\alpha_n}{n} \right\}.
\]
Clearly $B$ is an 
$\frac{\alpha_n}{n}$-image of $A'$. Hence there exist
$\beta_n \rightarrow 1$ and $\tau_n \rightarrow 0$ such that
\begin{align*}
\cexp{\bg{Y|X}{A}{\alpha_n}} &=\aexp{B} 
\geq 
\cexp{\g{Y|X}{A'}{\frac{\alpha_n}{n}}} 
\geq 
\cexp{\g{Y|X}{A'}{\beta_n}} - \tau_n
\end{align*}
by Lemma~\ref{lem:6.6} since 
$\frac{\logt n -\logt \alpha_n}{n} \rightarrow 0$.  Note that the same
$\beta_n$ and $\tau_n$ can be used uniformly for all
$\alpha_n \geq \frac{1}{n}$.

Further as $B$ is an $\alpha_n$-quasi-image of $A$, we have
\begin{align*}
\alpha_n \leq P_{Y^n}(B) 
&\leq \frac{\gamma_n}{\abs{A}} \sum_{x^n \in A} P^n_{Y|X} (B|x^n) \\
&= \frac{\gamma_n}{\abs{A}} \sum_{x^n \in A'} P^n_{Y|X} (B| x^n) +
\frac{\gamma_n}{\abs{A}} \sum_{x^n \in A \setminus A'} P^n_{Y|X}(B| x^n)  
\\
&< \gamma_n \frac{\abs{A'}}{\abs{A}}
+   \gamma_n\left(1- \frac{\abs{A'}}{\abs{A}} \right) 
\frac{\alpha_n}{n}
\end{align*}
which implies
$\frac{\abs{A'}}{\abs{A}} > (\frac{1}{\gamma_n}-\frac{1}{n}) \alpha_n$.
\end{IEEEproof}

\begin{lemma}\label{lem:daco}
  Suppose that $X^n$ is conditionally $\gamma_n$-uniform distributed
  on $A$ for some $\gamma_n \rightarrow 1$.  Then there exist
  $A^\ast \subseteq A$, $\varepsilon_n \rightarrow 0$, and
  $\beta_n \rightarrow 1$ satisfying
  $\frac{\abs{A^\ast}}{\abs{A}} \geq \frac{1}{2(K_{n}+1)}$ and
\[
\frac{1}{n}H(Y^n | X^n \in A^\ast) \geq \cexp{g^n_{Y|X}(A^\ast,\beta_n)} 
- 7.19 \delta_n - \varepsilon_n.
\]
 Neither $\varepsilon_n$ nor $\beta_n$ depends on $A$.
\end{lemma}

\begin{IEEEproof}
  Define $\eta_{k} \defn P_{Y^n} \left( \bigcup_{l=0}^{k} B_l \right)$ for
  $k\in [0:K_{n}]$ as in the proof of Lemma~\ref{lem:barg_bound}.
  Because the total number of sets in $\{B_k\}$ is $K_{n}+1$, we
  know that there exists at least one $k'_n \in [0:K_{n}]$ such
  that $P_{Y^n}(B_{k'_n}) \geq \frac{1}{K_{n}+1}$. Apply
  Lemma~\ref{lem:ApsubA} by choosing $\alpha_{n}\geq \eta_{k'_n} \geq
  \frac{1}{K_{n}+1}$ 
  to obtain $\tau_{n} \rightarrow 0$, $\beta_{n} \rightarrow 1$, and
  $A' \subseteq A$ that satisfy
\begin{align}  
\frac{\abs{A'}}{\abs{A}} 
&\geq 
\left(\frac{1}{\gamma_n}-\frac{1}{n} \right)\eta_{k'_n}
\geq \frac{1}{2(K_{n}+1)},
\label{eq:A/A'} \\
\cexp{\g{Y|X}{A'}{\beta_{n}}} &\leq
\cexp{\bg{Y|X}{A}{\eta_{k'_n}}} + \tau_{n}. 
\label{eq:g<bg}
\end{align}
Note that the $\beta_n$ and
$\tau_n$ above are the ones that work uniformly for all $\alpha_n \geq
\frac{1}{n}$ in Lemma~\ref{lem:ApsubA}.

First consider the case of $k'_n \leq c_n \defn 4.19 +
\frac{\tau_{n}}{\delta_n}$.  From \eqref{eq:g<bg},
\begin{align}
& \hspace*{-10pt} \cexp{\g{Y|X}{A'}{\beta_{n}}} 
\leq \cexp{\bg{Y|X}{A}{\eta_{k'_n}} } + \tau_{n} 
\stackrel{(a)}{=}  \aexp{ \bigcup_{k =0}^{k'_n}  B_{k}}  + \tau_{n} 
\leq (k'_n+2)\delta_n + \tau_{n} \label{e:bb} \\
&\leq 6.19 \delta_n + 2\tau_{n} \notag
\end{align}
where (a) is due to the fact that $\bigcup_{k =0}^{k'_n} B_{k}$ is the
$\eta_{k'_n}$-quasi-image of $A$ that achieves
$\bg{Y|X}{A}{\eta_{k'_n}}$ as shown in the proof of
Lemma~\ref{lem:barg_bound}. Since $H(Y^n | X^n \in A') \geq 0$,
the conclusions of the lemma are clearly satisfied.

It remains to consider the case of $k'_n > c_n$. To that end, let $k''_n
\defn \lfloor k'_n-c_n \rfloor$, and define the set $\tilde B =
\bigcup_{k=0}^{k''_n} B_{k}$. First assume that $P_{Y^n}(\tilde B) =
\eta_{k''_n} > \frac{1}{n}$. Apply Lemma~\ref{lem:ApsubA} again with
$\alpha_{n} = \eta_{k''_n} > \frac{1}{n}$ 
to obtain a set $A'' \subseteq A$ that satisfies
\begin{align}
\cexp{g^n_{Y|X}(A'',\beta_{n})} 
& 
\leq \cexp{\bar g^n_{Y|X}(A,\eta_{k''_n} )} + \tau_{n} \notag\\
&\stackrel{(a)}{\leq} (k''_n + 2)\delta_n  + \tau_{n} 
\leq (k'_n -2.19)\delta_n \notag \\
&\stackrel{(b)}{\leq} \aexp{B_{k'_n}} -1.19 \delta_n
\label{eq:bkpmg}
\end{align}
where (a) and (b) are due to Lemmas~\ref{lem:barg_bound} and
\ref{lem:bk_size}, respectively.

Let $\hat B$ be a $\beta_{n}$-image of $A''$ that achieves $\g{Y|X}{A''}{\beta_{n}}$. By definition, every $y^n \in B_{k'_n}$ has the property that $2^{-n(k'_n+1) \delta_n} < P_{Y^n}(y^n) \leq 2^{-n k'_n \delta_n}$. This implies 
\begin{align*}
P_{Y^n}(B_{k'_n} \setminus \hat B) 
&= P_{Y^n}(B_{k'_n})  - P_{Y^n}(B_{k'_n} \cap \hat B)  
\geq \frac{1}{K_{n}+1} -   2^{-n k'_n \delta_n} \left| B_{k'_n} \cap \hat B \right| 
\\
&
\geq \frac{1}{K_{n}+1} -  2^{-n k'_n \delta_n} g^n_{Y|X}(A'',\beta_{n})  
\geq  \frac{1}{K_{n}+1} - 2^{n \delta_n}
\frac{\g{Y|X}{A''}{\beta_{n}}}{\left| B_{k'_n} \right|} 
\geq  \frac{1}{K_{n}+1} - 2^{-0.19n\delta_n}
\end{align*}
where the second last and last inequalities are due to
Lemma~\ref{lem:bk_size} and \eqref{eq:bkpmg}, respectively. 
Continuing on,
\begin{align*}
& \hspace*{-10pt} 
\frac{1}{K_{n}+1} - 2^{-0.19n\delta_n} 
\leq P_{Y^n} (B_{k'_n} \setminus \hat B)  
\leq \frac{\gamma_n}{|A|} \sum_{x^n \in A} P^n_{Y|X}(B_{k'_n} \setminus \hat B | x^n)\\
&\stackrel{(a)}{=}
\frac{\gamma_n}{|A|} \sum_{x^n \in A'' } P^n_{Y|X}(B_{k'_n} \setminus \hat B | x^n ) 
+ \frac{\gamma_n}{|A|} \sum_{x^n \in A' \setminus A'' } 
P^n_{Y|X}( B_{k'_n} \setminus \hat B | x^n ) 
+ \frac{\gamma_n}{|A|} \sum_{x^n \in A \setminus (A' \cup A'')  } 
\hspace{-10pt} P^n_{Y|X}(B_{k'_n} \setminus \hat B | x^n) \\
&\stackrel{(b)}{\leq}
\gamma_n \left(1-\beta_{n}  + \frac{\abs{A' \setminus A'' }}{\abs{A}}
+ \frac{\eta_{k'_n}}{n} \right)
\end{align*}
where each term in (b) bounds the corresponding term in (a). In
particular, the first bound in (b) is due to the fact that each $x^n
\in A''$ satisfies $P^n_{Y|X}(\hat B^c | x^n)\leq 1- \beta_{n}$. On
the other hand, the third bound in (b) results from the fact that $\bigcup_{k=0}^{k'_n} B_{k}$ is
the unique minimum-size $\eta_{k'_n}$-quasi-image of $A$ (see the
proof of Lemma~\ref{lem:barg_bound}), and hence $A'$
contains all $x^n \in A$ that $P^n_{Y|X} \left( \bigcup_{k=0}^{k'_n}
  B_{k} \big| x^n \right) \geq \frac{\eta_{k'_n}}{n}$ as defined in the
proof of Lemma~\ref{lem:ApsubA}.  As a result, we have
\begin{align}
\frac{\abs{A' \setminus A''}}{\abs{A}}
& \geq 
\frac{1}{\gamma_n(K_{n}+1)} - \frac{2^{-0.19n\delta_n}}{\gamma_n}
-(1- \beta_{n}) - \frac{1}{n} 
\geq \frac{1}{2(K_{n}+1)}
\label{eq:Aminus}
\end{align}
for all sufficiently large $n$.  Now since $X^n$ is $\gamma_n$-uniform in $A$, we have
\begin{align}
P_{Y^n|X^n \in A'\setminus A''}(y^n) &\defn
\Pr( Y^n =y^n | X^n \in A'\setminus A'')
  \leq \frac{\gamma_n}{\left| A' \setminus A'' \right|} \sum_{x^n \in A' \setminus A'' }
P^n_{Y|X}(y^n | x^n) \notag \\
&\leq \frac{2\gamma_n(K_{n}+1)}{|A|} \sum_{x^n \in A} P^n_{Y|X}(y^n | x^n)
\leq 2\gamma_n(K_{n}+1) P_{Y^n}(y^n). \label{eq:ptyy}
\end{align}
Hence using \eqref{eq:ptyy} we get
\begin{align}
\frac{1}{n} H(Y^n | X^n \in A'\setminus A'') 
&\geq - \frac{1}{n} \sum_{y^n \notin \tilde B } P_{Y^n|X^n \in A'\setminus A''}(y^n) 
 \logt P_{Y^n|X^n \in A'\setminus A''}(y^n) 
\notag \\
&\geq -\frac{\logt 2\gamma_n (K_{n}+1)}{n}  - \frac{1}{n}
\sum_{k=k''_n}^{K_{n}} \sum_{y^n \in B_k} P_{Y^n|X^n \in A'\setminus A''}(y^n) \logt
P_{Y^n}(y^n)  \notag \\
&\geq -\frac{\logt 2\gamma_n (K_{n}+1)}{n} + \sum_{k=k''_n}^{K_{n}} 
P_{Y^n|X^n \in A'\setminus A''}(B_{k}) \cdot k \delta_n \notag \\
&\geq -\frac{\logt 2\gamma_n (K_{n}+1)}{n} + (k'_n-c_n-1)\delta_n 
  P_{Y^n|X^n \in A'\setminus A''}(\tilde B^c)
\notag \\
& \stackrel{(a)}{\geq}
  -\frac{\logt 2\gamma_n (K_{n}+1)}{n} + \bigg( \cexp{\g{Y|X}{A'}{\beta_{n}}} -
  7.19\delta_n 
 - 2\tau_{n} \bigg) \cdot P_{Y^n|X^n \in A'\setminus A''}(\tilde B^c) \notag \\
& \stackrel{(b)}{\geq}
-\frac{\logt 2\gamma_n (K_{n}+1)}{n} + \bigg( \cexp{\g{Y|X}{A' \setminus A''}{\beta_{n}}} 
- 7.19\delta_n - 2\tau_{n} \bigg) 
\cdot \left( 1 - \frac{\gamma_n\eta_{k''_n}}{n}  \right)  \notag \\
& \geq \cexp{\g{Y|X}{A' \setminus A''}{\beta_{n}}} 
  -\frac{\logt 2\gamma_n (K_{n}+1)}{n} 
   -\frac{ \gamma_n\logt \abs{\mcf{Y}}}{n} 
- 7.19\delta_n - 2\tau_{n} 
\label{eq:hty2}
\end{align}
where (a) is due to \eqref{e:bb} and (b) is due to the fact that
$A''$ contains all $x^n \in A$ that $P^n_{Y|X} \left( \tilde B|
  x^n \right) \geq \frac{\eta_{k''_n}}{n}$ (see the proof of
Lemma~\ref{lem:ApsubA}).  Clearly then the
conclusions of the lemma result from \eqref{eq:Aminus} and
\eqref{eq:hty2}.

Finally consider the case of $P_{Y^n}(\tilde B) \leq \frac{1}{n}$. Following a derivation similar to~\eqref{eq:ptyy},
\begin{equation}\label{eq:ptyy2}
P_{Y^n|X^n \in A'}(y^n) \leq 2\gamma_n(K_{n}+1)P_{Y^n}(y^n).
\end{equation}
Similarly, following the derivation of~(\ref{eq:hty2}a) with $A'$ in place of $A'\setminus A''$ gives
\begin{equation*}
\frac{1}{n} H(Y^n | X^n \in A')  \geq  -\frac{\logt 2\gamma_n(K_n+1)}{n} + \bigg( \cexp{\g{Y|X}{A'}{\beta_{n}}} -
  7.19\delta_n  - 2\tau_{n} \bigg) \cdot P_{Y^n|X^n \in A'}(\tilde B^c)
\end{equation*}
But by~\eqref{eq:ptyy2},
\begin{equation*}
P_{Y^n|X^n \in A'}(\tilde B^c) = 1 - P_{Y^n|X^n \in A'}(\tilde B) \geq 1 -2\gamma_n(K_n+1) P_{Y^n}(\tilde B),
\end{equation*}
and hence
\begin{equation*}
\frac{1}{n} H(Y^n | X^n \in A')  \geq    \cexp{\g{Y|X}{A'}{\beta_{n}}} -
  7.19\delta_n  - 2\tau_{n} -\frac{\logt 2\gamma_n(K_n+1)}{n} -\frac{2\gamma_n(K_n+1) \logt\abs{\mcf{Y}}}{n}. 
\end{equation*}
This, together with \eqref{eq:A/A'}, again gives the lemma.
\end{IEEEproof}

\begin{lemma}\label{lem:15.2}
There exists $\tau_n \rightarrow 0$ such that
\[
\cexp{\g{Y|X}{A'}{\eta}}  \geq H(Y^n|X^n \in A') - \tau_n - \frac{2}{n}
\]
for all $\eta \in (0,1)$ and $A' \subseteq \mcf{X}^n$.
\end{lemma}
\begin{IEEEproof}
This lemma is just a slightly strengthened version of the lower bound in \cite[Lemma 15.2, Eqn. (15.4)]{CK}. The same proof for \cite[Lemma 15.2]{CK} works here with the help of Lemma~\ref{lem:6.6}, which determines $\tau_n$.
\end{IEEEproof}

\begin{lemma}\label{lem:subset}
  For any $\mu_n \in (0,1]$ and $\eta \in (0,1)$,  let
  $A^\ast \subseteq \mcf{X}^n$ be such that $X^n$ is conditionally
  $\gamma_n$-uniform over $A^\ast$ and
\begin{equation}
\cexp{\g{Y|X}{A^\ast}{\eta}}  \leq \frac{1}{n}H(Y^n | X^n \in A^\ast)
+ \epsilon_n
\label{eq:sa}
\end{equation}
where $\epsilon_n \geq \tau_n+\frac{2}{n}$ with $\tau_n$ given in Lemma~\ref{lem:15.2} (or Lemma~\ref{lem:6.6}).
Then, for any $A' \subseteq A^\ast$ satisfying
$\frac{\abs{A'}}{\abs{A^\ast}} \geq \mu_n$, 
\[
\left| \frac{1}{n} H(Y^n | X^n \in A') - \frac{1}{n} H(Y^n | X^n \in
  A^\ast) \right| \leq  \frac{3\gamma_n \epsilon_n}{\mu_n}
\]
and
\[
\left| \frac{1}{n} H(Y^n | X^n \in A') - \cexp{\g{Y|X}{A'}{\eta}} \right| 
\leq  \frac{4\gamma_n \epsilon_n}{\mu_n}.
\]
\end{lemma}
\begin{IEEEproof}
Apply Lemma~\ref{lem:15.2} on $A'$ and $A^\ast \setminus A'$ to get  
\begin{align}
\frac{1}{n}H(Y^n | X^n \in A') 
&\leq \cexp{\g{Y|X}{A'}{\eta}} + \tau_n + \frac{2}{n}, 
\label{eq:sa1} \\
\frac{1}{n}H(Y^n | X^n \in A^\ast \setminus A') 
&\leq \cexp{\g{Y|X}{A^\ast \setminus A'}{\eta}} + \tau_n + \frac{2}{n}
\leq \cexp{\g{Y|X}{A^\ast}{\eta}} + \epsilon_n 
\leq \frac{1}{n}H(Y^n | X^n \in A^\ast)  + 2\epsilon_n  \label{eq:sa2}
\end{align}
where the last inequality is due to \eqref{eq:sa}.

Now let $S$ be in the indicator random variable of the
event that $X^n \in A'$. We have
\begin{align}
&\hspace*{-10pt} H(Y^n | X^n \in A^\ast)  \notag \\
&=  I(S ; Y^n | X^n \in A^\ast) + H(Y^n | S, X^n \in A^\ast) \notag\\
& \leq 1+H(Y^n | X^n \in A') \cdot \Pr\{X^n \in A' |X^n \in A^\ast\} + H(Y^n
  | X^n \in A^\ast \setminus A') \cdot
  \Pr\{ X^n \in A^\ast \setminus A' |X^n \in A^\ast\}
\notag \\
& \leq 1 + H(Y^n | X^n \in A') \cdot \Pr\{X^n \in A' |X^n \in A^\ast \} 
  +  \left[ H(Y^n | X^n \in A^\ast) + 2n \epsilon_n \right] \cdot
 \left[ 1 -  \Pr\{X^n \in A' |X^n \in A^\ast\}\right]
\label{eq:sa3}
\end{align}
where the last inequality is due to \eqref{eq:sa2}.  Since $X^n$ is
conditionally $\gamma_n$-uniform over $A^\ast$,
\[
\Pr\{X^n \in A' |X^n \in A^\ast\} = \frac{\Pr\{X^n \in A' \}}{\Pr\{X^n \in A^\ast\}}
\geq \frac{\abs{A'}}{\gamma_n \abs{A^\ast}} \geq \frac{\mu_n}{\gamma_n}.
\]
As a result, we can rearrange \eqref{eq:sa3} to get
\begin{align}
\frac{1}{n}H(Y^n | X^n \in A')
&\geq
\frac{1}{n}  H(Y^n | X^n \in A^\ast) - \left(2\epsilon_n  + \frac{1}{n}\right)
  \frac{1}{\Pr\{X^n \in A' |X^n \in A^\ast\}}
\notag \\
&\geq
\frac{1}{n}  H(Y^n | X^n \in A^\ast) - \frac{3\gamma_n \epsilon_n}{\mu_n}
\notag \\
&\geq
 \cexp{\g{Y|X}{A^\ast}{\eta}}  -\frac{3\gamma_n \epsilon_n}{\mu_n}
 - \epsilon_n
\notag \\
&\geq
 \cexp{\g{Y|X}{A'}{\eta}}  
-\frac{4\gamma_n \epsilon_n}{\mu_n}
\label{eq:sa4}
\end{align}
where the second last inequality is again due to
\eqref{eq:sa}. Finally we obtain the lemma by combining \eqref{eq:sa1}
and \eqref{eq:sa4}.
\end{IEEEproof}

\section{Equal-image-size source partitioning}\label{sec:mt}

In this section, we develop the nearly equal-image-size source partitioning result previously described in section~\ref{sec:preview}. We start by specifying the exponent of the minimum image size in terms of entropy in Lemma~\ref{thm:main}. Using this specification, we obtain in Lemma~\ref{thm:partition} a simple source partition with the same specification of the minimum image size in terms of entropy for each partitioning subsets.  Finally by applying this simple partitioning to every source subset indexed by a message, we arrive at the main result, in Theorem~\ref{thm:cond}, of equal-image-size source partitioning.

\begin{lemma}\label{thm:main}
  Fix any $\eta\in (0,1)$. Let $X^n$ be $\gamma_n$-uniform distributed over any
  $A \subseteq \mcf{X}^n$ for some $\gamma_n \rightarrow 1$. For
  $k \in [1:K]$, suppose that $Y^n_k$ is conditionally distributed
  according to the channel $P^n_{Y_k|X}$ given $X^n$.  Then there
  exist $\mu \in \left(0,\frac{1}{2} \right)$, $\epsilon_n \rightarrow 0$, 
  and $A'\subseteq A$ such that
\begin{enumerate}
\item $ \frac{\abs{A'}}{\abs{A}} \geq \frac{\mu}{n}$,
\item $\aexp{A'} - \epsilon_n \leq \frac{1}{n}H(X^n | X^n \in A') 
  \leq \aexp{A'}$, and
\item $\left|\frac{1}{n}H(Y_k^n| X^n \in A')
    -\cexp{g^n_{Y_k|X}(A',\eta) }\right| \leq \epsilon_n$ for $k \in
  [1:K]$,
\end{enumerate}
whenever $n \geq N$ for some large $N$. Note that $N$, $\mu$, and $\epsilon_n$
work uniformly for all $A \subseteq \mcf{X}^n$.
\end{lemma}

\begin{IEEEproof}
  Fix any $\delta>0$. Choose $\delta_{1,n} = n^{-\frac{1}{K+1}}$.
  Apply Lemma~\ref{lem:daco} based on the $(\delta_{1,n},\delta)$-entropy
  spectrum partition of $\mcf{Y}_1^n$ to obtain $A_1 \subseteq A$ and
  $\varepsilon_{1,n} \rightarrow 0$ such that
\begin{align}
 \frac{\abs{A_1}}{\abs{A}}  &\geq  \frac{1}{2 [K_{\delta_{1,n},\delta}(\mcf{Y}_1^n)+1]}
\geq \frac{\delta_{1,n}}{3(\delta + \logt \abs{\mcf{Y}_1}) } 
\label{eq:A1} \\
\cexp{\g{Y_1|X}{A_1}{\eta}} 
  &\leq \frac{1}{n}H(Y_1^n | X^n \in A_1) + \varepsilon_{1,n} +
  7.19\delta_{1,n},
\label{eq:gA1<}
\end{align}
for all sufficiently large $n$. On the other hand, 
by Lemma~\ref{lem:15.2}
\begin{equation}
\frac{1}{n}H(Y_1^n | X^n \in A_1) \leq
\cexp{\g{Y_1|X}{A_1}{\eta}} + \frac{2}{n} + \tau_{1,n}
\label{eq:gA1>}
\end{equation}
for some $\tau_{1,n} \rightarrow 0$. 
Moreover since $X^n$ is also conditionally $\gamma_n$-uniform on
$A_1$, applying Lemma~\ref{lem:htoa} gives us
\begin{equation}
\aexp{A_1} - \frac{1}{n}\logt\gamma_n 
\leq \frac{1}{n} H(X^n|X^n \in A_1) \leq \aexp{A_1}.
\label{eq:A1h}
\end{equation}
Note that \eqref{eq:A1}, \eqref{eq:gA1<}, \eqref{eq:gA1>} and
\eqref{eq:A1h} together establish the theorem for the case of $K=1$.

Next, choose $\delta_{2,n} \rightarrow 0$ satisfying
$\frac{\max\left\{\delta_{1,n}, \varepsilon_{1,n}, \tau_{1,n} \right\}}{\delta_{2,n}} \rightarrow 0$.
Apply Lemma~\ref{lem:daco} based on the $(\delta_{2,n},\delta)$-entropy
spectrum partition of $\mcf{Y}_2^n$ and Lemma~\ref{lem:15.2} again to
obtain $A_2 \subseteq A_1$, $\varepsilon_{2,n} \rightarrow 0$, and $\tau_{2,n} \rightarrow 0$ such that
\begin{align}
\frac{\abs{A_2}}{\abs{A_1}}  
&\geq  \frac{1}{2 [K_{\delta_{2,n},\delta}(\mcf{Y}_2^n)+1]}
\geq \frac{\delta_{2,n}}{3 \left( \delta + \logt \abs{\mcf{Y}_2} \right)}
\label{eq:A2} \\
\cexp{\g{Y_2|X}{A_2}{\eta}} &\leq 
\frac{1}{n}H(Y_2^n | X^n \in A_2) + \varepsilon_{2,n} + 7.19\delta_{2,n},
\label{eq:gA2<} \\
\frac{1}{n}H(Y_2^n | X^n \in A_2) &\leq
\cexp{\g{Y_2|X}{A_2}{\eta}} + \frac{2}{n} + \tau_{2,n},
\label{eq:gA2>}
\end{align}
whenever $n$ is sufficiently large. Furthermore,
applying Lemma~\ref{lem:subset}
with \eqref{eq:A2} and \eqref{eq:gA1<} gives us
\begin{equation}
\left| \frac{1}{n}H(Y_1^n | X^n \in A_2) - \cexp{\g{Y_1|X}{A_2}{\eta}}
\right|
\leq 
 \frac{12 \gamma_n \left( \delta + \logt \abs{\mcf{Y}_2} \right)}{\delta_{2,n}}
\cdot \max\left\{\varepsilon_{1,n} + 7.19\delta_{1,n}, \tau_{1,n}+
\frac{2}{n} \right\}. 
\label{eq:gA12>} 
\end{equation}
Again $X^n$ is conditionally $\gamma_n$-uniform on $A_2$, applying
Lemma~\ref{lem:htoa} to $A_2$ gives us the desired bounds on
$H(X^n|X^n \in A_2)$ as in \eqref{eq:A1h}, simply with $A_1$ replaced
by $A_2$. 
Finally putting this, \eqref{eq:A1}, \eqref{eq:A2},
\eqref{eq:gA2<}, \eqref{eq:gA2>}, and \eqref{eq:gA12>} together,
we get the theorem for the case of $K=2$. 

The proof naturally extends for $K>2$ by induction. In specific, we have $A' = A_K$ with
\[
\frac{\abs{A'}}{\abs{A}} 
\geq \prod_{k=1}^{K} \frac{\delta_{k,n}}{3(\delta + \logt\abs{\mcf{Y}_k})}
\geq \frac{1}{n} \prod_{k=1}^{K} \frac{1}{3(\delta + \logt\abs{\mcf{Y}_k})}
\]
where the last inequality is due to the required choice of $\prod_{k=1}^K \delta_{k,n} \geq \frac{1}{n}$ in the induction process. It is clear now that $\mu \defn \prod_{k=1}^{K} \frac{1}{3(\delta + \logt\abs{\mcf{Y}_k})}$ is within the interval $\left(0,\frac{1}{2} \right)$. 
\end{IEEEproof}

\begin{lemma} \label{thm:partition} Fix any $\eta\in (0,1)$ and  
  $\gamma_n \geq 1$ satisfying $n(\gamma_n-1) \rightarrow 0$.  
  Let $X^n$ be $\gamma_n$-uniform distributed over $A \subseteq \mcf{X}^n$.
  For $k \in [1:K]$, suppose that $Y^n_k$ is
  conditionally distributed according to the channel $P^n_{Y_k|X}$
  given $X^n$.  Then there exist a constant $\Gamma >0$, 
  $\epsilon_n \rightarrow 0$, and a
  partitioning index $V$ of $A$ w.r.t. $P_{X^n}$ ranging
  over $[1:\Gamma n^{2}]$ such that
\begin{enumerate}
\item
  $\aexp{A_{V=v}} - \epsilon_n\leq \frac{1}{n}H(X^n |V=v) \leq
  \aexp{A_{V=v}}$ and
\item $\left|\frac{1}{n}H(Y_k^n| V=v)
    -\cexp{g^n_{Y_k|X}(A_{V=v},\eta) }\right| \leq \epsilon_n$ for $k \in
  [1:K]$,
\end{enumerate}
for all $v \in [1:\Gamma n^2]$. Note that
both $\Gamma$ and $\epsilon_n$ work uniformly for all $A \subseteq \mcf{X}^n$.
\end{lemma}
\begin{IEEEproof}
Let $\mu$ and $N$ be as they are in Lemma~\ref{thm:main}. Consider
  $n \geq N $ to be large enough that
  $\gamma_n \left(1-\frac{\mu}{n} \right) \leq 1 - \frac{\mu}{2n}$.  Note that this is possible because $n(\gamma_n-1) \rightarrow 0$.

Using Lemma~\ref{thm:main} on $A$, we immediately obtain
  $A_1 \subseteq A$ that satisfies 1) and 2). In addition,
\[
\Pr\{ X^n \in A \setminus A_1 | X^n \in A\} 
\leq \gamma_n \left(1-\frac{\mu}{n} \right) \leq 1 - \frac{\mu}{2n}.
\]
Next apply Lemma~\ref{thm:main} again on $A\setminus A_1$, we get
  $A_2 \subseteq A\setminus A_1$ satisfying 1), 2), and
\[
\Pr \{ X^n \in A \setminus (A_1 \cup A_2) | X^n \in A \setminus
  A_1\} 
\leq \gamma_n \left(1-\frac{\mu}{n} \right) 
\leq 1 - \frac{\mu}{2n}.
\]
Repeat this process $m-2$ more times to get
$A_v \subseteq A \setminus \bigcup_{j=1}^{v-1} A_j$ satisfying 1),
  2), and
\[
\Pr \bigg\{ X^n \in A\setminus \bigcup_{j=1}^{v} A_j ~\bigg|~ X^n \in
A\setminus \bigcup_{j=1}^{v-1} A_j \bigg\}\leq \gamma_n \left(1-\frac{\mu}{n} \right) 
\leq 1 - \frac{\mu}{2n}
\]
for $v \in [3:m]$. Write
$\tilde A \defn A\setminus \bigcup_{j=1}^{m} A_j$. Then combining the
conditional probability bounds above, we have
$\Pr\{X^n \in \tilde A \} \leq \left(1-\frac{\mu}{2n}
\right)^m$.
Since $X^n$ is $\gamma_n$-uniform over $A$,
$\Pr\{X^n \in \tilde A \} \geq \frac{1}{\gamma_n \abs{\tilde A}} \geq
\frac{1}{\gamma_n \abs{\mcf{X}}^n}$
for all non-empty $\tilde A \subseteq A$. 
In other words, if $\tilde A$ is non-empty, then
\[
m \leq \frac{\logt \gamma_n + n \logt \abs{\mcf{X}}}{ - \logt \left(1-\frac{\mu}{2n}\right)}.
\]
Therefore by picking the smallest $m$ that satisfies
\[
m > \frac{2\ln 2 ( 1+\card{X})n^2}{\mu}
\geq
\frac{\logt \gamma_n + n \logt \abs{\mcf{X}}}{ - \logt \left(1-\frac{\mu}{2n}\right)},
\]
$\tilde A$ must be empty, and  
hence $A = \bigcup_{j=1}^{m} A_j$. Now define the random index $V$ over
$\mcf{V} = [1:m]$ by setting $V = v$ if $X^n \in A_v$. Then $V$ is
clearly a partitioning index of $A$ w.r.t. $P_{X^n}$ and $A_{V=v} = A_v$.
\end{IEEEproof}

\begin{lemma}\label{lem:quick_fix}
Let $E$ be any random variable over a discrete alphabet $\mcf{E}$ and $S$ be any binary random variable over $\set{0,1}$. If $P_S(1) \geq p$, then
\[
\abs{\frac{1}{n} H(E) - \frac{1}{n}H(E|S=1) } \leq \frac{1}{n} + \frac{1-p}{n}\card{E}.
\]
\end{lemma}
\begin{IEEEproof}
First noting that $H(E|S) = H(E|S=1) ( 1 - P_{S}(0)) + H(E|S=0)P_{S}(0)$, we have
\begin{equation*}
\abs{\frac{1}{n} H(E) - \frac{1}{n} H(E|S=1) + \frac{1}{n} P_{S}(0) \left( H(E|S=1) - H(E|S=0) \right)}  = \abs{\frac{1}{n} H(E) - \frac{1}{n}H(E|S)} \leq \frac{1}{n}.
\end{equation*}
This together with the triangular inequality imply
\begin{equation*}
\abs{\frac{1}{n}H(E) - \frac{1}{n} H(E|S=1) } \leq \frac{1}{n} + \frac{1}{n} P_{S}(0) \cdot  \abs{H(E|S=1) - H(E|S=0)} \leq \frac{1}{n} + \frac{1-p}{n}\card{E}.
\end{equation*}
\end{IEEEproof}

\begin{theorem}\label{thm:cond} 
\textbf{(Equal-image-size source partitioning theorem)} 
Fix any $\eta \in (0,1)$.  Let $X^n$ be $\gamma_n$-uniform distributed over
  any $A \subseteq \mcf{X}^n$ for some
  $\gamma_n \geq 1$ satisfying $n(\gamma_n-1) \rightarrow 0$. 
  Suppose that $Y_k^n$ are conditionally distributed according to the
  channel $P^n_{Y_k|X}$ given $X^n$ for $k = [1:K]$, and 
  $M_1, M_2, \dots, M_{J}$ be $J$ are partitioning
  indices of $A $ w.r.t. $P_{X^n}$. Then there exist $\lambda_n \rightarrow 0$ and
  a partitioning index $V^*$ of $A$ w.r.t. $P_{X^n}$, over an index set $\mcf{V}^*$ 
  with cardinality satisfying $\frac{\abs{\mcf{V}^*}}{n^2} \rightarrow 0$,
  such that for every $S \subseteq [1:J]$ and $v \in \mcf{V}^*$: 
\begin{enumerate} 
\item $\cexp{\g{Y_k|X}{A_{M_S=m_S,V^*=v}}{\eta}} \leq \frac{1}{n}H(Y^n_k|M_S,V^*=v) + \lambda_n$ for all $k \in [1:K]$ \\
 \mbox{~~~}and  $m_S \in \mcf{M}_S(v) \defn \{ m_S \in \mcf{M}_S : P_{M_S,V^*}(m_S,v) > 0\}$,
\end{enumerate}
and there exists a $\mcf{\tilde M}_S(v) \subseteq \mcf{M}_S(v)$ satisfying
\begin{enumerate}  \setcounter{enumi}{1}
\item $\abs{\frac{1}{n} \logt\abs{\mcf{\tilde M}_S(v)} - \frac{1}{n} \logt\abs{\mcf{M}_S(v) } } \leq \lambda_n $ 
\item $\abs{\frac{1}{n} H(M_S | V^*=v) - \frac{1}{n} \logt\abs{\mcf{\tilde M}_S(v)}}  \leq  \lambda_n $ 
\item $\abs{ \frac{1}{n}H(Y_k^n|M_S,V^*=v) - \cexp{\g{Y_k|X}{A_{M_S=m_S,V^*=v}}{\eta}} }\leq \lambda_n $ for all $k \in [1:K]$ and $m_S \in \mcf{\tilde M}_S(v)$.
\end{enumerate}
Note that $\lambda_n$ applies
uniformly for all $A \subseteq \mcf{X}^n$.  
For the special case of $S = \emptyset$,
$\mcf{M}_{\emptyset} = \{ m_{\emptyset}\}$ is singleton,
$P_{M_\emptyset}( m_{\emptyset}) = 1$,
$A_{M_{\emptyset}=m_{\emptyset}} = A$, and we may drop the
conditioning notation $M_\emptyset = m_{\emptyset}$ when stating the
results above.
\end{theorem}
\begin{IEEEproof}
  First, pick any $S \subseteq [1:J]$ to focus on. Since $M_{[1:J]}$
  is a partitioning index of $A$, $M_{S}$ is also a partitioning index
  of $A$.  For each $m_{S} \in \mcf{M}_{S}$, apply
  Lemma~\ref{thm:partition} to $A_{M_S=m_S}$ to obtain the
  partitioning index $U(m_S)$ of $A_{M_S=m_S}$. Note that $U(m_S)$
  ranges over $\mcf{U} \defn [1:\Gamma n^{2}]$, and
\begin{align}
\aexp{A_{M_S = m_s, U(m_S) =u }} -\epsilon_n \leq \frac{1}{n}H(X^n | M_S = m_s,  U(m_S)=u) ) 
&\leq \aexp{A_{M_S = m_s, U(m_S) =u }} 
\label{eq:cutting_chiX} \\
\left|\frac{1}{n}H(Y_k^n| M_S = m_s, U(m_{S}) = u )
  -\cexp{g^n_{Y_k|X}(A_{M_S = m_s, U(m_S)=u},\eta) }\right| 
&\leq \epsilon_n, ~~~~k \in [1:K] 
\label{eq:cutting_chi}
\end{align}
whenever $n$ is sufficiently large, uniformly for all $m_S \in \mcf{M}_S$ and $u \in \mcf{U}$ such that $\Pr\{M_S=m_s, U(m_s) = u\} > 0$. We may assume with no loss of generality that $\epsilon_n \geq \max_{k \in [1:K]} \tau_{k,n} + \frac{2}{n}$, where $\tau_{k,n}$'s are as stated in the proof of Lemma~\ref{thm:main}, in addition to $\epsilon_n \rightarrow 0$ as provided by Lemma~\ref{thm:partition}. Thus we also have $n\sqrt{\epsilon_n} \rightarrow \infty$ . Pick any $\delta_n \rightarrow 0$ that satisfies 
$\sqrt{\epsilon_n} \delta_n^{-2^{J+2}(K+2)} \rightarrow 0$.  Consider the $(K+1)$-dimensional lattice $\mcf{I}_{\delta_n} \defn \left[0: \left\lceil \frac{\card{X}}{\delta_n}
  \right\rceil \right] \times \left[0: \left\lceil
    \frac{\logt\abs{\mcf{Y}_1}}{\delta_n} \right\rceil\right] \times \dots \times
\left[0: \left\lceil \frac{\logt\abs{\mcf{Y}_K}}{\delta_n}
  \right\rceil\right]$.
For any
$\mbf{i} = (\mbf{i}(0), \mbf{i}(1),\ldots, \mbf{i}(K)) \in
\mcf{I}_{\delta_n}$, define the index bin
\begin{align*}
\mcf{B}_{\delta_n}(m_S; \mbf{i}) &\defn  \Big\{ u \in \mcf{U}:
  -\frac{\delta_n}{2} \leq \frac{1}{n}H(X^n|M_S=m_S, U(m_S) = u) - \mbf{i}(0)\delta_n  <
    \frac{\delta_n}{2} 
\\ 
& \hspace{80pt} \text{ and } -\frac{\delta_n}{2} \leq
    \frac{1}{n}H(Y_k^n|M_S=m_S, U(m_{S}) = u) - \mbf{i}(k) \delta_n  <
    \frac{\delta_n}{2}, ~~~k \in [1:K] \Big\}.
\end{align*}
Denote the collection of all possible subsets of $[1:J]$ as
$\mcf{S}$. Let $\mcf{V}$ be an index set having the same number of
elements as $\mcf{I}_{\delta_n}^{\abs{\mcf{S}}}$. That is,
\begin{align}
\card{V} 
&= 
\logt \left[\left(\left\lceil\frac{\card{X}}{\delta_n} \right\rceil+1\right)
  \prod_{k=1}^{K} \left(\left\lceil\frac{\logt \abs{\mcf{Y}_k}}{\delta_n}\right\rceil +1\right)
  \right]^{2^J}
\notag \\
&\leq 
  2^{J} \left[ \logt (\card{X} +2\delta_n)+ 
  \sum_{k=1}^{K} \logt (\logt\abs{\mcf{Y}_k}+2\delta_n) -
    (K+1) \logt \delta_n  \right]
\notag \\
& \leq
-2^J (K+2) \logt \delta_n 
\label{eq:thm3_0}
\end{align}
for all sufficiently large $n$. This implies that 
\[ 
\frac{\abs{\mcf{V}}}{n} \leq \frac{1 }{n \delta_n^{2^{J}(K+2)}} \leq  \frac{ 1}{n  \sqrt{\epsilon_n} } \rightarrow 0  
\]
because$\sqrt{\epsilon_n}\delta_n^{-2^{J+2}(K+2)} \rightarrow 0$ and $n\sqrt{\epsilon_n} \rightarrow \infty$.
Consider a one-to-one mapping 
$q: \mcf{I}_{\delta_n}^{\abs{\mcf{S}}} \rightarrow \mcf{V}$. Index the
coordinates of an element in $\mcf{I}_{\delta_n}^{\abs{\mcf{S}}}$ by the
subsets in $\mcf{S}$. For $v \in \mcf{V}$, if
$q(\ldots, \mbf{i}_S, \ldots) = v$, then
$q_{S}^{-1}(v) \defn \mbf{i}_{S}$. For convenience, we also write
$q_{S}^{-1}(v,k) \defn \mbf{i}_{S}(k)$ for each $k \in [0:K]$. 
For each $v \in \mcf{V}$, define
\[
A_{v} \defn \bigcap_{S \in \mcf{S} } \bigcup_{m_{S} \in \mcf{M}_{S}}
\bigcup_{u \in \mcf{B}_{\delta_n}(m_S;q_{S}^{-1}(v)) } A_{M_S=m_S, U(m_{S}) = u}.
\]
and the random index $V$ by setting $V =v$ if $X^n \in A_v$
for $v \in \mcf{V}$. 
Clearly $V$ is a partitioning index of $A$ with $A_{V=v} = A_v$, and
thus $(M_S,V)$ is a partitioning index of $A$. 
Furthermore, for each $m_S \in \mcf{M}_S$ and $u \in \mcf{U}$, define $A_{m_S,u,v} \defn A_{M_S=m_S, U(m_S)=u} \cap A_v$. Then it is easy to verify that $A_{m_S,v} \defn \bigcup_{u \in \mcf{B}_{\delta_n}(m_S;q^{-1}_S(v))} A_{m_S, u,v} = A_{M_S=m_S,V=v}$.
Furthermore for each $\mbf{i} \in \mcf{I}_{\delta_n}$, define
\[
A_{m_S,\mbf{i}} \defn \bigcup_{v': q^{-1}_S(v') = \mbf{i}} A_{M_S=m_S,V=v'} = \bigcup_{u \in \mcf{B}_{\delta_n}(m_S;\mbf{i})} A_{M_S=m_S, U(m_S)=u}.  
\]
Then, by \eqref{eq:cutting_chiX}, for each $v \in \mcf{V}$ such that $A_{m_S, q^{-1}_S(v)} \neq \emptyset$,
\begin{equation}
\abs{A_{M_S=m_S, U(m_{S})=u}} 
\geq 2^{ n\left[ q^{-1}_S(v,0) \delta_n - \frac{\delta_n}{2} \right]}
\label{eq:amcv_max}
\end{equation}
for all $u \in \mcf{B}_{\delta_n}( m_{S}; q^{-1}_S(v))$.
On the other hand, note that $A_{m_S,u,v} \subseteq A_{M_S=m_S, U(m_S)=u}$. Again~\eqref{eq:cutting_chiX} implies
\begin{equation}
\abs{A_{m_S,u,v}} \leq \abs{A_{M_S=m_S,U(m_S)=u}} \leq 
2^{ n \left[ q^{-1}_S(v,0) \delta_n + \frac{\delta_n}{2} +
\epsilon_n \right]}.
\label{eq:amcv_min}
\end{equation}
Hence combining \eqref{eq:amcv_max} and \eqref{eq:amcv_min} gives
\begin{equation}\label{eq:amcv_ratio}
\frac{\max_{(m_S,u) \in \mcf{M}_S\times \mcf{U} } \abs{A_{m_{S},u,v}}}{\min_{m_{S} \in
  \mcf{M}_S:A_{m_S, q^{-1}_S(v)} \neq \emptyset } \min_{u \in \mcf{B}_{\delta_n}( m_{S}; q^{-1}_S(v))} 
\abs{A_{M_S=m_S,U(m_S)=u}}} 
\leq 2^{n \left( \delta_n + \epsilon_n \right)}
\end{equation}
for each $v \in \mcf{V}$.

Now for each $v \in \mcf{V}$, define the index sets
\begin{align*}
\mcf{ M}_S(v) &\defn \left\{ m_S \in \mcf{M}_S: P_{M_S,V}(m_S,v) >0\right\} \\
\Omega_S(v) &\defn \left\{ (m_S,u) \in \mcf{M}_S \times \mcf{B}_{\delta_n}(m_S; q^{-1}_S(v)): 
 A_{M_S=m_S, U(m_S)=u} \neq \emptyset \text{ and }
  \frac{\abs{A_{m_S,u,v}}}{\abs{A_{M_S=m_S, U(m_S)=u}}}
              \geq \sqrt{\epsilon_n} \right\} \\
  \mcf{\hat M}_S(v) & \defn \left\{ m_S \in \mcf{M}_S: \text{ there exists some } u \text{ such that } (m_S,u) \in \Omega_S(v) \right\} \\
  \mcf{\tilde M}_S(v) &\defn \left\{ m_S \in \mcf{\hat M}_S(v): \sum_{u: (m_S,u) \in \Omega_S(v)} \frac{\abs{A_{m_S,u,v}} }{\abs{A_{m_S,v} } }  \geq  1- \delta_n \right\} .
\end{align*}
It is easy to see that $\mcf{\tilde M}_S(v) \subseteq \mcf{\hat M}_S(v) \subseteq \mcf{M}_S(v)$. In addition, since $P_{M_S,V}(m_S,v)>0$ is equivalent to $A_{m_s,v} \neq \emptyset$ and $A_{m_s,v} \subseteq A_{m_S,q_S^{-1}(v)}$, $\mcf{ M}_S(v) \subseteq \{ m_S \in \mcf{M}_S : A_{m_S,q_S^{-1}(v)} \neq \emptyset\}$.
Intuitively $\mcf{ M}_S(v)$ is the set containing all possible values of $m_S \in \mcf{M}_S$ for a particular $v \in \mcf{V}$. In comparison the set $\mcf{\hat M}(v)$ contains all sufficiently probable $m_S$, while $\mcf{\tilde M}_S(v)$ is the subset of $\mcf{\hat M}_S(v)$ which has the properties listed in the theorem statement. Our first goal is to show that $\abs{\mcf{\tilde M}_S(v)}$ is on the same exponential order as $\abs{\mcf{M}_S(v)}$. This is important primarily for the reason that any bound on $\aexp{\mcf{\tilde M}_S(v)}$ will then be a bound for $\aexp{\mcf{M}_S(v)}$. Following this we will establish the properties listed in the theorem.
Toward this end, note that
\begin{align}
P_V(v) 
&=
\sum_{(m_S,u) \in (\mcf{M}_S \times \mcf{U}) \setminus \Omega_S(v)} \Pr\{M_S=m_S, U(m_S) = u, V=v\} +
\sum_{(m_S,u) \in \Omega_S(v)} \Pr\{M_S=m_S, U(m_S) = u, V=v\} 
\notag \\
&\stackrel{\tiny{(a)}}{\leq}
\sum_{(m_S,u) \in (\mcf{M}_S \times \mcf{U}) \setminus \Omega_S(v)} \frac{\gamma_n\sqrt{\epsilon_n}\abs{A_{M_S=m_S,U(m_S)=u}}}{\abs{A}}
    + \sum_{(m_S,u) \in \Omega_S(v)} \frac{\gamma_n\abs{A_{m_S,u,v}}}{\abs{A}}
    \notag \\
&\stackrel{\tiny{(b)}}{\leq}
\gamma_n \sqrt{\epsilon_n} +
\gamma_n^2 P_V(v) \sum_{m_S \in \mcf{\hat M}(v)} \frac{ \sum_{u:(m_S,u) \in \Omega_S(v)}\abs{A_{m_S,u,v}}}{\abs{A_v}}
\notag\\
&\leq \gamma_n \sqrt{\epsilon_n} +
\gamma_n^2 P_V(v) \left(\sum_{m_S \in \mcf{\tilde M}_S(v)} \frac{ \abs{A_{m_S,v}}}{\abs{A_v}}  + 
(1-\delta_n)\sum_{m_S \in \mcf{\hat M}_S(v) \setminus \mcf{ \tilde M}_S(v)}   \frac{\abs{A_{m_S,v}}}{\abs{A_v}}  \right) 
\notag \\
&\stackrel{\tiny{(c)}}{\leq}
\gamma_n \sqrt{\epsilon_n} +
\gamma_n^2 P_{V}(v) \left(  1 - \delta_n + \delta_n  \sum_{m_S \in \mcf{\tilde M}_S(v)} \frac{\abs{A_{m_S,v}}}{\abs{A_{v}}} \right)
\notag \\
& \leq \gamma_n \sqrt{\epsilon_n} + \gamma_n^2(1-\delta_n)P_V(v) +
 \gamma_n^3\delta_n  \sum_{m_S \in \mcf{\tilde M}_S(v)} \frac{\abs{A_{m_S,v}}}{\abs{A}}
\notag \\
& \leq \gamma_n \sqrt{\epsilon_n} + \gamma_n^2(1-\delta_n)P_V(v) +
\frac{\gamma_n^3\delta_n \sum_{m_S \in \mcf{ \tilde M}_S(v)}\sum_{u \in \mcf{B}_{\delta_n}(m_S; q^{-1}_S(v))}\abs{A_{m_S,u,v}}}{
  \sum_{m_S \in \mcf{ M}_S(v)} \sum_{u \in \mcf{B}_{\delta_n}(m_S; q^{-1}_S(v))} \abs{A_{M_S=m_s, U(m_S)=u}}}
\notag \\
& \leq \gamma_n \sqrt{\epsilon_n} + \gamma_n^2(1-\delta_n)P_V(v) +
\frac{\gamma_n^3\delta_n \Gamma n^2 \abs{\mcf{\tilde M}_S(v)} \cdot \max_{m_S \in \mcf{\tilde M}_S(v)} \max_{u \in \mcf{B}_{\delta_n}(m_S; q^{-1}_S(v))} \abs{A_{m_s, u,v}}}{\abs{\mcf{ M}_S(v)} \cdot \min_{m_S \in \mcf{ M}_S(v)} \min_{u \in \mcf{B}_{\delta_n}(m_S; q^{-1}_S(v))} \abs{A_{M_S=m_S, U(m_S)=u}}}
\notag \\
&\stackrel{\tiny{(d)}}{\leq}
\gamma_n \sqrt{\epsilon_n} + \gamma_n^2(1-\delta_n)P_V(v) +
\gamma_n^3\delta_n 2^{n(\delta_n+\epsilon_n)} \Gamma n^2 \cdot 
  \frac{\abs{\mcf{\tilde M}_S(v)}}{\abs{\mcf{M}_S(v)} }
\label{eq:thm3_4_a}
\end{align}
where $(d)$ is from \eqref{eq:amcv_ratio}, $(b)$ results since $M_S$ partitions $A$ and $U(m_S)$ partitions $A_{M_S=m_S}$, and $(a)$ results from the following fact: 
If $ (m_S,u) \in (\mcf{M}_S \times \mcf{U}) \setminus \Omega_S(v)$, then
\begin{align*} 
\Pr\{M_S=m_S, U(m_S) = u, V=v\} 
& \leq
\begin{cases}
\frac{\gamma_n \abs{A_{m_S,u,v}}}{\abs{A}} 
  & \text{if } u \in \mcf{B}_{\delta_n}(m_S; q^{-1}_S(v)) \\
0 & \text{otherwise}
\end{cases}
\\
&\leq
\begin{cases}
  \frac{\gamma_n\sqrt{\epsilon_n}\abs{A_{M_S=m_S,U(m_S)=u}}}{\abs{A}} 
&\text{if } u \in \mcf{B}_{\delta_n}(m_S; q^{-1}_S(v)) \text{ and } A_{M_S=m_S,U(m_S)=u} \neq  \emptyset
 \\
 0 & \text{otherwise}.
\end{cases}
\end{align*}
Rearranging (\ref{eq:thm3_4_a}d), we obtain 
\begin{equation}\label{eq:cond_1}
  \aexp{\mcf{ M}_S(v) } - \aexp{\mcf{\tilde M}_S(v)} \leq  \delta_n + \epsilon_n + \cexp{\gamma_n} + \frac{\logt \Gamma n^2}{n} - \cexp{P_V(v)} - \cexp{\left(1 - \chi_n(v) \right)}
\end{equation}
where 
\[
\chi_n(v) \defn \frac{1}{\delta_n}\left( 1 - \frac{1}{\gamma_n^2} \right) + \frac{\sqrt{\epsilon_n}}{\gamma_n \delta_nP_{V}(v)}. 
\] 
For all $v \in \mcf{V}$ such that $P_V(v) \geq \frac{1}{\abs{\mcf{V}}^2}$, 
we have $-\frac{1}{n} \logt P_{V}(v) \leq -\frac{1}{2n} \logt\epsilon_n \rightarrow 0$, and for sufficiently large $n$, 
\begin{equation} \label{eq:chi_n}
\chi_n(v)  \leq \frac{1}{n\delta_n} + \frac{\sqrt{\epsilon_n}}{\delta_n P_{V}(v)}
\leq \frac{1}{n\delta_n} + \frac{\sqrt{\epsilon_n}}{\delta_n^{2^{J+1}(K+2)+1}} \rightarrow 0
\end{equation}
by way of~\eqref{eq:thm3_0}.
Hence the right hand side of inequality~\eqref{eq:cond_1} vanishes as $n \rightarrow \infty$ uniformly for all for $v \in \mcf{V}$ such that $P_V(v) \geq \frac{1}{\abs{\mcf{V}}^2}$. Similarly, rearranging (\ref{eq:thm3_4_a}c), we obtain 
\begin{equation}\label{eq:thm3_7}
  \sum_{m_S \in \mcf{\tilde M}_S(v)} \frac{ \abs{A_{m_S,v}}}{\abs{A_v}}  \geq
1 - \chi_n(v)  \rightarrow 1
\end{equation}
uniformly for all $v \in \mcf{V}$ such that $P_V(v) \geq \frac{1}{\abs{\mcf{V}}^2}$. This also implies that
\begin{equation}\label{eq:3_7_pv_bound}
  \Pr \left\{ M_S \in \mcf{\tilde M}_S(v) \middle| V = v\right\} 
\geq \frac{1}{\gamma_n} \sum_{m_S \in \mcf{\tilde M}_S(v)}\frac{\abs{A_{m_S,v} }}{\abs{A_v} }
\geq 1- \left( 1 - \frac{1}{\gamma_n} \right) - \frac{\chi_n(v)}{\gamma_n}
\geq 1 - \frac{1}{n} - \chi_n(v) 
\rightarrow 1
\end{equation}
uniformly for all $v \in \mcf{V}$ such that $P_V(v) \geq \frac{1}{\abs{\mcf{V}}^2}$.
This lower bound on 
$\Pr \left\{ M_S \in \mcf{\tilde M}_S(v) \middle| V = v\right\}$ will be useful later in the proof.


Next let us concentrate on bounding $\abs{\frac{1}{n} H(M_S| V=v) - \aexp{\mcf{\tilde M}_S(v)}}$.
Fix any $v \in \mcf{V}$. For each $m_S \in \mcf{\tilde M}_S(v)$, let $u' = \arg\max_{u:(m_S,u) \in \Omega_S(v)} \abs{A_{m_S,u,v}}$. Then
\begin{equation} \label{eq:pms|v}
\frac{\sqrt{\epsilon_n}\abs{A_{M_S=m_S,U(m_S)=u'}}}{\gamma_n\abs{A_v}}
\leq 
\frac{\abs{A_{m_S,u',v}}}{\gamma_n\abs{A_v}} 
\leq
P_{M_S|V}(m_S|v) = \Pr\{X^n \in A_{m_S,v} | X^n \in A_v\} 
\leq
\frac{\gamma_n \abs{A_{m_S,v}}}{\abs{A_v}}
\leq
\frac{\gamma_n \abs{\mcf{U}} \abs{A_{m_S,u',v}}}{\abs{A_v}}. 
\end{equation}
Thus, combining~\eqref{eq:pms|v} and~\eqref{eq:amcv_ratio} shows that
\begin{align} 
\frac{\max_{m_S \in \mcf{\tilde M}_S(v)} P_{M_S|V}(m_S|v)}{\min_{m_S \in \mcf{\tilde M}_S(v)} P_{M_S|V}(m_S|v)}
&  \leq \frac{\abs{\mcf{U}} \gamma_n^2}{\sqrt{\epsilon_n}} \cdot \frac{\max_{(m_S,u) \in \Omega_S(v)}
  \abs{A_{m_S,u,v}} }{\min_{m_S \in \mcf{ M}_S(v)} \min_{u \in \mcf{B}_{\delta_n}(m_S; q^{-1}_S(v)) }\abs{A_{M_S=m_S, U(m_S)=u}}} 
\notag \\
& \leq 2^{n\left( \delta_n + \epsilon_n + \frac{2}{n}\logt\gamma_n-
    \frac{1}{2n}\logt\epsilon_n + \aexp{\mcf{U}} \right)}.
\label{eq:thm3_1}
\end{align}
Next for all $m_S \in \mcf{\tilde M}_S(v)$ we have $\Pr \left\{ M_S = m_S \middle| V=v \right\} = \Pr \left\{ M_S = m_S \middle| V=v, M_{S} \in \mcf{\tilde M}_S(v) \right\} \Pr \left\{ M_S \in \mcf{\tilde M}_S(v) \middle| V=v \right\}$. Thus because equation~\eqref{eq:thm3_1} only considers $m_S \in \mcf{\tilde M}_S(v)$ in the ratio, we have
\[
\frac{\max_{m_S \in \mcf{\tilde M}_S(v)} P_{M_S|V}(m_S|v)}{\min_{m_S \in \mcf{\tilde M}_S(v)} P_{M_S|V}(m_S|v)} = \frac{\max_{m_S \in \mcf{\tilde M}_S(v)} \Pr \left\{ M_S = m_S | V=v , M_S \in \mcf{ \tilde M}_S(v) \right\} }{\min_{m_S \in \mcf{\tilde M}_S(v)} \Pr \left\{ M_S =  m_S | V=v , M_S \in \mcf{ \tilde M}_S(v) \right\}}.
\]
Then applying Lemma~\ref{lem:htoa} directly leads to
\begin{align}
\abs{\frac{1}{n} H(M_S | V=v, M_S \in \mcf{\tilde M}_S(v) ) - \aexp{\mcf{ \tilde M}_S(v)} }&\leq  \delta_n + \epsilon_n + \frac{2}{n}\logt\gamma_n-
    \frac{1}{2n}\logt\epsilon_n + \frac{1}{n} \logt  \abs{\mcf{U}}   \label{eq:thm_3_7_c_2} 
\end{align}  
It remains to bound the difference between $H(M_S|V=v,M_S \in \mcf{\tilde M}_S(v))$ and $H(M_S|V=v)$. Recall that 
\[
\Pr \left\{ M_S \notin \mcf{\tilde M}_S(v) \middle| V=v\right\} \leq  \frac{1}{n} + \chi_n(v)
\]
by~\eqref{eq:3_7_pv_bound} and that $\abs{\mcf{M}_S} \leq \abs{\mcf{X}}^n$ because $M_S$ partitions $A \subset \mcf{X}^n$. Thus by Lemma~\ref{lem:quick_fix}
\begin{equation}\label{eq:thm_3_7_c_1}
\abs{ \frac{1}{n} H(M_S|V=v) - \frac{1}{n} H(M_S|V=v, M_S \in \mcf{\tilde M}_S(v)) } \leq \frac{1}{n} + \left( \frac{1}{n} + \chi_n(v) \right) \card{X}.
\end{equation}
Thus from the triangle inequality and equations~\eqref{eq:thm_3_7_c_2} and \eqref{eq:thm_3_7_c_1}, we obtain
\begin{equation}\label{eq:cond_2}
\abs{ \frac{1}{n} H(M_S|V=v) -  \aexp{\mcf{\tilde M}_S(v)} } \leq   \delta_n + \epsilon_n + \frac{2}{n} \logt \gamma_n + \frac{1}{n} - \frac{1}{2n} \logt \epsilon_n  + \frac{1}{n}\logt\Gamma n^2 + \left(\frac{1}{n}  + \chi_n(v) \right) \card{X} \rightarrow 0
\end{equation}
for all $v \in \mcf{V}$ such that $P_V(v) \geq \frac{1}{\abs{\mcf{V}}^2}$.

Finally we turn our attention to upper bounding $\frac{1}{n} H(Y_k^n|M_S, V=v) - \cexp{ \g{Y_k|X}{A_{m_S,v}}{\eta} }$ for all $m_S \in \mcf{ M}_S(v)$ and $\abs{\frac{1}{n} H(Y_k^n|M_S, V=v) - \cexp{ \g{Y_k|X}{A_{m_S,v}}{\eta} }}$ for all $m_S \in \mcf{\tilde M}_S(v)$.
To obtain bounds for these differences, we will first bound the following terms:
\begin{enumerate}
\item $\abs{\frac{1}{n} H(Y_k^n|M_S, V=v)  -  \frac{1}{n} H(Y_k^n|M_S , M_S \in \mcf{\tilde M}_S(v), V=v)}$ ,
\item $\abs{\frac{1}{n} H(Y_k^n|M_S, M_S \in \mcf{\tilde M}_S(v), V=v) - \frac{1}{n} H(Y_k^n|M_S , U(M_S), M_S \in \mcf{\tilde M}_S(v),V=v ) }$
\item $\abs{\frac{1}{n} H(Y_k^n|M_S, U(M_S),M_S \in \mcf{\tilde M}_S(v), V=v) - \frac{1}{n} H(Y_k^n|M_S ,U(M_S), M_S \in \mcf{\tilde M}_S(v), (M_S,U(M_S)) \in \Omega_S(v), V=v) }$
\item $\abs{\frac{1}{n} H(Y_k^n|M_S ,U(M_S),M_S \in \mcf{\tilde M}_S(v), (M_S,U(M_S)) \in \Omega_S(v), V=v) - q_{S}^{-1}(v,k)\delta_n }$
\item $\cexp{\g{Y_k|X}{A_{m_S,v}}{\eta}} - q_S^{-1}(v,k)\delta_n$ for all $m_S \in \mcf{ M}_S(v)$
\item $\abs{\cexp{\g{Y_k|X}{A_{m_S,v}}{\eta}} -q_S^{-1}(v,k)\delta_n }$ for all $m_S \in \mcf{\tilde M}_S(v)$
\end{enumerate}
and then repeatedly apply the triangle inequality to obtain the desired results.
The bound for 1),
\begin{equation}\label{eq:3_7_ineq_1}
\abs{\frac{1}{n} H(Y_k^n|M_S, V=v)  -  \frac{1}{n} H(Y_k^n|M_S , M_S \in \mcf{\tilde M}_S(v), V=v)} \leq \frac{1}{n} + \left( \frac{1}{n} + \chi_n(v) \right) \logt \abs{\mcf{Y}_k},
\end{equation}
is a result of \eqref{eq:3_7_pv_bound} and Lemma~\ref{lem:quick_fix}. The bound for 2) clearly follows as
\begin{equation}\label{eq:3_7_ineq_2}
\abs{\frac{1}{n} H(Y_k^n|M_S , M_S \in \mcf{\tilde M}_S(v), V=v) - \frac{1}{n} H(Y_k^n|M_S ,U(M_S), M_S \in \mcf{\tilde M}_S(v),V=v ) } \leq \frac{1}{n} \card{U} \leq \frac{1}{n} \logt\Gamma n^2.
\end{equation}
The bound for 3), 
\begin{align}
&\abs{\frac{1}{n} H(Y_k^n|M_S ,U(M_S),  M_S \in \mcf{\tilde M}_S(v) , V=v) - \frac{1}{n} H(Y_k^n|M_S ,U(M_S), M_S \in \mcf{\tilde M}_S(v), (M_S,U(M_S)) \in \Omega_S(v), V=v) } \notag \\
&\hspace{50pt}\leq \frac{1}{n} + \left( \frac{1}{n} + \delta_n \right) \logt \abs{\mcf{Y}_k} \label{eq:3_7_ineq_3} .
\end{align}
also follows from Lemma~\ref{lem:quick_fix} because 
\[
\Pr\left\{ U(m_S) \in \mcf{U}: (m_S,U(m_S)) \in \Omega_S(v) \middle|M_S = m_S, V =v \right\} \geq \frac{1}{\gamma_n} \sum_{u: (m_S,u) \in \Omega_S(v)} \frac{\abs{A_{m_S,u,v} }}{\abs{A_{m_S,v} }} 
\geq \frac{1 - \delta_n}{\gamma_n} 
\geq 1  - \frac{1}{n}  - \delta_n,
\]
for all $m_S\in \mcf{\tilde M}_S(v)$. The bound for 4)
\begin{equation}\label{eq:3_7_ineq_4}
\left| \frac{1}{n} H(Y_k^n|M_S , U(M_S), M_S \in \mcf{\tilde M}_S(v), (M_S,U(M_S)) \in \Omega_S(v) , V=v)  - q_{S}^{-1}(v,k)\delta_n \right| \leq \frac{\delta_n}{2} +  3 \gamma_n \sqrt{\epsilon_n}
\end{equation}
follows from
\begin{equation}\label{eq:HYU=HYMV1_n}
\left| \frac{1}{n} H(Y_k^n |  M_S=m_S, U(m_S) = u )  - \frac{1}{n} H(Y_k^n |  M_S=m_S, U(m_S) = u,V=v )\right| \leq 3 \gamma_n \sqrt{\epsilon_n},
\end{equation}
and
\begin{equation}\label{eq:2414124}
\left| \frac{1}{n} H(Y_k^n |  M_S=m_S, U(m_S) = u )  - q_{S}^{-1}(v,k)\delta_n \right| \leq \frac{\delta_n}{2},
\end{equation}
for all $(m_S,u) \in \Omega_S(v)$. We can obtain~\eqref{eq:HYU=HYMV1_n} from Lemma~\ref{lem:subset}
because of~\eqref{eq:cutting_chi} and $\frac{\abs{A_{m_S,u,v}}}{\abs{A_{M_S=m_S, U(m_S)=u}}} \geq \sqrt{\epsilon_n}$ for all $(m_S,u) \in \Omega_S(v)$ while~\eqref{eq:2414124} is a direct consequence of the construction of $A_{v}$. Now  using the triangle inequalities with~\eqref{eq:3_7_ineq_1}--\eqref{eq:3_7_ineq_4}, we get
\begin{equation} \label{eq:3_7(1-4)}
\abs{\frac{1}{n}H(Y^n_k|M_S,V=v) - q_S^{-1}(v,k)\delta_n} \leq \mu_n
\end{equation}
for all $v \in \mcf{V}$ such that $P_V(v) \geq \frac{1}{\abs{\mcf{V}}^2}$,
where $\mu_n \rightarrow 0$ because of~\eqref{eq:chi_n}.

Continuing on to establish the bounds 5) and 6), notice that 
\begin{align}
\max_{u: (m_S,u) \in \Omega_S(v) } \cexp{\g{Y_k|X}{A_{m_S,u,v}}{\eta}} 
&\leq 
\cexp{\g{Y_k|X}{A_{m_S,v}}{\eta}} 
\notag \\
&\leq 
\frac{1}{n}\card{U} + \max_{u: (m_S,u) \in \Omega_S(v) } \cexp{\g{Y_k|X}{ A_{m_S,u,v}}{\eta}}
\notag \\
&\leq 
\cexp{\Gamma n^2} + \max_{u: (m_S,u) \in \Omega_S(v) }\cexp{\g{Y_k|X}{A_{M_S=m_S,U(m_S)=u}}{\eta}}.
\label{eq:3_7_ISbounds}
\end{align}
Together with~\eqref{eq:cutting_chi} and the definition of $\mcf{B}_{\delta_n}(m_S;q^{-1}_S(v))$, the upper bound in~\eqref{eq:3_7_ISbounds} gives us the bound 5):
\begin{equation}\label{eq:3_7_ineq_5}
\cexp{\g{Y_k|X}{A_{m_S,v}}{\eta}} - q_S^{-1}(v,k)\delta_n \leq 
\frac{\delta_n}{2} + \epsilon_n + \frac{1}{n} \logt\Gamma n^2
\end{equation}
for all $m_S \in \mcf{ M}_S(v)$.
Furthermore, combining ~\eqref{eq:2414124} and again the application of Lemma~\ref{lem:subset} based on~\eqref{eq:cutting_chi} and the restriction that $\frac{\abs{A_{m_S,u,v}}}{\abs{A_{M_S=m_S, U(m_S)=u}}} \geq \sqrt{\epsilon_n}$ for all $(m_S,u) \in \Omega_S(v)$ , we get  
\begin{equation*}
\abs{\max_{u: (m_S,u) \in \Omega_S(v) } \cexp{\g{Y_k|X}{A_{m_S,u,v}}{\eta}} - q_S^{-1}(v,k) \delta_n}
\leq \frac{\delta_n}{2} + 4\gamma_n\sqrt{\epsilon_n}.
\end{equation*}
for all $m_S \in \mcf{\tilde M}_S(v)$. This together with~\eqref{eq:3_7_ISbounds} give us the bound 6):
\begin{equation}\label{eq:3_7_ineq_6}
\abs{\cexp{\g{Y_k|X}{A_{m_S,v}}{\eta}} - q_S^{-1}(v,k)\delta_n } \leq 
\frac{\delta_n}{2} + 4\gamma_n\sqrt{\epsilon_n} + \frac{1}{n} \logt\Gamma n^2
\end{equation}
for all $m_S \in \mcf{\tilde M}_S(v)$.

Now putting~\eqref{eq:3_7(1-4)} and~\eqref{eq:3_7_ineq_5} together, we arrive at
\begin{equation}\label{eq:cond_4}
\cexp{\g{Y_k|X}{A_{m_S,v}}{\eta}} \leq \frac{1}{n}H(Y^n_k|M_S,V=v) + \mu_n + \frac{\delta_n}{2} + \epsilon_n + \frac{1}{n} \logt\Gamma n^2
\end{equation}
for all $m_S \in \mcf{ M}_S(v)$. On the other hand, putting~\eqref{eq:3_7(1-4)} and~\eqref{eq:3_7_ineq_6} together gives us 
\begin{equation}
\abs{\frac{1}{n}H(Y^n_k|M_S,V=v) - \cexp{\g{Y_k|X}{A_{m_S,v}}{\eta}}} \leq \mu_n  + \frac{\delta_n}{2} + 4\gamma_n\sqrt{\epsilon_n} + \frac{1}{n} \logt\Gamma n^2
\label{eq:cond_3}
\end{equation}
for all $m_S \in \mcf{\tilde M}_S(v)$. 

Summarizing~\eqref{eq:cond_1},~\eqref{eq:cond_2},~\eqref{eq:cond_3}, and~\eqref{eq:cond_4}, by letting $\mcf{\tilde V} \defn \{ v \in \mcf{V} : P_{V}(v) \geq \frac{1}{\abs{\mcf{V}}^2} \}$ there exist a $\lambda_n \rightarrow 0$ such that for all $v \in \mcf{\tilde V}$,
\begin{align}
\abs{ \aexp{\mcf{M}_S(v) } - \aexp{\mcf{\tilde M}_S(v)}} &\leq  \lambda_n \notag \\
\abs{ \frac{1}{n} H(M_S|V=v) -  \aexp{\mcf{\tilde M}_S(v)} } &\leq  \lambda_n \notag \\
\cexp{\g{Y_k|X}{A_{M_S=m_S,V=v}}{\eta}} - \frac{1}{n}H(Y^n_k|M_S,V=v)&\leq \lambda_n 
~~~~\text{for all } m_S \in \mcf{M}_S(v) \text{ and all } k \in [1:K] .
\notag \\
\abs{\frac{1}{n}H(Y^n_k|M_S,V=v) - \cexp{\g{Y_k|X}{A_{M_S=m_S,V=v}}{\eta}}} &\leq \lambda_n ~~~~\text{for all } m_S \in \mcf{\tilde M}_S(v) \text{ and all } k \in [1:K] .
\label{eq:part_bounds}                                                                       
\end{align}
Note that $\lambda_n$ works uniformly for all $A  \subseteq \mcf{X}^n$. 
Let $\tilde A \defn \bigcup_{v \in \mcf{\tilde V}} A_{V=v}$. Since $V$ is a partitioning index of $A$ w.r.t. $P_{X^n}$, 
\begin{equation}\label{eq:part_prob}
P_{X^n}(\tilde A) = P_V(\mcf{\tilde V}) \geq 1 - \frac{1}{\abs{\mcf{V}}}.
\end{equation}

To complete the proof, we will apply~\eqref{eq:part_bounds} and~\eqref{eq:part_prob} iteratively as below. First apply the result on $A$ to obtain the partitioning index $V_1$ of $A$ w.r.t. $P_{X^n}$, $\mcf{V}_1 \subseteq \mcf{V}$, and $A_1 = \bigcup_{v \in \mcf{V}_1} A_{V_1=v}$ such that~\eqref{eq:part_bounds} holds for all $v \in \mcf{V}_1$ and $P_{X^n}(A_1) \geq 1 - \frac{1}{\abs{\mcf{V}}}$. Next apply the result on $A \setminus A_1$ to obtain the partitioning index $V_2$ of $A \setminus A_1$ w.r.t. $P_{X^n|X^n \in A \setminus A_1}$, $\mcf{V}_2 \subseteq \mcf{V}$, and $A_2 = \bigcup_{v \in \mcf{V}_2} A_{V_2=v}$ such that~\eqref{eq:part_bounds} holds for all $v \in \mcf{V}_2$ and $P_{X^n|X^n \in A \setminus A_1}(A_2) \geq 1 - \frac{1}{\abs{\mcf{V}}}$. Repeat the process $m-2$ times to get, for each $i \in [3:m]$, the partitioning index $V_i$ of $A \setminus \bigcup_{l=1}^{i-1} A_l$ w.r.t. $P_{X^n| X^n \in A \setminus \bigcup_{l=1}^{i-1} A_l}$, $\mcf{V}_i \subseteq \mcf{V}$, and $A_i  = \bigcup_{v \in \mcf{V}_i} A_{V_i=v}$ such that~\eqref{eq:part_bounds} holds for all $v \in \mcf{V}_i$ and $P_{X^n| X^n \in A \setminus \bigcup_{l=1}^{i-1} A_l}(A_i) \geq 1 - \frac{1}{\abs{\mcf{V}}}$. Then 
\[
P_{X^n}\left(A \setminus \bigcup_{l=1}^{m} A_l\right) 
\leq \left( \frac{1}{\abs{\mcf{V}}} \right)^m.
\]
Thus, as in the proof of Lemma~\ref{thm:partition}, if we pick $m \geq 2n\card{X} \geq \frac{n (\logt\gamma_n + \card{X}) }{\card{V}}$, $A \setminus \bigcup_{l=1}^{m} A_l$ will be empty, i.e., $A =  \bigcup_{l=1}^{m} A_l$. Finally, define and set the random index $V^* = (l,v)$ if $X^n \in A_l \cap A_{V_l = v}$. Then $V^*$ partitions $A$ w.r.t. to $P_{X^n}$ and~\eqref{eq:part_bounds} holds for all $v \in \mcf{V}^*$, where $\mcf{V}^*$ denotes the alphabet of $V^*$. It is also clear that $\abs{\mcf{V}^*}$ is at most $(2n\card{X}) \abs{\mcf{V}}$.   
\end{IEEEproof}

\section{Strong Fano's inequalities} \label{sec:fanos}

In this section, we develop stronger versions of Fano's inequality for 
the multi-terminal communication scenario in which a set of $J$ messages $M_1, M_2, \ldots, M_J$ ranging over $\mcf{M}_1, \mcf{M}_2, \ldots, \mcf{M}_J$, respectively, are to be sent to $K$ receivers through a set of $K$ DMCs $P_{Y_1|X}, P_{Y_2|X}, \ldots, P_{Y_K|X}$.  We will see that these strong Fano's inequalities are direct consequences of the source partitioning results obtained in Theorem~\ref{thm:cond}. More specifically, let $A$, which can be any subset of $\mcf{X}^n$, denote the set of possible codewords. Let $S_1, S_2, \ldots, S_K$ be any $K$ non-empty subsets of $[1:J]$ with the interpretation that the $k$th receiver is to decode the message $M_{S_k}$.  Let $F^n : \mcf{M}_1 \times \dots \times \mcf{M}_J \rightarrow A$ denote the encoding function and $\Phi^n_{k}: Y_k^n \rightarrow \mcf{M}_{S_k}$ denote the decoding function employed by the $k$th receiver, for $k \in [1:K]$. We allow the encoding and decoding functions to be stochastic with $F$ and $\Phi_k$ specified by the conditional distributions $P_{X^n|M_{[1:J]}}$ and $P_{\Phi^n_k|Y^n_k}$, respectively. As mentioned before, distributed encoding is allowed in this model. For example, if there are $L$ distributed encoders, each generates a codeword in $\mcf{X}_l^n$ for $l \in [1:L]$, we may set $\mcf{X}^n = \mcf{X}_1^n \times \cdots \times \mcf{X}_L^n$, $A = A_1 \times \cdots \times A_L$, where $A_l \subseteq \mcf{X}_l^n$, and disjointly distribute the $J$ messages to the $L$ encoders. The Cartesian product of these $L$ distributed encoders then forms the encoding function $F$. If $P_{X^n|M_{[1:J]}}$ is degenerative, the encoding function is deterministic.
Similarly if $P_{\Phi^n_k|Y^n_k}$ is degenerative, the $k$th decoding function is deterministic.

We consider two decoding error criteria, namely, the average and maximum error probabilities. For any encoder-decoder pair $(F^n,\Phi^n_k)$, the average error probability is $\Pr\{ \Phi^n_k(Y^n_k) \neq M_{S_k} \}$ for the $k$th receiver. On the other hand, the maximum error probability is defined as 
\[
\max_{(m_{S_k},x^n) \in \mcf{M}_{S_k} \times A: P_{M_{S_k}, X^n}(m_{S_k},x^n) > 0} \Pr\{ \Phi^n_k(Y^n_k) \neq M_{S_k} | X^n = x^n, M_{S_k} = m_{S_k}\}.
\]
If the encoding function is deterministic (and one-to-one), the maximum error probability reduces to \[
\max_{m_{[1:J]} \in \mcf{M}_{[1:J]}} \Pr\{ \Phi^n_k(Y^n_k) \neq M_{S_k} | M_{[1:J]} = m_{[1:J]}\}.
\]
Note that this maximum-error criterion is stricter than one using the probability $\max_{m_{S_k} \in \mcf{M}_{S_k}} \Pr\{ \Phi^n_k(Y^n_k) \neq M_{S_k} | M_{S_k} = m_{S_k}\}$. We argue that the former more naturally conveys the notion of maximum error in multi-terminal settings. This is because the former truly describes the maximum decoding error probability at the $k$th receiver over all possible transmitted codewords, while the latter is a somewhat unnatural mix between maximum (over all messages that are intended for the $k$th receiver) and average (over all messages that are not) error probabilities.  

The stronger versions of Fano's inequality are first derived under the maximum error criterion in Theorem~\ref{thm:maxerr_fano}--Corollary~\ref{cor:max_fano}. The stronger versions of Fano's inequality under the average error criterion are then obtained as consequences of Theorem~\ref{thm:maxerr_fano} in Theorem~\ref{thm:isfd}--Corollary~\ref{cor:stable}.

\begin{theorem}\label{thm:maxerr_fano} 
  Consider a set of $K$ DMCs
  $P_{Y_1|X}, P_{Y_2|X}, \ldots, P_{Y_K|X}$ and a set of $J$ messages
  $M_1, M_2, \ldots, M_J$ ranging over
  $\mcf{M}_1, \mcf{M}_2, \ldots, \mcf{M}_J$, respectively. Let $A$ be
  any subset of $\mcf{X}^n$ and $S_1, S_2, \ldots, S_K$ be any $K$
  non-empty subsets of $[1:J]$. Then
  for any encoding function
  $F^n : \mcf{M}_1 \times \dots \times \mcf{M}_J \rightarrow A$ and
  decoding functions
  $\Phi^n_{k}: \mcf{Y}_k^n \rightarrow \mcf{M}_{S_k}$, $k \in [1:K]$, that
  satisfy the maximum-error conditions that
  \begin{equation*}
  \frac{-1}{n} \logt \left(1-\max_{(m_{S_k},x^n) \in \mcf{M}_{S_k} \times A :
      P_{M_{S_k},X^n}(m_{S_k},x^n) > 0}\Pr\left\{
    \Phi^n_k(Y^n_k) \neq M_{S_k} \,\middle|\,
    M_{S_k} = m_{S_k},X^n = x^n \right\} \right)  \rightarrow 0 
  \end{equation*}
  for all $k \in [1:K]$, there exist an index set $\mcf{Q}$ with cardinality no larger than $\Gamma n^{5}$ for some $\Gamma>0$, $\zeta_n \rightarrow 0$,
  and a partitioning index $Q$, which ranges over $\mcf{Q}$, of $\mcf{M}_{[1:J]} \times A$ w.r.t. $P_{M_{[1:J]},X^n}$ such that for all $k \in [1:K]$, $Q \markov X^n \markov Y^n_k$ and
\begin{equation} \label{eq:max_err_fano}
\aexp{\mcf{M}_{S_k}(q)} \leq \frac{1}{n}I(M_{S_k};Y_k^n | Q=q) + \zeta_n
\end{equation}
for each $q \in \mcf{Q}$ except an element $q_0$ satisfying
$P_{Q}(q_0) \leq \frac{1}{\abs{\mcf{X}}^n}$, where
$\mcf{M}_{S_k}(q) \defn \{ m_{S_k} \in \mcf{M}_{S_k}: P_{M_{S_k},Q}(m_{S_k},q) > 0 \}$.
Note that the same $\Gamma$ and $\zeta_n$ apply uniformly for all $A \subseteq \mcf{X}^n$ and all $\mcf{M}_{[1:J]}$.
In addition, if $F^n$ is such that $M_{[1:J]}$ partitions $A$ w.r.t. $P_{X^n}$, then 
$Q$ also partitions $A$ w.r.t. $P_{X^n}$.
\end{theorem}

\begin{IEEEproof}
First, let us restrict the encoding function $F^n$ be such that $M_{[1:J]}$ partitions $A$ w.r.t. $P_{X^n}$.
For convenience, write for each $k\in [1:K]$
\begin{align}
\alpha_{k,n} 
&\defn 
\min_{(m_{S_k},x^n) \in \mcf{M}_{S_k} \times A: P_{M_{S_k}, X^n}(m_{S_k},x^n) > 0}
  \Pr\{ \Phi^n_k(Y^n_k)= M_{S_k} | X^n = x^n, M_{S_k} = m_{S_k}\} 
\notag \\
&=
\min_{m_{S_k} \in \mcf{M}_{S_k}: A_{M_{S_k}=m_{S_k}} \neq \emptyset} \, \min_{x^n \in A_{M_{S_k}=m_{S_k}} } 
\Pr\{ \Phi^n_k(Y^n_k) = m_{S_k} \,|\, X^n = x^n\}
\label{eq:alpha}
\end{align}
where the equality is due to the assumption that $M_{S_k}$ partitions $A$, which also implies $\{ M_{S_k} = m_{S_k}\} = \{ X^n \in A_{M_{S_k}=m_{S_k}} \}$.  Note that $1-\alpha_{k,n}$ is the maximum conditional decoding error probability of the decoder $\Phi^n_k$.

By Lemma~\ref{lem:x_slice}, we can find an index set $\mcf{W}$ of $\Gamma n^{3}$ elements for some $\Gamma>0$, a partitioning index $W$ of $A$ w.r.t.  $P_{X^n}$, ranging over $\mcf{W}$, and an element $w_0 \in \mcf{W}$ with $P_W(w_0) \leq \frac{1}{\abs{\mcf{X}}^n}$ such that $X^n$ is conditional $2^{\frac{2}{n^2}}$-uniformly distributed given $W=w$ for each $w \in \mcf{W}$ except $w_0$. Note that $n(2^{\frac{2}{n^2}} -1) \rightarrow 0$.
Also note that $(M_{[1:J]}, W)$ and hence $(M_{S_k},W)$ for each $k\in[1:K]$ are partitioning indices of $A$ w.r.t.  $P_{X^n}$, and $A_{M_{S_k}=m_{S_k}, W=w} =A_{M_{S_k}=m_{S_k}} \cap A_{W=w}$. Thus for each $w \in \mcf{W}$ such that $P_W(w)>0$, $M_{[1:J]}$ and hence $M_{S_k}$ for $k\in[1:K]$ are partitioning indices of $A_{W=w}$ w.r.t. $P_{X^n|W=w}$. It is important to point out that $A_{M_{S_k}=m_{S_k}, W=w}$ may be empty for some $m_{S_k} \in \mcf{M}_{S_k}$. 

For convenience, write $\mcf{\tilde W} \defn \mcf{W} \setminus \{w_0\}$. 
Next, for any fixed $\eta \in (0,1)$ and $w \in \mcf{\tilde W}$, apply Theorem~\ref{thm:cond} to the partitioning index $M_{[1:J]}$ of $A_{W=w}$ w.r.t. $P_{X^n|W=w}$, we can find a partitioning index $V(w)$ of $A_{W=w}$ w.r.t. $P_{X^n|W=w}$, ranging over $\mcf{V}$ with $\frac{\abs{\mcf{V}}}{n^2} \rightarrow 0$. Defining for each $k\in [1:K]$, $w \in \mcf{\tilde W}$, and $v \in \mcf{V}$,
\[
\mcf{M}_{S_k}(w,v) \defn \{ m_{S_k} \in \mcf{M}_{S_k}: A_{M_{S_k}=m_{S_k}, W=w,V(w)=v} \neq \emptyset\}
= \{ m_{S_k} \in \mcf{M}_{S_k}: \Pr\{M_{S_k}=m_{S_k}, W=w,V(w)=v\} > 0\},
\]
Theorem~\ref{thm:cond} guarantees the existence of a $\lambda_n \rightarrow 0$ and $\mcf{\tilde M}_{S_k}(w,v) \subseteq \mcf{M}_{S_k}(w,v)$ for each $w \in \mcf{\tilde W}$, $v \in \mcf{V}$, and $k \in [1:K]$ that
\begin{align}\label{eq:use thm3_6}
\abs{\aexp{\mcf{\tilde M}_{S_k}(w,v)}- \aexp{\mcf{M}_{S_k}(w,v)}}
&\leq \lambda_n, \\
\label{eq:isfd_msk}
\abs{\aexp{\mcf{\tilde M}_{S_k}(w,v)}- \frac{1}{n} H(M_{S_k} | W=w,V(w) = v) }
& \leq \lambda_n, \\ 
\label{eq:HY|VW}
\left|\frac{1}{n} H(Y_k^n|W=w, V(w)=v ) -   \cexp{g^n_{Y_k|X}(A_{W=w, V(w)=v}, \eta)} \right|
& \leq \lambda_n, 
\end{align}
and 
\begin{equation} \label{eq:HY|VW_2}
\left|\frac{1}{n} H(Y_k^n|M_{S_k},W=w, V(w)=v ) - 
  \cexp{g^n_{Y_k|X}(A_{M_{S_k}=m_{S_k},W=w, V(w)=v}, \eta)} \right|
\leq \lambda_n \text{~~~~for all } m_{S_k} \in \mcf{\tilde M}_{S_k}(w,v).
\end{equation}
Note that~\eqref{eq:HY|VW} is obtained from Theorem~18 with $S= \emptyset$.

Our strategy for the proof is to take advantage of the source partitioning result of equivalence between minimum image size and entropy described by~\eqref{eq:use thm3_6}--\eqref{eq:HY|VW_2} in each source partition set $A_{W=w, V(w)=v}$, for $w \in \mcf{\tilde W}$ and $v \in \mcf{V}$.  Correspondingly for the source set $A_{W=w, V(w)=v}$, we want to construct decoding sets 
$C_{k}(m_{S_k},w,v) \subseteq \mcf{Y}^n_k$ for $m_{S_k} \in  \mcf{M}_{S_k}(w,v)$ from the original decoder $\Phi^n_k$ that satisfy the property that the total size of these decoding sets is bounded by minimum image sizes as below:
\begin{align}
& \hspace*{-10pt}
  2^{-n\lambda_n} \abs{\mcf{M}_{S_k}(w,v)}  \cdot 
\min_{m_{S_k} \in \mcf{\tilde M}_{S_k}(v,w) } 
g^n_{Y_k|X}\left(A_{M_{S_k}=m_{S_k}, W=w,V(w)=v}, \frac{\alpha_{k,n}}{4} \right) 
\notag \\
&\leq
\sum_{m_{S_k} \in \mcf{M}_{S_k}(w,v)} \abs{C_k(m_{S_k},w,v)} 
\leq
\frac{2}{\alpha_{k,n}}  g^n_{Y_k|X}\left(A_{W=w,V(w)=v}, 1 - \frac{\alpha_{k,n}}{4} \right).
  \label{eq:c_prop}
\end{align}
To accomplish this we set $C_{k}(m_{S_k},w,v) = B_k(w,v) \cap \tilde B_{k}(m_{S_k})$, where $B_k(w,v) \subseteq \mcf{Y}_k^n$ is a $\left( 1 - \frac{\alpha_{k,n}}{4} \right)$-image of $A_{W=w,V(w)=v}$ that achieves the minimum size $g^n_{Y_k|X}\left( A_{W=w,V(w)=v}, 1 - \frac{\alpha_{k,n}}{4} \right)$ and   
\[
\tilde B_k(m_{S_k}) \defn \left\{ y^n_k : P_{\Phi^n_k|Y^n_k}(m_{S_k} |
  y^n_k) \geq \frac{\alpha_{k,n} }{2} \right\},
\]
for each $k \in [1:K]$ and $m_{S_k} \in \mcf{M}_{S_k}(w,v)$. 

To verify the upper bound in~\eqref{eq:c_prop}, note that 
\[
1 \geq \sum_{m_{S_k}: y_k^n \in  C_k(m_{S_k},w,v)} P_{\Phi^n_k|Y^n_k}(m_{S_k}|y^n_k) \geq  \abs{ \{m_{S_k} \in \mcf{M}_{S_k}(w,v): y^n_k \in C_{k}(m_{S_k},w,v) \}} \cdot \frac{\alpha_{k,n}}{2}
\]
where the last inequality is due to the implication that  
if $y_k^n \in C_{k}(m_{S_k},w,v) \subseteq \tilde B_k(m_{S_k})$, then $P_{\Phi^n_k|Y^n_k}(m_{S_k}|y^n_k) \geq \frac{\alpha_{k,n}}{2}$. Hence for each $y^n_k \in \mcf{Y}_k^n$,
\begin{equation}\label{eq:10100101}
\abs{ \{m_{S_k} \in \mcf{M}_{S_k}(w,v): y^n_k \in C_{k}(m_{S_k},w,v) \}} \leq \frac{2}{\alpha_{k,n}}.
\end{equation}
By direct counting now
\begin{align*}
\sum_{m_{S_k} \in \mcf{M}_{S_k}(v,w)} \abs{C_k(m_{S_k},w,v)} 
&=  \sum_{y_k^n \in \bigcup_{m_{S_k} \in \mcf{M}_{S_k}(w,v)} C_k(m_{S_k},w,v)} 
  \abs{ \{m_{S_k} \in \mcf{M}_{S_k}(w,v): y^n_k \in C_{k}(m_{S_k},w,v) \}} \\
&\stackrel{(a)}{\leq} \frac{2}{\alpha_{k,n}} 
  \cdot \abs{\bigcup_{m_{S_k} \in \mcf{M}_{S_k}(w,v)} C_k(m_{S_k},w,v)} \\
&\stackrel{(b)}{\leq} \frac{2}{\alpha_{k,n}} \cdot 
  g^n_{Y_k|X}\left( A_{W=w,V(w)=v}, 1 - \frac{\alpha_{k,n}}{4} \right)
\end{align*}
where $(a)$ is due to~\eqref{eq:10100101} and $(b)$ results because $\bigcup_{m_{S_k} \in \mcf{M}_{S_k}(w,v)} C_k(m_{S_k},w,v) \subseteq B_k(w,v)$. Thus we have confirmed the upper bound in~\eqref{eq:c_prop}.

For the lower bound in~\eqref{eq:c_prop}, we first show that $P_{Y_k^n|X^n} \left( C_{k}(m_{S_k},w,v) \middle| x^n\right) \geq \frac{\alpha_{k,n}}{4}$ for all $x^n \in A_{M_{S_k} = m_{S_k},W=w,V(w)=v}$. 
To this end, first observe that
\[
\alpha_{k,n} 
\leq \sum_{y^n_k \in \mcf{Y}^n_k}
P_{\Phi^n_k|Y^n_k}(m_{S_k}|y^n_k) P^n_{Y_k|X}(y^n_k|x^n)
\leq \frac{\alpha_{k,n}}{2} \left[ 1 - P^n_{Y_k|X}(\tilde
  B_k(m_{S_k})|x^n )
\right] + P^n_{Y_k|X}(\tilde B_k(m_{S_k})|x^n)
\]
and hence
\begin{equation} \label{eq:sto_decoder1}
P^n_{Y_k|X}\left(\tilde B_k(m_{S_k})\middle|x^n
  \right) 
\geq
\frac{\alpha_{k,n}}{2-\alpha_{k,n}} \geq \frac{\alpha_{k,n}}{2}
\end{equation}
for all $x^n \in A_{M_{S_k} = m_{S_k}}$ and all $m_{S_k} \in
  \mcf{M}_{S_k}$.
Next because $B_k(w,v)$ is also a $\left(1-\frac{\alpha_{k,n}}{4}\right)$-image of $A_{M_{S_k}=m_{S_k}, W=w,  V(w)=v}$ for each $m_{S_k} \in \mcf{M}_{S_k}(v,w)$, we have
\begin{equation}\label{eq:PYB}
\Pr \left\{ Y_k^n \in B_k(w,v) | X^n = x^n, M_{S_k}=m_{S_k}, W=w, V(w)=v\right\} 
= P_{Y^n_k|X^n}(B_k(w,v) | x^n) 
\geq 1 - \frac{\alpha_{k,n}}{4}
\end{equation}
for all $x^n \in A_{M_{S_k}=m_{S_k}, W=w, V(w)=v}$. Then for
all $m_{S_k} \in \mcf{M}_{S_k}(w,v)$ and
$x^n \in A_{M_{S_k}=m_{S_k}, W=w, V(w)=v}$,
\begin{align}
P_{Y^n_k|X^n}\left(C_{k}(m_{S_k},w,v) | x^n \right) 
& =
\Pr \left\{ Y_k^n \in C_{k}(m_{S_k},w,v) | X^n =x^n, M_{S_k}=m_{S_k}, W=w,
  V(w)=v \right\} 
\notag \\
&\geq 1 - 
\Pr \left\{ Y_k^n \notin B_k(w,v) | X^n = x^n, M_{S_k}= m_{S_k}, W=w, V(w)=v \right\} 
\notag \\
& \hspace*{17pt}
  - \Pr \left\{ Y_k^n \notin \tilde B_k(m_{S_k}) | 
  X^n=x^n, M_{S_k}= m_{S_k}, W=w, V(w)=v\right\} 
\notag \\
&\geq 1 - \left(1 - \frac{\alpha_{k,n}}{2}\right)  - \frac{\alpha_{k,n}}{4}
\geq \frac{\alpha_{k,n}}{4} \label{eq:prob_tag}
\end{align}
where the second to last inequality is due to \eqref{eq:PYB} and
\eqref{eq:sto_decoder1}. Hence
$C_k(m_{S_k},w,v)$  is an $\frac{\alpha_{k,n}}{4}$-image of $A_{M_{S_k} = m_{S_k},W=w,V(w)=v}$ and therefore 
\[
\abs{ C_{k}(m_{S_k},w,v) } \geq g^n_{Y_k|X}\left(A_{M_{S_k}=m_{S_k},W=w,V(w)=v} , \frac{\alpha_{k,n}}{4} \right).
\] 
Hence follows the lower bound in~\eqref{eq:c_prop} since
\begin{align}
\sum_{m_{S_k} \in \mcf{ M}_{S_k}(w,v)} \abs{C_{k}(m_{S_k},w,v)} &\geq \sum_{m_{S_k} \in \mcf{\tilde M}_{S_k}(w,v)} \abs{C_{k}(m_{S_k},w,v)} \notag \\
&\geq \abs{\mcf{\tilde M}_{S_k}(w,v)} \cdot
\min_{m_{S_k} \in \mcf{\tilde M}_{S_k}(w,v) } g^n_{Y_k|X}\left(A_{M_{S_k}=m_{S_k}, W=w,V(w)=v}, \frac{\alpha_{k,n}}{4} \right) \notag\\
&\stackrel{(a)}{\geq} 2^{-n\lambda_n} \abs{\mcf{M}_{S_k}(w,v)}  \cdot
\min_{m_{S_k} \in \mcf{\tilde M}_{S_k}(w,v) } 
  g^n_{Y_k|X}\left(A_{M_{S_k}=m_{S_k}, W=w,V(w)=v}, \frac{\alpha_{k,n}}{4} \right)
\end{align}
where $(a)$ is due to~\eqref{eq:use thm3_6}. 

Now, taking $\logt$ on the upper and lower bounds of~\eqref{eq:c_prop} and then rearranging gives
\begin{align*}
\aexp{\mcf{M}_{S_k}(w,v)} 
&\leq \cexp{g^n_{Y_k|X}\left(A_{W=w,V(w)=v}, 1 - \frac{\alpha_{k,n}}{4} \right)}
\notag \\
& \hspace{20pt}
-  \min_{m_{S_k} \in \mcf{\tilde M}_{S_k}(v,w) } \cexp{g^n_{Y_k|X}\left(A_{M_{S_k}=m_{S_k}, W=w,V(w)=v}, \frac{\alpha_{k,n}}{4} \right)} + \lambda_n + \cexp{ \frac{2}{\alpha_{k,n}}}.
\end{align*}
From whence it follows that for all $w \in \mcf{\tilde W}$ and $v \in \mcf{V}$
\begin{equation} \label{eq:max_err_fano_final}
\aexp{\mcf{M}_{S_k}(w,v)} \leq \frac{1}{n} I(M_{S_k};Y_k^n|W=w, V(w)=v ) + 2 \tau_n + 3 \lambda_n + \cexp{ \frac{2}{\alpha_{k,n}}}
\end{equation}
because of Lemma~\ref{lem:6.6}, \eqref{eq:HY|VW}, and~\eqref{eq:HY|VW_2}, where $\tau_n \rightarrow 0$ is from Lemma~\ref{lem:6.6}. Let $\mcf{Q} \defn \mcf{\tilde W} \times \mcf{V} \cup \{w_0\}$. Define the random index $Q$ over $\mcf{Q}$ by setting $Q= \begin{cases} 
  (w,v) & \text{if } X^n \in A_{W=w,V(w)=v} \text{ for } w \neq w_0\\
  w_0 & \text{if } X^n \in A_{W=w_0}
\end{cases}$. Noting that $Q$ partitions $A$ w.r.t. $P_{X^n}$ and $\zeta_n \defn 2\tau_n + 3 \lambda_n + \cexp{ \frac{2}{\alpha_{k,n}}} \rightarrow 0$, the theorem statement is established by~\eqref{eq:max_err_fano_final}.

We can now extend inequality~\eqref{eq:max_err_fano} to any encoder $F^n$ by slightly expanding the length of the codewords. The main idea is to construct an equivalent encoder $\tilde F^n$ that gives the same error probabilities by appending extra symbols to the codewords as below.

First we claim that since 
\[
\alpha_{k,n} = \min_{(m_{S_k}, x^n) \in \mcf{M}_{S_k} \times A: P_{M_{S_k},X^n}(m_{S_k},x^n) >0}
\Pr\{ \Phi^n_k(Y^n_k) = M_{S_k} \,|\, X^n = x^n, M_{S_k} = m_{S_k} \}
\]
for each $k \in [1:K]$, we have for each $x^n \in A$
\begin{equation}
\left|\{m_{S_k} \in \mcf{M}_{S_k} : P_{M_{S_k},X^n}(m_{S_k},x^n) >0 \}\right| 
\leq \frac{1}{\alpha_{k,n} }.
\label{eq:not_many}
\end{equation}
Indeed, fix any $x^n \in A$, we have
\begin{align*}
& \hspace*{-10pt}
\alpha_{k,n} \cdot \left| \{m_{S_k} \in \mcf{M}_{S_k} : P_{M_{S_k},X^n}(m_{S_k},x^n) >0 \}\right| 
\notag \\
& \leq 
\sum_{m_{S_k} \in \mcf{M}_{S_k}: P_{M_{S_k},X^n}(m_{S_k},x^n) >0}  
\Pr\{ \Phi^n_k(Y^n_k) = M_{S_k} \,|\, X^n=x^n, M_{S_k} = m_{S_k} \} 
\notag \\
& \stackrel{(a)}{=} 
  \sum_{m_{S_k} \in \mcf{M}_{S_k}: P_{M_{S_k},X^n}(m_{S_k},x^n) >0}  
  \Pr\{ \Phi^n_k(Y^n_k) = m_{S_k} \,|\, X^n=x^n \}
  \leq 1
\end{align*}
where $(a)$ is due to the fact that $M_{S_k} \markov X^n \markov \Phi^n_k(Y^n_k)$. From \eqref{eq:not_many}, we know that there can only be at most $\prod_{k=1}^{K} \left\lceil \frac{1}{\alpha_{k,n}} \right\rceil$ messages $m_{[1:J]} \in\mcf{M}_{[1:J]}$ associated with each $x^n \in A$ such that $P_{M_{[1:J]},X^n}(m_{[1:J]},x^n) >0$. Thus by appending 
$\tilde n \defn \log_{\abs{\mcf{X}}} \left(\prod_{k=1}^{K} \left\lceil \frac{1}{\alpha_{k,n}} \right\rceil \right)$ symbols to the end of each $x^n \in A$, we obtain $\tilde A \subseteq \mcf{X}^{n+\tilde n}$ and an encoding function $\tilde F^n: \mcf{M}_{[1:J]} \rightarrow \tilde A$ such that the first $n$ symbols of $\tilde F^n(M_{[1:J]})$ is the same as $F^n(M_{[1:J]})$ and $M_{[1:J]}$ partitions $\tilde A$. Also note that $\frac{\tilde n}{n} \rightarrow 0$. Moreover, we can construct the decoder $\tilde \Phi^n_k: \mcf{Y}^{n+\tilde n}_k \rightarrow \mcf{M}_{S_k}$ by setting $\tilde \Phi^n_k(Y^{n+\tilde n}_k) = \Phi^n_k (Y^n_k)$ and hence giving the same error probabilities as $\Phi^n_k$. 
Next application of~\eqref{eq:max_err_fano} on the encoder-decoder pairs $(\tilde F^n, \tilde \Phi^n_k)$ over the codeword set $\tilde A$ yields
\begin{equation}
  \frac{1}{n+\tilde n} \logt \abs{\mcf{M}_{S_k}(\tilde q)} \leq \frac{1}{n+\tilde n} I(M_{S_k}; Y^{n+\tilde n}_k | \tilde Q = \tilde q) + \zeta_n
  \label{eq:gse_max}
\end{equation}
where $\tilde Q$ partitions $\tilde A$ or equivalently $\mcf{M}_{[1:J]} \times A$. Note that $\tilde Q$ may not partition $A$, but we still have $\tilde Q \markov X^n \markov Y^n_k$.
Moreover from \eqref{eq:gse_max}, we have
\begin{align*}
 \frac{1}{n} \logt \abs{\mcf{M}_{S_k}(\tilde q)}
  &\leq
    \frac{1}{n} I(M_{S_k}; Y^{n+\tilde n}_k | \tilde Q = \tilde q) + \left(1 + \frac{\tilde n}{n}\right) \zeta_n
  \leq
    \frac{1}{n} I(M_{S_k}; Y^{n}_k | \tilde Q = \tilde q) +\frac{\tilde n}{n} \logt \abs{\mcf{Y}_k} +  \left(1 + \frac{\tilde n}{n}\right) \zeta_n.
\end{align*}
Hence we obtain back the desired inequality~\eqref{eq:max_err_fano} since $\frac{\tilde n}{n} \rightarrow 0$. 
\end{IEEEproof}

\begin{cor}\label{cor:conditional}
Under the conditions of Theorem~\ref{thm:maxerr_fano}, in addition to the stated results, there also exists $\zeta_n \rightarrow 0$ such that 
 all every $k \in [1:K]$ and $\bar S_k \subseteq [1:J] \setminus S_k$,
\begin{equation*}
\aexp{\mcf{M}_{S_k \cup \bar S_k}(q)} - \aexp{\mcf{M}_{\bar S_k}(q)} 
\leq \frac{1}{n}I(M_{S_k};Y_k^n | M_{\bar S_k}, Q=q) + \zeta_n
\end{equation*}
for all $q \in \mcf{Q} \setminus \{q_0\}$. 
\end{cor}
\begin{IEEEproof}
It suffices to prove the corollary under the restriction that $M_{[1:J]}$ partitions $A$. Extending the result to the case of general encoding function $F^n$ follows the same argument of appending extra symbols used at the end of the proof of Theorem~\ref{thm:maxerr_fano}. 

Recall the proof of  Theorem~\ref{thm:maxerr_fano}. 
For any $k \in [1:K]$, fix any $\bar S_k \subseteq [1:J] \setminus S_k$. Consider $\alpha_{k,n}$ as defined in~\eqref{eq:alpha}. For any $(x^n, m_{S_k}, m_{\bar S_k})$ such that $P_{X^n,M_{S_k}, M_{\bar S_k}}(x^n, m_{S_k}, m_{\bar S_k}) >0$, 
\begin{align}
\Pr\{ \Phi^n_k(Y^n_k) = M_{S_k} \,|\, X^n=x^n, M_{S_k} = m_{S_k},  M_{\bar S_k} = m_{\bar S_k}\} 
& \stackrel{(a)}{=} 
\Pr\{ \Phi^n_k(Y^n_k) = m_{S_k} \,|\, X^n=x^n \}
\notag \\
& \stackrel{(b)}{=} 
\Pr\{ \Phi^n_k(Y^n_k) = M_{S_k} \,|\, X^n=x^n, M_{S_k} = m_{S_k}\}
\notag \\
&\geq \alpha_{k,n}
\label{eq:err_bM}
\end{align}
where $(a)$ and $(b)$ are both due to the fact that $(M_{S_k}, M_{\bar S_k}) \markov X^n \markov \Phi^n_k(Y^n_k)$. For each $w \in \mcf{\tilde W}$ and $v \in \mcf{V}$, define
\begin{align*}
\mcf{M}_{\bar S_k}(w,v) &\defn \left\{m_{\bar S_k} \in \mcf{M}_{\bar S_k}
: \Pr\{ M_{\bar S_k} = m_{\bar S_k}, W=w, V(w)=v\} >0 \right\},
\\
\mcf{M}_{S_k \cup \bar S_k}(w,v) &\defn \left\{ (m_{S_k},m_{\bar S_k}) \in \mcf{M}_{S_k \cup \bar S_k}
: \Pr\{M_{S_k}=m_{S_k}, M_{\bar S_k} = m_{\bar S_k}, W=w, V(w)=v\} >0 \right\}.
\end{align*}
Applying Theorem~\ref{thm:cond}, we obtain, in addition to \eqref{eq:use thm3_6}--\eqref{eq:HY|VW_2}, 
$\mcf{\tilde M}_{\bar S_k}(w,v) \subseteq \mcf{M}_{\bar S_k}(w,v)$ and 
$\mcf{\tilde M}_{S_k \cup \bar S_k}(w,v) \subseteq \mcf{M}_{S_k \cup \bar S_k}(w,v)$ for each $w \in \mcf{\tilde W}$, $v \in \mcf{V}$, and $k \in [1:K]$ that
\begin{align}
\label{eq:bmsk}
\abs{\aexp{\mcf{\tilde M}_{\bar S_k}(w,v)}- \aexp{\mcf{M}_{\bar S_k}(w,v)}}
&\leq \lambda_n, \\
\label{eq:bmsk_H}
\abs{\aexp{\mcf{\tilde M}_{\bar S_k}(w,v)}- \frac{1}{n} H(M_{\bar S_k} | W=w,V(w) = v) }
& \leq \lambda_n, \\
\cexp{g^n_{Y_k|X}(A_{M_{\bar S_k}=m_{\bar S_k},W=w, V(w)=v}, \eta)} 
- \frac{1}{n} H(Y_k^n|M_{\bar S_k},W=w, V(w)=v ) 
&\leq \lambda_n 
\notag \\
\label{eq:HY|bM}
& \hspace*{-40pt}\text{for all } m_{\bar S_k} \in \mcf{M}_{\bar S_k}(w,v).
\\
\label{eq:mskbmsk}
\abs{\aexp{\mcf{\tilde M}_{S_k \cup \bar S_k}(w,v)}- \aexp{\mcf{M}_{S_k \cup \bar S_k}(w,v)}}
&\leq \lambda_n, \\
\label{eq:mskbmsk_H}
\abs{\aexp{\mcf{\tilde M}_{S_k \cup \bar S_k}(w,v)}- \frac{1}{n} H(M_{S_k}, M_{\bar S_k} | W=w,V(w) = v) }
& \leq \lambda_n, \\
\left|\frac{1}{n} H(Y_k^n|M_{S_k},M_{\bar S_k},W=w, V(w)=v ) - 
  \cexp{g^n_{Y_k|X}(A_{M_{S_k}=m_{S_k},M_{\bar S_k}=m_{\bar S_k},W=w, V(w)=v}, \eta)} \right|
&\leq \lambda_n 
 \notag \\
\label{eq:HY|MVW}
& \hspace*{-80pt}\text{for all } (m_{S_k},m_{\bar S_k}) \in \mcf{\tilde M}_{S_k \cup \bar S_k}(w,v). 
\end{align}

For each $k \in [1:K]$ and $(m_{S_k},m_{\bar S_k}) \in \mcf{M}_{S_k \cup \bar S_k}(w,v)$, define
$C_{k}(m_{S_k},m_{\bar S_k},w,v) \defn B_k(m_{\bar S_k},w,v) \cap \tilde B_{k}(m_{S_k})$, where $B_k(m_{\bar S_k},w,v)$ is a $\left( 1 - \frac{\alpha_{k,n}}{4} \right)$-image of $A_{M_{\bar S_k}=m_{\bar S_k},W=w,V(w)=v}$ that achieves $g^n_{Y_k|X}\left( A_{M_{\bar S_k}=m_{\bar S_k},W=w,V(w)=v}, 1 - \frac{\alpha_{k,n}}{4}\right)$ and $\tilde B_k(m_{S_k})$ is as before.
Following the arguments in the proof of Theorem~\ref{thm:maxerr_fano}, we get
\begin{align}
\sum_{(m_{S_k},m_{\bar S_k}) \in \mcf{M}_{S_k \cup \bar S_k}(w,v)} \abs{C_{k}(m_{S_k},m_{\bar S_k},w,v) }
&\leq
\frac{2}{\alpha_{k,n}} \sum_{m_{\bar S_k} \in \mcf{M}_{\bar S_k}(w,v)} g^n_{Y_k|X}\left( A_{M_{\bar S_k}=m_{\bar S_k},W=w,V(w)=v}, 1 - \frac{\alpha_{k,n}}{4}\right)\notag \\
&\leq
\frac{2}{\alpha_{k,n}} \abs{ \mcf{M}_{\bar S_k}(w,v) } 
\cdot 2^{n\left[\frac{1}{n} H(Y^n|M_{\bar S_k}, W=w,V(w)=v) + \lambda_n + \tau_n\right]}
\label{eq:hergxZsaf}
\end{align}
using~\eqref{eq:HY|bM} and Lemma~\ref{lem:6.6} ($\tau_n \rightarrow 0$), and
\begin{align}
&\hspace*{-40pt}
  \sum_{(m_{S_k},m_{\bar S_k}) \in \mcf{M}_{S_k \cup \bar S_k}(w,v)} \abs{C_{k}(m_{S_k},m_{\bar S_k},w,v) }
\notag \\
&\geq
2^{-n\lambda_n} \abs{\mcf{M}_{S_k \cup \bar S_k}(w,v)} \min_{(m_{S_k},m_{\bar S_k}) \in \mcf{\tilde M}_{S_k \cup \bar S_k}(w,v)} g^n_{Y_k|X}\left( A_{M_{S_k}=m_{S_k},M_{\bar S_k}=m_{\bar S_k},W=w,V(w)=v}, 1 - \frac{\alpha_{k,n}}{4}\right)\notag\\
&\geq
2^{-n\lambda_n} \abs{\mcf{M}_{S_k \cup \bar S_k}(w,v)} 
\cdot 2^{ n \left[\frac{1}{n} H(Y^n|M_{S_k},M_{\bar S_k}, W=w, V(w) = v) - \lambda_n - \tau_n\right]}
\label{eq:0214lm}
\end{align}
using \eqref{eq:mskbmsk_H}, \eqref{eq:HY|MVW}, and Lemma~\ref{lem:6.6}. 
Then for all $w \in \mcf{\tilde W}$ and $v \in \mcf{V}$,
\begin{equation}
\cexp{\frac{\abs{\mcf{M}_{S_k \cup \bar S_k}(w,v)}}{\abs{\mcf{M}_{\bar S_k}(w,v)}} } \leq \frac{1}{n} I\left(M_{S_k};Y_k^n\middle| M_{\bar S_k}, W=w,V(w)=v \right) + 3\lambda_n + 2 \tau_n + \frac{1}{n}\logt \frac{2}{\alpha_{k,n}},
\end{equation}
which clearly follows~\eqref{eq:hergxZsaf} and~\eqref{eq:0214lm}. The corollary is hence proven since $2 \tau_n + 3\lambda_n + \frac{1}{n}\logt \frac{2}{\alpha_{k,n}} \rightarrow 0$.
\end{IEEEproof}

\begin{remark} \label{rem:lbs}
Although both Theorem~\ref{thm:maxerr_fano} and Corollary~\ref{cor:conditional} are written as upper bounds on the sizes of partitioned message sets, obtaining corresponding lower bounds is trivial. Indeed for every $q \neq q_0$,
\[
\aexp{\mcf{M}_{S_k}(q)} \geq \frac{1}{n} H(M_{S_k}|Q=q), 
\]
and similarly 
\begin{align*}
\aexp{\mcf{M}_{S_k \cup \bar S_k}(q)} - \aexp{\mcf{M}_{\bar S_k}(q)} 
&\stackrel{(a)}{\geq} 
\frac{1}{n} H\left( M_{S_k}, M_{\bar S_k}| Q=q\right) - \frac{1}{n} H\left(  M_{\bar S_k} \middle| Q=q\right)  - \lambda_n 
\\
&=
\frac{1}{n} H\left( M_{S_k} \middle| M_{\bar S_k}, Q=q\right) - \lambda_n
\end{align*}
where $(a)$ is due to \eqref{eq:bmsk} and \eqref{eq:bmsk_H}.
\end{remark}


\begin{cor}\label{cor:max_fano} \textbf{(Strong maximum-error Fano's inequality)} 
Under the conditions of Theorem~\ref{thm:maxerr_fano}, there exists $\zeta'_n \rightarrow 0$ 
such that 
\begin{align*}
\frac{1}{n}H(M_{S_k}|M_{\bar S_k} ) &\leq \frac{1}{n}I(M_{S_k};Y_k^n|M_{\bar S_k}) + \zeta'_n,
\end{align*}
for all $\bar S_k \subseteq [1:J] \setminus S_k$ and $k \in [1:K]$.
\end{cor}
\begin{IEEEproof}
To verify the corollary, revisiting the proof of Theorem~\ref{thm:maxerr_fano} and note that
\begin{align*}
\frac{1}{n} H(M_{S_k} |M_{\bar S_k}, W=w_0) 
\leq \frac{1}{n} H(M_{S_k} | W=w_0) 
& \leq 
\aexp{\{ m_{S_k} \in \mcf{M}_{S_k} : P_{M_{S_k},W}(m_{S_k},w_0)>0 \}}
\notag \\
&= 
\aexp{\{ m_{S_k} \in \mcf{M}_{S_k} : \sum_{x^n \in A_{W=w_0}} P_{M_{S_k},X^n}(m_{S_k},x^n)>0 \}}
\notag \\
& \leq 
\cexp{ \sum_{x^n \in A_{W=w_0}} \abs{\{ m_{S_k} \in \mcf{M}_{S_k} : P_{M_{S_k},X^n}(m_{S_k},x^n)>0 \}}}
\notag \\
&\stackrel{(a)}{\leq} 
\cexp{\frac{\abs{A_{W=w_0}}}{\alpha_{k,n}}}
\leq \card{X} - \cexp{\alpha_{k,n}}
\end{align*}
where $(a)$ is due to~\eqref{eq:not_many}. This means that 
\[
\frac{1}{n} H(M_{S_k} |M_{\bar S_k}, Q=q_0)  \cdot P_Q(q_0)  \leq
\frac{1}{n} H(M_{S_k} | Q=q_0) \cdot P_Q(q_0) \leq \frac{1}{\abs{\mcf{X}}^n} \left(\card{X}  - \cexp{\alpha_{k,n}}\right) \rightarrow 0.
\]
As a result, the corollary follows from Remark~\ref{rem:lbs} and the fact that
\begin{align*}
\abs{\frac{1}{n} H(M_{S_k}|M_{\bar S_k}) - \frac{1}{n} H(M_{S_k}|M_{\bar S_k}, Q) } 
&\leq \aexp{\mcf{Q}} \leq \cexp{\Gamma n^5}.
\end{align*}
\end{IEEEproof}

\begin{theorem}\label{thm:isfd} 
  Consider a set of $K$ DMCs 
  $P_{Y_1|X}, P_{Y_2|X}, \ldots, P_{Y_K|X}$ and a set of $J$ messages
  $M_1, M_2, \ldots, M_J$ ranging over
  $\mcf{M}_1, \mcf{M}_2, \ldots, \mcf{M}_J$, respectively. Let $A$ be
  any subset of $\mcf{X}^n$ and $S_1, S_2, \ldots, S_K$ be any $K$
  non-empty subsets of $[1:J]$. Then
  for any encoding function
  $F^n : \mcf{M}_1 \times \dots \times \mcf{M}_J \rightarrow A$ and
decoding functions
$\Phi^n_{k}: \mcf{Y}_k^n \rightarrow \mcf{M}_{S_k}$, $k \in [1:K]$, that satisfy the average-error
conditions that
\[
-\frac{1}{n} \logt \left(1-\Pr\{ \Phi^n_k(Y^n_k) \neq M_{S_k}\} \right)
\rightarrow 0
\]
for all $k \in [1:K]$, there exist $\zeta_n \rightarrow 0$, $\delta_n \in \left(0, \frac{1}{2} - \frac{\Pr\{ \Phi^n_k(Y^n_k) \neq M_{S_k}\}}{2} \right)$ satisfying $\delta_n \rightarrow 0$, and a partitioning index $Q$ of $\mcf{M}_{[1:J]}\times A$ w.r.t. $P_{M_{[1:J]}, X^n}$, over an index set $\mcf{Q}$ with cardinality no larger than $\Gamma n^{5}$ for some $\Gamma>0$, such that 
\begin{align*}
& \hspace*{-20pt}
P_{Q}\left(\set{ q \in \mcf{Q} : \aexp{\mcf{M}_{S_k \cup \bar S_k}(q)} - \aexp{\mcf{M}_{\bar S_k}(q) }  \leq \frac{1}{n} I(M_{S_k};Y^n_k |M_{\bar S_k},Q =q) + \zeta_n } \right) 
\\
&\geq 1  - \Pr\{ \Phi^n_k(Y^n_k) \neq M_{S_k}\} - \delta_n
\end{align*}
for all $\bar S_k \subseteq [1:J] \setminus S_K$ and $k\in [1:K]$, where
$\mcf{M}_{S_k \cup \bar S_k}(q) \defn \{ (m_{S_k},m_{\bar S_k}) \in \mcf{M}_{S_k \cup \bar S_k} : P_{M_{S_k}, M_{\bar S_k},Q}(m_{S_k},m_{\bar S_k},q) >0 \}$ and $\mcf{M}_{\bar S_k}(q) \defn \{ m_{\bar S_k} \in \mcf{M}_{\bar S_k} : P_{M_{\bar S_k},Q}(m_{\bar S_k},q) >0 \}$. Moreover, it is also true that $Q \markov X^n \markov Y^n_k$ for all $k\in [1:K]$.
Note that the same $\Gamma$, $\zeta_n$, and $\delta_n$ apply uniformly for all
$\mcf{M}_{[1:J]} \times A \subseteq \mcf{M}_{[1:J]}\times\mcf{X}^n$. 
\end{theorem}
\begin{IEEEproof}
It suffices to establish the special case where $\bar S_k = \emptyset$ for all $k \in [1:K]$ using Theorem~\ref{thm:maxerr_fano}. The general case follows immediately by using Corollary~\ref{cor:conditional} instead.

Set $\alpha_n = \frac{1-\Pr\{ \Phi^n_k(Y^n_k) \neq M_{S_k}\}}{\logt n}$.
 Since $-\frac{1}{n} \logt (1-\Pr\{ \Phi^n_k(Y^n_k) \neq M_{S_k}\}) \rightarrow 0$, 
$\alpha_n \rightarrow 0$ but $\frac{-\logt\alpha_n}{n} \rightarrow 0$. 
Let $\mcf{T}$ denote the collection of all subsets of $[1:K]$. For each
  $T \in \mcf{T}$, define the following subset of $\mcf{X}^n \times M_{[1:J]}$:
\begin{align*}
\Omega_T &\defn \Big\{ (m_{[1:J]},x^n) \in \mcf{M}_{[1:J]} \times A:
  \Pr\left\{\Phi^n_k(Y^n_k) = m_{S_k} \middle| X^n =x^n, M_{[1:J]} = m_{[1:J]}
  \right\} \geq \alpha_n \text{ for all } k \in T, \\
& \hspace*{60pt}  \text{ and }
   \Pr\left\{\Phi^n_k(Y^n_k) = m_{S_k} \middle| X^n =x^n, M_{[1:J]} = m_{[1:J]}
  \right\} < \alpha_n
\text{ for all } k \notin T\Big\}.
\end{align*}
Consider a one-to-one mapping $\sigma: \mcf{T} \rightarrow \mcf{U} \defn [1:2^K]$. Define the random index $U$ over $\mcf{U}$ by setting $U=\sigma(T)$ if $(M_{[1:J]},X^n) \in \Omega_T$. Clearly $U$ is a partitioning index of $\mcf{M}_{[1:J]} \times A$ w.r.t. $P_{M_{[1:J]},X^n}$ with $\Omega_{U=\sigma(T)} =\Omega_T$. Further for each $k \in [1:K]$, define $\mcf{U}_k \defn \{ \sigma(T): k \in T \subseteq \mcf{T}\}$.
For every value $u \in \mcf{U}$ except for $\sigma^{-1}(u) = \emptyset$, consider the encoder $F^n$ restricted to $\Omega_{U=u}$ specified by $P_{M_{[1:J]},X^n | (M_{[1:J]},X^n) \in \Omega_{U=u}} = P_{M_{[1:J]},X^n |U=u}$, and the decoder $\Phi^n_k$ decoding only to the projection of $\Omega_{U=u}$ onto $\mcf{M}_{S_k}$ for each $k \in \sigma^{-1}(u)$. The maximum conditional error probability for these encoder-decoder pairs is $1-\alpha_n$. Now apply Theorem~\ref{thm:maxerr_fano} to obtain a random index $V(u)$, over an index set $\mcf{V}$ with cardinality of at most $\Gamma n^5$, and $\zeta_n \rightarrow 0$ such that
\begin{equation}\label{eq:use_maxerr_fano}
P_{V(u)|U=u} \left( \left\{v \in \mcf{V}: \aexp{\mcf{M}_{S_k}(u,v)} \leq \frac{1}{n} I(M_{S_k};Y_k^n|U=u,V(u) = v) + \zeta_n \text{ for all } k \in \sigma^{-1}(u) \right\}  \right) \geq 1 -\frac{1}{\abs{\mcf{X}}^n}
\end{equation}
where $\mcf{M}_{S_k}(u,v) \defn \left\{ m_{S_k} \in \mcf{M}_{S_k} : \Pr\{ M_{S_k} =m_{S_k}, U=u, V(u)=v\} > 0 \right\}$.
Define $\mcf{Q} = \mcf{U}\times \mcf{V}$ and $Q = (U,V(U)) \in \mcf{Q}$. 
Then, for each $k \in [1:K]$, we have 
\begin{align}
& \hspace*{-10pt} 
\Pr\{ \Phi^n_k(Y^n_k) = M_{S_k} \} 
\notag \\
&=
 \sum_{u \in \mcf{U}} \sum_{(m_{[1:J]},x^n) \in \Omega_{U=u}}
 \Pr\left\{\Phi^n_k(Y_k^n) = m_{S_k} \middle| X^n =x^n, M_{[1:J]} = m_{[1:J]}
  \right\}  P_{M_{[1:J]}, X^n}(m_{[1:J]},x^n)
\notag \\
&\leq 
\alpha_n +  \sum_{u \in \mcf{U}_k} P_{U}(u)
\notag \\
&\stackrel{(a)}{\leq} 
\alpha_n + \frac{1}{\abs{\mcf{X}}^n}
+ \sum_{u \in \mcf{U}_k} \Pr \left\{ U=u, V(u) \in \left\{ v \in \mcf{V} : \aexp{\mcf{M}_{S_k}(u,v)} \leq \frac{1}{n} I(Y_k^n;M_{S_k}|U=u,V(u) = v) + \zeta_n \right\} \right\}  \notag \\
&\leq \alpha_n + \frac{1}{\abs{\mcf{X}}^n}
+ P_Q \left( \left\{ q \in \mcf{Q} : \aexp{\mcf{M}_{S_k}(q)} \leq \frac{1}{n} I(M_{S_k};Y_k^n|Q=q) + \zeta_n \right\} \right)  
\label{eq:PnotUk}
\end{align}
where $(a)$ is due to \eqref{eq:use_maxerr_fano}.
Observe that $Q$ partitions $\mcf{M}_{[1:J]} \times A$ w.r.t $P_{M_{[1:J]},X^n}$ and $Q \markov X^n \markov Y^n_k$. Thus~\eqref{eq:PnotUk} gives the theorem.
\end{IEEEproof}

\begin{cor}\label{cor:uniform} \textbf{(Strong average-error Fano's
    inequality for nearly uniform messages)} 
If the encoder-decoder pair $(F^n,\Phi^n_k)$ in Theorem~\ref{thm:isfd} satisfies the more stringent average-error condition 
\[
\frac{\logt n}{n \left(1-\Pr\{ \Phi^n_k(Y^n_k) \neq M_{S_k}\} \right)}
\rightarrow 0
\]
and $M_{S_k}$ is $\gamma_n$-uniformly distributed with 
\[
\frac{\logt\gamma_n}{n(1-\Pr\{ \Phi^n_k(Y^n_k) \neq M_{S_k}\}) } \rightarrow 0,
\]
then there exist $\mcf{Q}_k^* \subseteq \mcf{Q}$ satisfying
\[
P_Q(\mcf{Q}_k^*) \geq \frac{1}{4} \left(1-\Pr\{ \Phi^n_k(Y^n_k) \neq M_{S_k}\} \right)
\]
and $\mu_n \rightarrow 0$
such that for all $\bar S_k \subseteq [1:J] \setminus S_k$
\begin{equation*}
\aexp{\mcf{M}_{S_k}} \leq \frac{1}{n}I(M_{S_k};Y_k^n |M_{\bar S_k}, Q=q)
+\mu_n,
\end{equation*}
for all $q \in \mcf{Q}_k^*$.
\end{cor}
\begin{IEEEproof}
Apply Theorem~\ref{thm:isfd} with 
\[
\mcf{Q}_k \defn \set{ q \in \mcf{Q} : \aexp{\mcf{M}_{S_k \cup \bar S_k}(q)} 
  - \aexp{\mcf{M}_{\bar S_k}(q)}\leq \frac{1}{n} I(M_{S_k};Y^n_k | M_{\bar S_k}, Q =q) + \zeta_n },
\]
we have $P_Q(\mcf{Q}_k) \geq \frac{1}{2}\Pr\{\Phi^n_k(Y^n_k) = M_{S_k}\}$ for all $k \in [1:K]$.
Set
\[
\delta_n \defn \frac{4\left(\logt \Gamma n^5+ \logt\gamma_n \right)}{n\Pr\{\Phi^n_k(Y^n_k) = M_{S_k}\}} \rightarrow 0,
\]
and define 
\[
\mcf{Q}^*_k \defn \set{ q \in \mcf{Q}_k : \aexp{\mcf{M}_{S_k}} \leq  
  \frac{1}{n} H(M_{S_k}| M_{\bar S_k},Q =q) + \delta_n }.
\]
Then
\begin{align*}
& \hspace*{-10pt}
\aexp{\mcf{Q}}
\notag \\
&\geq
\frac{1}{n} H(M_{S_k}|M_{\bar S_k}) -\frac{1}{n} H(M_{S_k} |M_{\bar S_k},Q) 
\notag \\
& \geq
\frac{1}{n} H(M_{S_k},M_{\bar S_k}) - \frac{1}{n} H(M_{\bar S_k})
  -\sum_{q \in \mcf{Q}_k} \frac{1}{n} H(M_{S_k} |M_{\bar S_k},Q=q) P_Q(q) - \aexp{\mcf{M}_{S_k}} \cdot \left[1 - P_Q(\mcf{Q}_k)  \right]
\notag \\
&\stackrel{\tiny{(a)}}{\geq} 
\aexp{\mcf{M}_{S_k \cup \bar S_k}} - \cexp{\gamma_n} - \aexp{\mcf{M}_{\bar S_k}}
  - \sum_{q \in \mcf{Q}_k} \frac{1}{n} H(M_{S_k} |M_{\bar S_k},Q=q) P_Q(q) - \aexp{\mcf{M}_{S_k}} \cdot \left[1 - P_Q(\mcf{Q}_k)  \right]
\notag \\
& =
\sum_{q \in \mcf{Q}_k} \left[ \aexp{\mcf{M}_{S_k}} - \frac{1}{n} H(M_{S_k} |M_{\bar S_k},Q=q) \right] P_Q(q)
  - \cexp{\gamma_n}
\notag \\
&\geq 
\delta_n \left[ P_Q(\mcf{Q}_k) - P_Q(\mcf{Q}_k^*) \right]   - \cexp{\gamma_n}
\end{align*}
where in $(a)$ results from Lemma~\ref{lem:htoa} as $M_{S_k}$ is $\gamma_n$-uniform. This implies that 
\begin{equation} \label{eq:uniform_base}
P_Q(\mcf{Q}_k^*)  
\geq  
P_Q(\mcf{Q}_k) - \frac{1}{n\delta_n} \left( \logt \Gamma n^5 + \logt\gamma_n\right)
= \frac{1}{4} \Pr\{\Phi^n_k(Y^n_k) = M_{S_k}\}.
\end{equation}
Moreover, for each $q \in  \mcf{Q}_k^*$,
\begin{align*} 
\aexp{\mcf{M}_{S_k}} 
& \leq
\frac{1}{n} H(M_{S_k} |M_{\bar S_k},Q=q) + \delta_n
\notag \\
& \leq
\frac{1}{n} H(M_{S_k}, M_{\bar S_k}|Q=q) - \frac{1}{n} H(M_{\bar S_k}|Q=q) + \delta_n
\notag \\
& \stackrel{(a)}{\leq}
\aexp{\mcf{M}_{S_k\cup \bar S_k}(q)} -\aexp{\mcf{M}_{\bar S_k}(q)} + 2\lambda_n 
+ \delta_n
\notag \\
& \leq 
\frac{1}{n} I(M_{S_k};Y^n_k |M_{\bar S_k}, Q =q) + \underbrace{\zeta_n  + 2\lambda_n
+\delta_n}_{\mu_n \rightarrow 0}
\end{align*}
where $(a)$ is due to~\eqref{eq:bmsk} and \eqref{eq:bmsk_H} with $\lambda_n \rightarrow 0$.
This together with~\eqref{eq:uniform_base} establish the corollary.
\end{IEEEproof}
\begin{remark}\label{rem:stable}
By Lemma~\ref{lem:htoa} as $M_{S_k}$ is $\gamma_n$-uniform, it is easy to show that
\begin{equation} \label{eq:info_stable}
1 - \frac{\cexp{\gamma_n}}{\aexp{\mcf{M}_{S_k}}} 
\leq
\frac{-\cexp{P_{M_{S_k}}(m_{S_k})}}{\frac{1}{n} H(M_{S_k})}
\leq 
1 + \frac{\frac{2}{n} \logt\gamma_n}{\aexp{\mcf{M}_{S_k}} - \cexp{\gamma_n}} 
\end{equation}
for all $m_{S_k} \in \mcf{M}_{S_k}$. Note that the condition on $\gamma_n$ in Corollary~\ref{cor:uniform} implies that $\cexp{\gamma_n} \rightarrow 0$. Thus we have from~\eqref{eq:info_stable} that $M_{S_k}$ is information stable~\cite{dobrushin1963general}, provided that $\aexp{\mcf{M}_{S_k}}> R$ for some $R>0$ (otherwise the corollary becomes trivial).

On the other hand, the converse is not true. However, if $M_{S_k}$ is information stable and $H(M_{S_k}) \leq n R$ for some $R>0$, then there is a $\mcf{\tilde M}_{S_k} \subseteq \mcf{M}_{S_k}$ with $P_{M_{S_k}}(\mcf{\tilde M}_{S_k} ) \rightarrow 1$ that $M_{S_k}$ is conditionally $\gamma_n$-uniform given $M_{S_k} \in \mcf{\tilde M}_{S_k}$, where $\gamma_n = 2^{2n\epsilon_n R}$ for some $\epsilon_n \rightarrow 0$. Thus further if $\frac{\epsilon_n}{1 - \Pr\{ \Phi^n_k(Y^n_k) \neq M_{S_k}\}} \rightarrow 0$, then Corollary~\ref{cor:uniform} is still applicable to the restricted message subset $\mcf{\tilde M}_{S_k}$. 
\end{remark}

\begin{cor}\label{cor:stable} \textbf{(Strong average-error Fano's
    inequality for information-stable messages)} 
If the encoder-decoder pair $(F^n,\Phi^n_k)$ in Theorem~\ref{thm:isfd} has average error
$\Pr\{ \Phi^n_k(Y^n_k) \neq M_{S_k}\} \leq \epsilon < 1$
and $M_{S_k}$ is information stable with $\aexp{\mcf{M}_{S_k}} \leq R$ for some $R>0$,
then there exist $\kappa_n \rightarrow 0$ and $\mcf{Q}_k^* \subseteq \mcf{Q}$ satisfying
$P_Q(\mcf{Q}_k^*) \geq \frac{1-\epsilon}{4} - \kappa_n$
such that for all $\bar S_k \subseteq [1:J] \setminus S_k$
\begin{equation*}
\frac{1}{n}H(M_{S_k}) \leq \frac{1}{n}I(M_{S_k};Y_k^n |M_{\bar S_k}, Q=q)
+\kappa_n,
\end{equation*}
for all $q \in \mcf{Q}_k^*$.
\end{cor}
\begin{IEEEproof}
Restrict to $\mcf{\tilde M}_{S_k}$ as defined in Remark~\ref{rem:stable}, and consider the message $\tilde M_{[1:J]}$ distributed according to $P_{M_{[1:J]}|M_{S_k} \in \mcf{\tilde M}_{S_k}}$ using the same encoder and decoders. Then the average error archived is
$\Pr\{ \Phi^n_k(Y^n_k) \neq M_{S_k} | M_{S_k} \in \mcf{\tilde M}_{S_k} \}
\leq \frac{\epsilon}{P_M(\mcf{\tilde M}_{S_k})}$,
where $P_M(\mcf{\tilde M}_{S_k}) \rightarrow 1$.
Hence by Remark~\ref{rem:stable} we may apply Corollary~\ref{cor:uniform} to obtain $\mu_n$, $Q$, $\mcf{Q}^*_k$ with $P_Q(\mcf{Q}^*_k) \geq \frac{1}{4} \left(1-\frac{\epsilon}{P_M(\mcf{\tilde M}_{S_k})} \right)$, and
\begin{equation} \label{eq:last}
\aexp{\mcf{\tilde M}_{S_k}} \leq \frac{1}{n}I(M_{S_k};Y_k^n |M_{\bar S_k}, M_{S_k} \in \mcf{\tilde M}_{S_k}, Q=q) + \mu_n
\end{equation}
for all $q \in \mcf{Q}^*_k$.
But by Lemma~\ref{lem:quick_fix}, we have
\[
\frac{1}{n} H(M_{S_k}) 
\leq  \frac{1}{n} H(M_{S_k}| M_{S_k} \in \mcf{\tilde M}_{S_k}) + \frac{1}{n} + \left[1- P_M(\mcf{\tilde M}_{S_k}) \right]R
  \leq     \aexp{\mcf{\tilde M}_{S_k}}       + \frac{1}{n} + \left[1- P_M(\mcf{\tilde M}_{S_k}) \right]R               
\]
and 
\[
\frac{1}{n}I(M_{S_k};Y_k^n |M_{\bar S_k},Q=q)
\geq  \frac{1}{n} I(M_{S_k}|M_{\bar S_k}, M_{S_k} \in \mcf{\tilde M}_{S_k}, Q=q) - \frac{1}{n} - \left[1- P_M(\mcf{\tilde M}_{S_k}) \right]R.
\]
Putting these back into~\eqref{eq:last}, we obtain the corollary with 
\[
\kappa_n \defn \max \left\{\mu_n + \frac{2}{n} + 2\left[1- P_M(\mcf{\tilde M}_{S_k}) \right]R, \, \frac{\epsilon}{4} \left( \frac{1}{P_M(\mcf{\tilde M}_{S_k})} -1 \right) \right\} \rightarrow 0.
\]
\end{IEEEproof}

\section{Concluding remarks} \label{sec:conclusion}

We constructed a solution to the image size characterization problem in a multi-message, multi-terminal DMC environment. The solution is referred to as equal-image-size source partitioning, in which a source set is partitioned into at most polynomially many subsets. 
Over every partitioning subset, the exponential orders of the minimum image sizes for nearly all messages are equal to the same entropy quantity and all these messages are nearly uniformly distributed. While we believe the method of equal-image-size source partitioning has many applications, we first used it to establish new necessary conditions for reliable communications over multi-terminal DMCs under the maximum and average decoding error criteria, respectively. These necessary conditions were specialized to give stronger, but still easy to use, versions of Fano's inequality that can be directly used on codes with non-vanishing decoding error probabilities.   

The strong versions of Fano's inequality immediately apply to proving strong converses of coding theorems for many multi-terminals DMCs. We gave an application example showing how the strong converse of the general DM-WTC with decaying leakage can be readily obtained using our results. While this example alone, as the strong converse of the general DM-WTC with decaying leakage had been an open problem, might be enough to justify the development of this new tool, it hardly did justice in illustrating the real powerfulness of our results. Revisiting the example again, one would recognize that the DM-WTC did not present a ``true'' multi-terminal problem because only the legitimate receiver was imposed upon with a decoding error constraint. The constraint imposed upon the wiretapper was, on the other hand, the amount of information leakage. This constraint was easily taken care of due to the polynomial number of partitioning subsets. In order to demonstrate the  power of the tool of equal-image-size source partitioning in  ``true'' multi-terminal communications as well as joint source-channel coding scenarios, in a direct sequel to this paper we will use our results here to characterize the $\epsilon$-transmissible regions~\cite[Section~3.8]{han2003} of the following DMCs: 
\begin{itemize}
\item the degraded broadcast channel,
\item the multiple access channel\footnote{For independent information stable sources},
\item the composite channel, and
\item the wiretap channel with non-decaying leakage.
\end{itemize} 
At best only partial results about the $\epsilon$-transmissible regions are currently available for these channels. We expect the results in this paper will help addressing many open issues in the characterization of the $\epsilon$-transmissible regions of these channels.

\bibliographystyle{ieeetr}
\bibliography{this} 

\end{document}